\documentclass[a4paper,11pt]{article}
\usepackage{caption}
\usepackage{float}

\renewcommand{\thetable}{\Roman{table}}

\pdfoutput=1 

\usepackage{jheppub} 
\usepackage[T1]{fontenc} 
\usepackage{comment}
\usepackage{lineno,hyperref}
\usepackage{subcaption}
\usepackage{soul}
\usepackage[normalem]{ulem}
\usepackage{makecell}
\preprint{MPP-2026-126}

\title{\boldmath The VORTEX cavity for the RADES axion haloscope}

\author[*,a,b]{Jose María Garc\'ia-Barcel\'o,}
\author[b]{Cristian Cogollos,}
\author[b]{Babette Döbrich,}
\author[b]{David Kittlinger,}
\author[b]{Diego Nicolas Lobato-De La Cruz,}
\author[c]{and Walter Wuensch}

\affiliation[a]{Center for Astroparticles and High Energy Physics (CAPA), Universidad de Zaragoza, 50009 Zaragoza, Spain}
\affiliation[b]{Max-Planck-Institut f\"{u}r Physik (Werner-Heisenberg-Institut), Boltzmannstraße 8, 85748 Garching, Germany}
\affiliation[c]{CERN - European Organization for Nuclear Research, Geneva, Switzerland}

\affiliation[*]{Corresponding author}

\emailAdd{josemaria.garcia@unizar.es}

\abstract
{One of the major challenges in axion dark matter haloscope searches is a loss-less tuning mechanism that is able to cover a significant frequency range around the haloscope's central frequency. In this article, we report on the implementation and performance of an axion haloscope dubbed 
Vertical-cut Optimised Resonant Tunable cavity for dark matter EXploration (VORTEX)
centred at $8.5$~GHz with a tuning range of around $800$~MHz. The performance of this setup is measured at temperatures in the mK range and compared to simulation. In addition, we test the cavity-mode structure of this cavity type directly via the `bead-pull method' and observe satisfactory agreement with expectations. The arrangement is poised to be used in an upcoming RADES (Relic Axion Detection Exploratory Setup) data-taking campaign employing a $12$~T solenoid magnet.}

\begin{document}

\maketitle
\flushbottom

\section{Introduction}
\label{s:Introduction}

The axion is a pseudoscalar particle that appears as part of the solution to the strong CP problem of quantum chromodynamics (QCD)~\cite{Peccei:1977hh,Peccei:1977ur,Weinberg:1977ma,Wilczek:1977pj}. Axions are expected to be ultralight feebly interacting particles that feature production mechanisms that could create them abundantly in the early universe, see e.g. \cite{OHare:2024nmr}.
This makes them ideal candidates for dark matter~\cite{Preskill:1982cy,Abbott:1982af,Dine:1982ah}. 

Unfortunately, as of today, the axion remains hypothetical; on the bright side, attempts to discover it are becoming more numerous, more versatile and more sensitive. A number of recent introductory texts on axions \cite{Irastorza:2021tdu,DiLuzio:2020wdo,Sikivie:2024isv,Yu:2023gdq} and their current experimental landscape \cite{Irastorza:2018dyq,Adams:2022pbo,Baryakhtar:2025jwh,Dobrich:2025oso} provide topical context to understand the boost in worldwide searches for axions.

\begin{figure} [htb]
    \centering
    \includegraphics[width=0.65\textwidth]{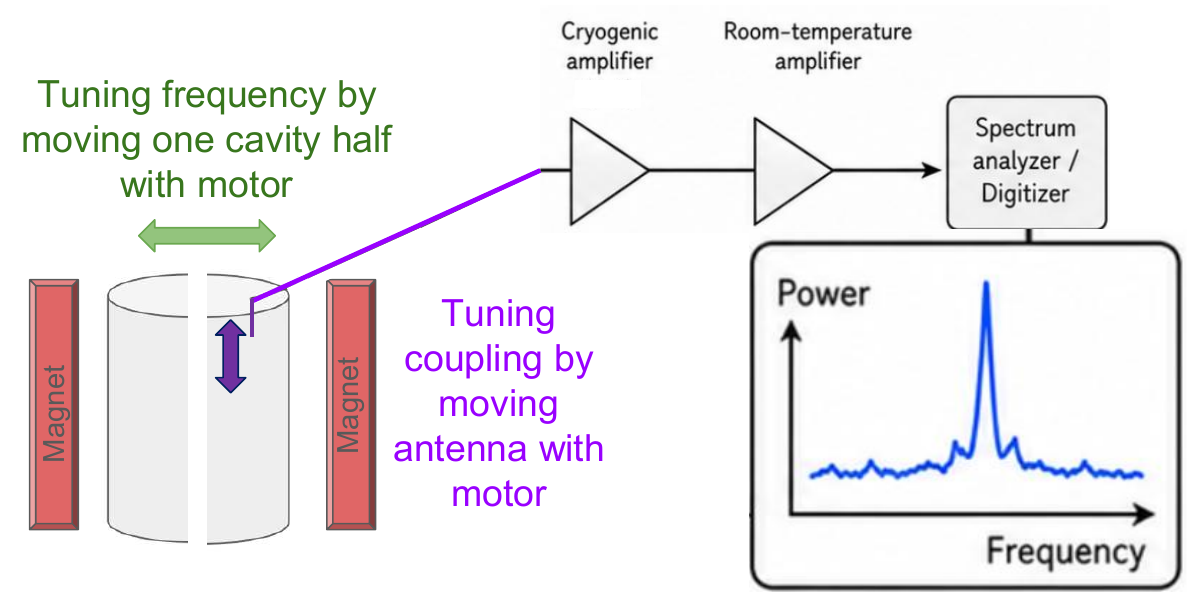}
    \caption{Sketch of the haloscope setup discussed in this paper: A microwave cavity (grey cylinder) is embedded between the coils of a magnet (red). Tuning of the cavity resonant frequency here is achieved by varying the distance of the cavity halves (green), while the antenna coupling can be varied through another motor (pink). The signal is read out through a chain of cryogenic and room-temperature amplifiers, connected to a DAQ. The observable axion-induced power peak corresponding to its mass will typically only be observed after some integration time and signal processing/analysis.}
    \label{fig:sketch_cyc}
\end{figure}

The most established technique to search for the axion as dark matter experimentally is the axion `haloscope'. Its idea was first proposed by Pierre Sikivie in 1983 \cite{Sikivie:1983ip}. Sikivie suggested that converting axions into photons in the presence of an external magnetic field, via the ``Primakoff effect'' would be a promising experimental strategy.
In a haloscope experiment, an electromagnetic resonator is placed inside a strong magnetic field, which enhances the conversion of relic axions coupled to an appropriate mode of the resonator into detectable microwave photons. The tiny signal with a comparably narrow width $Q_a=10^{6}$, must then be amplified in several stages (conceptually drawn in Figure~\ref{fig:sketch_cyc}). Through established filtering and analysis techniques, the axion particle could eventually be revealed. This approach targets axions with masses around the $\mu$eV scale and larger.

The smallest coupling of axions to photons $g_{a\gamma\gamma}$ at a given mass $m_a$ for a certain signal-to-noise ratio $S/N$ that can be probed,
\begin{equation}
\label{eq:ga}
    g_{a\gamma\gamma} \, = \,  \left(\frac{\frac{S}{N} \, k_B \, T_\mathrm{sys} \, \left(1+\beta\right)^2}{\rho_a \, C \, V \, \beta \, Q_0}\right)^{\frac{1}{2}}\frac{1}{B_e}\left(\frac{m_a^3}{Q_a \, \Delta t}\right)^{\frac{1}{4}}, \text{provided that } Q_a \gg \frac{Q_0}{1+\beta},
\end{equation}
depends on a set of experimental parameters:
the biggest lever-arm is the external magnetic field $B_e$. Further, such experiments require large cavity quality factors $Q_0$, large volumes $V$ and low system temperatures $T_\mathrm{sys}$. Note that the latter has contributions from the cavity physical temperature itself but also depends on the effective temperature of the amplifier chain.
Other parameters that experimentalists can control are the cavity coupling $\beta$ and the time spent at each frequency step $\Delta t$. The two remaining factors appearing in the equation are $\rho_a$, the local density of relic axions, and the Boltzmann constant $k_B$.

Another, arguably more useful way to assess the experimental reach (since the axion mass is a priori unknown) is via the scan rate:
\begin{equation}
\label{eq:dmadt}
    \frac{dm_a}{dt} = Q_aQ_0\frac{\beta^2}{\left(1+\beta\right)^3} \, g_{a\gamma\gamma}^4\left(\frac{\rho_a}{m_a}\right)^2B_e^4C^2V^2\left(\frac{S}{N}k_BT_\mathrm{sys}\right)^{-2} \ ,
\end{equation}
where $C$ is the form factor, which depends on the electric field of the resonant mode and the external magnetic field, a more detailed treatment can be found at the end of this section.

Thus, for the analysis of the cavity performance, the following figure of merit is studied:
\begin{equation}
\label{eq:FoM}
FoM = Q_0V^2C^2.
\end{equation}

After the pioneering efforts, particularly by ADMX, the past few years have seen significant advancements in haloscope experiments, driven by improved technology and a vastly increasing interest in axion dark matter. ADMX has successfully excluded axion-photon couplings in the $2.66-4.2$~$\mu$eV range at sensitivity down to the DFSZ benchmark~\cite{Du:2018uak,Braine:2019fqb,ADMX:2021nhd}, and then extended to $5.41$~$\mu$eV~\cite{ADMX:2025vom}. More recently, CAPP in Korea has emerged as a strong player \cite{CAPP:2023pace,CAPP:2023capp18t,CAPP:2024max} in this region of the parameter space.

At higher frequencies, above 10s of GHz, conventional axion haloscopes suffer a reduction in sensitivity. Most experimental efforts in this context target compensating for smaller cavity volumes by enhancing other experimental parameters (more intense magnetic fields, very high $Q_0$ factors, noise reduction, superconducting coatings to increase $Q_0$~\cite{Alesini:2019ajt,Ahyoune:2024klt}, higher-order modes~\cite{Boutan:2018uoc}, and different versions of multi-cavity systems~\cite{Melcon:2020xvj,CAST:2020rlf,Jeong:2020cwz}).

A particular challenge in high-frequency haloscopes is the implementation of loss-less but wide-range tuning: the most commonly used tuning system is the insertion and movement of a rod inside the resonant cavity, which can, however, reduce the quality factor. This is most true in systems employing High-Temperature Superconducting (HTS) coating or tapes to improve the quality factor, due to the large $Q_0$ factor that can be obtained in such systems.

For this reason, another route to tune the cavity was proposed in the Relic Axion Detection Exploratory Setup (RADES) \cite{Melcon:2018dba,Melcon:2020xvj,Ahyoune:2024klt}. In this case, tuning is achieved 
with a technique where the haloscope radius is varied by mechanically moving the cavity halves split symmetrically, as first developed in \cite{ArguedasCuendis:2019swy} and later studied in \cite{Golm:2023iwe,Garcia-Barcelo:2025hci} by the same collaboration. It is sketched in Figure~\ref{fig:sketch_cyc}: as the cavity is frequency-tuned through a motor (sketched in green), a readjustment of the antenna is also needed if optimal coupling should be maintained (sketched in pink).
A similar technique, the ``clamshell technique'' was employed in \cite{Braggio:2023gnd} but with an opening procedure that is asymmetric.
Such a procedure was recently also tested successfully at $4$~K temperatures in \cite{Maiello:2026cxu} employing a Nb$_{3}$Sn superconducting cavity but without a sliding mechanism system for physics data taking at a mK ambient. Finally, a related idea to drive a wedge to tune a thin-shell resonator is discussed in \cite{Dyson:2024vhx}.

The use of the proposed tuning mechanism with a vertical cut avoids the insertion of additional elements inside the cavity, and therefore the reduction of $Q_0$ with tuning only varies relative to the gap for any given electrical conductivity. This work reports a pioneering mechanical implementation of a continuous sliding system and the performance of such a setup at room temperature, at $4$~K temperature, and in a mK environment for the first time. Also, the implementation of an antenna tuning system to control the $\beta$ parameter has been conducted.

In Eq.~\ref{eq:dmadt}, the form factor $C$ was introduced; this parameter is defined as the overlap of the electric field of a single resonant mode and the external magnetic field, integrated over the whole volume of the resonator
\begin{equation}
\label{eq:C}
    C \, = \, \frac{|\int _V \, \vec{E} \cdot \vec{B}_e \, dV|^2}{\int_V \, |\vec{B}_e|^2 \, dV \int_V \, \varepsilon \, |\vec{E}|^2 \, dV},
\end{equation}
where $\varepsilon$ represents the relative permittivity of the medium filling the cavity. The form factor quantifies the coherence between the resonant mode and the electric field generated by axion conversion inside the cavity.

In axion search experiments, the value of the form factor is often extracted by taking into account only simulations of the resonant modes. However, in this work, a method to measure this value empirically was implemented. This measurement allows experimental verification of the value obtained in simulations and a better characterisation of the experimental setup\footnote{We stress the even more increased relevance of such measurements for dielectric haloscopes; see, e.g. \cite{MADMAX:2026dsh}.}.
In this regard, it will be shown in this work that another advantage of a split-tuning technique is that the electrical field inside the cavity can be experimentally measured easily: with closed cavities, measuring their real-life performance requires small frequency-cut-off holes that must be drilled in the housing of the cavity through which the thread is passed. An open cavity can be probed directly through a bead-pull method, as reported in this paper.

The paper is structured as follows:
in Section~\ref{s:CavityProperties}, the principle of the vertical-cut tuning is reviewed, including considerations on manufacturing and the effects of misalignment. The concept of the two cryogenic tuning systems employed here is shown: the tuning of the cavity by adjusting the distance of the housing halves, as well as the movement of the antenna.
In Section~\ref{s:Cavity_measurements}, the measurement results in different ambient conditions are reported.
Section~\ref{s:Bead-pull_analysis} is devoted to verifying experimentally the simulated mode structure: this constitutes one of the few instances in which such a comparison is available in axion search literature (see \cite{Egge:2023cos,Vora:2025,Lewis:2024dyn} for other instances). To conclude, in Section~\ref{s:Conclusions}, prospects of the sensitivity reach for the full setup are given.

\section{Tunable cavity design and implementation}
\label{s:CavityProperties}

\subsection{Cavity design}
\label{ss:CylindricalCavityDesign}

\begin{figure} [htb]
    \centering
    \includegraphics[width=0.8\textwidth]{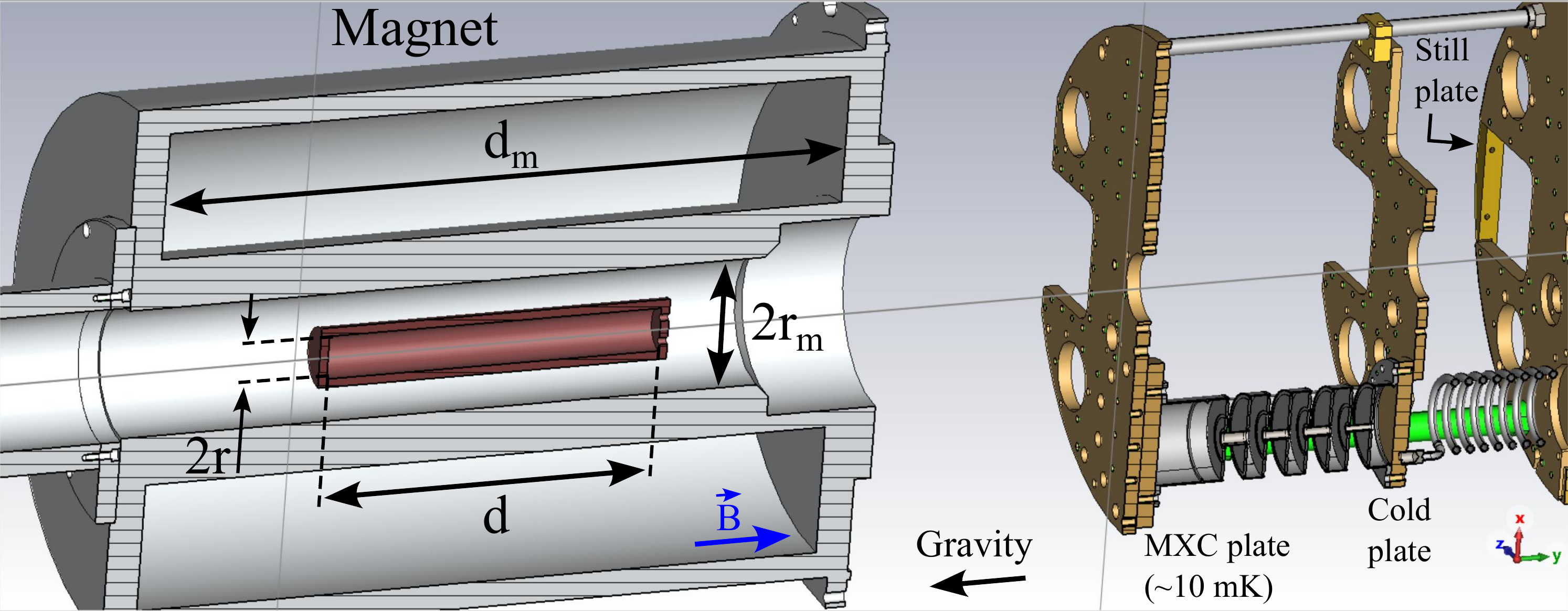}
    \caption{Longitudinal cut of the 3D model of the dilution refrigerator system comparing the size of the cavity (brown colour) with the $12$~T solenoid magnet (grey colour). On the right side (top part, considering the direction of gravity), in golden colour, the thermalisation plates of the dilution refrigerator system can be observed, with the last one to which the cavity is attached. It has a thermal link (not shown here) for arriving at mK temperatures.}
    \label{fig:CavityDesignAtBF}
\end{figure}

In this work, the haloscope Vertical-cut Optimised Resonant Tunable cavity for dark matter EXploration (VORTEX) was developed, which focuses on exploiting a maximum of the available space of the magnet installed in a Bluefors LD-250 dilution fridge (see Figure~\ref{fig:CavityDesignAtBF}) and covering a spectral range around $8.5$~GHz to search for dark matter axions.
For this purpose, an elongated cylindrical cavity operating with the $TM_{010}$ resonant mode at a frequency of $f_r=9$~GHz has been designed, giving internal radius and length values of $r = 12.75$~mm and $d = 200$~mm\footnote{In principle, a longer cavity could be employed, but less weight was desired for these prototypes to have a more conservatively chosen load for the piezoelectrics of the tuning system.}, respectively. This mode of operation is commonly used in cylindrical haloscopes installed in solenoid magnets, as it provides the best match between the cavity electric field and the magnetic field of the magnet, thereby maximising the form factor (see Eq.~\ref{eq:C}). The space where the cavity can be installed at mK temperatures is limited by the solenoid magnet bore, which is based on a radius and length of $r_m=38$~mm and $d_m=410$~mm, respectively. As can be seen in Figure~\ref{fig:CavityDesignAtBF}, the cavity is installed longitudinally inside the magnet, making its $TM_{010}$ electric field ($\vec{E}_\mathrm{cav}$) parallel to the external static magnetic field ($\vec{B_e}$) of the solenoid. These dimensions allow for the installation of auxiliary equipment such as tuning mechanisms and temperature sensors.

Unloaded quality and form factor values of $Q_0^{cryo}=72850$\footnote{These values, as well as the one indicated for the form factor and quality factor at room temperature, have been obtained after implementing an inner radius of $2$~mm in the upper and lower corner of the cylindrical cavity, which is necessary for the manufacturing stage.} (with an electrical conductivity of $\sigma_c = 10^9$~S/m, this is copper at cryo) and $C=0.682$, respectively, are obtained for the $TM_{010}$ mode with this cylindrical cavity (considering no tuning for the moment), from simulations employing Computer Simulation Technology (CST) Studio Suite software \cite{CST}. Then, since the cavity volume is $V=103$~mL, making use of Eq.~\ref{eq:FoM}, a figure of merit of $FoM = 359$~L$^2$ is obtained. For room temperature copper conductivity ($\sigma_c = 5.8\times10^7$~S/m) a value of $Q_0^{RT} = 17550$ is obtained\footnote{For cylindrical cavities of shorter lengths (where $r$ is similar to $d$), the quality factor is usually half these values \cite{Garcia-Barcelo:2023iri}.}.

\begin{figure} [htb]
    \centering
    \includegraphics[width=0.95\textwidth]{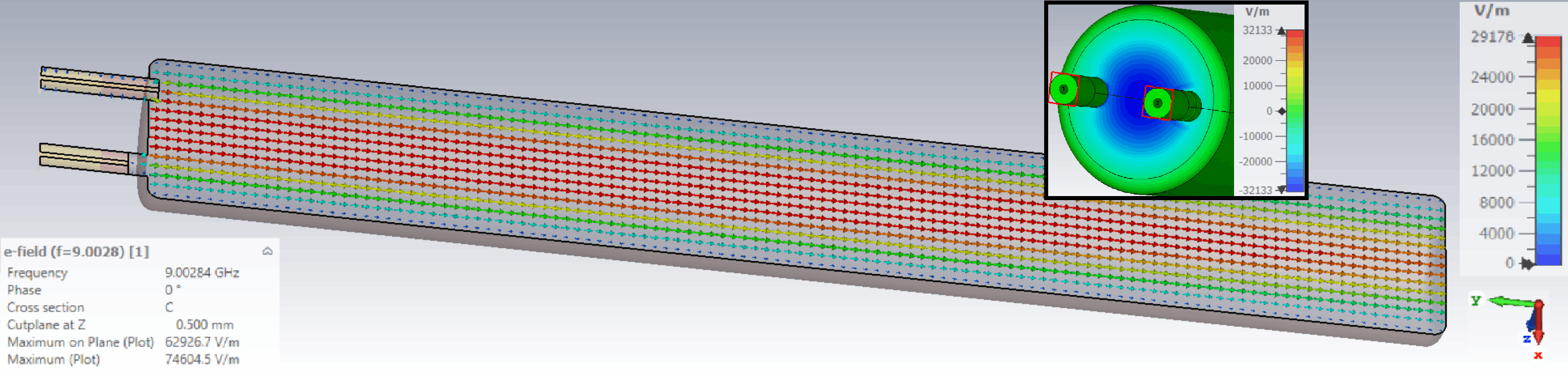}
    \caption{Cavity electric field for the $TM_{010}$ resonant mode  on the vertical cut plane. The inset shows the longitudinal component of the electric field ($E_y$) in the upper region of the cavity, verifying that it is the right mode: E-field pattern with no variation in the angular axis, with one maximum in the radial axis, and no variation in the longitudinal axis (so, 0-1-0). Furthermore, it has been observed that the longitudinal component of the magnetic field is $H_y = 0$, so it can be confirmed that it belongs to the TM mode family.}
    \label{fig:CylCav_TM010modeEfield_WithInset}
\end{figure}

This haloscope employs a vertical-cut mechanical tuning system (a technique previously used by the RADES collaboration for a multicavity structure in \cite{Golm:2023iwe}), whose details are given in the following section. The housing is manufactured from two parts, separated symmetrically in the longitudinal direction. Separating the cavity along the longitudinal symmetry plane means that power is not coupled into the resulting gap, thus maintaining the quality factor until a limit (see the $TM_{010}$ mode electric field lines in Figure~\ref{fig:CylCav_TM010modeEfield_WithInset}, which are parallel to the cavity cut).

Two ports are positioned at the top (longitudinal axis) of the cylindrical cavity, as space allows. This facilitates access to the radiofrequency (RF) lines of the dilution fridge system. One of the ports is intended for extracting the axion induced signal, and hence it is adjusted to be critically coupled for maximising Eq.~\ref{eq:ga}. The second one is weakly coupled and used for characterising the cavity via transmission measurements\footnote{This second port was also used to test a signal injection scheme similar to the one described in \cite{Zhu:2022cuw}.}.

\subsection{Vertical cut frequency tuning}
\label{ss:Vertical_cut_frequency_tuning}

\begin{figure}[h]
\centering
\begin{subfigure}[b]{0.7\textwidth}
         \centering
         \includegraphics[width=1\textwidth]{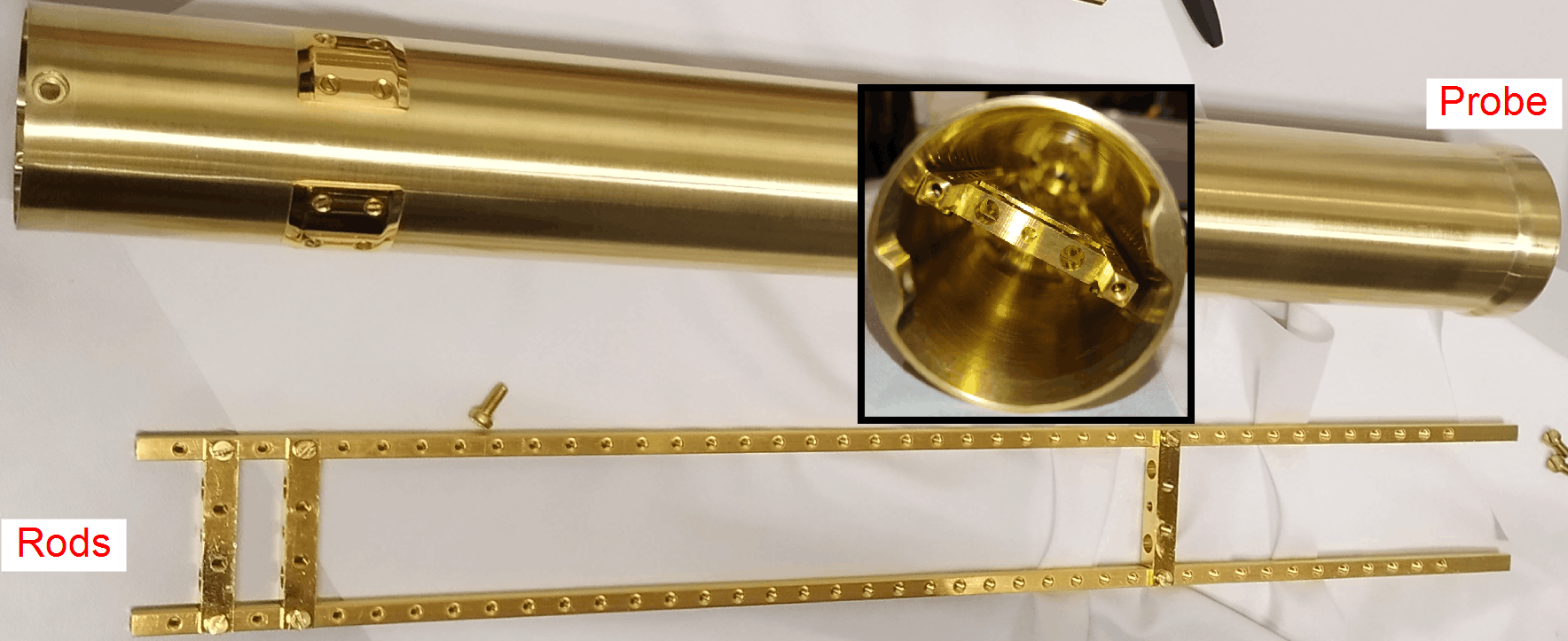}
         \caption{}
         \label{fig:ProbeAndRods}
\end{subfigure}
\hfill
\begin{subfigure}[b]{0.66\textwidth}
         \centering
         \includegraphics[width=1\textwidth]{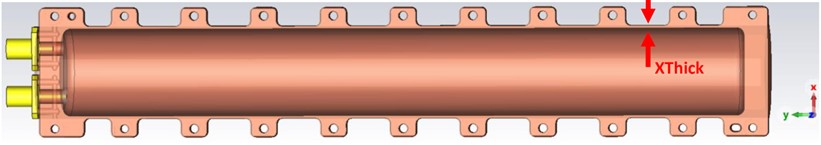}
         \caption{}
         \label{fig:CylCavHousingDesign_XThick}
\end{subfigure}
\hfill
\begin{subfigure}[b]{0.33\textwidth}
         \centering
         \includegraphics[width=1\textwidth]{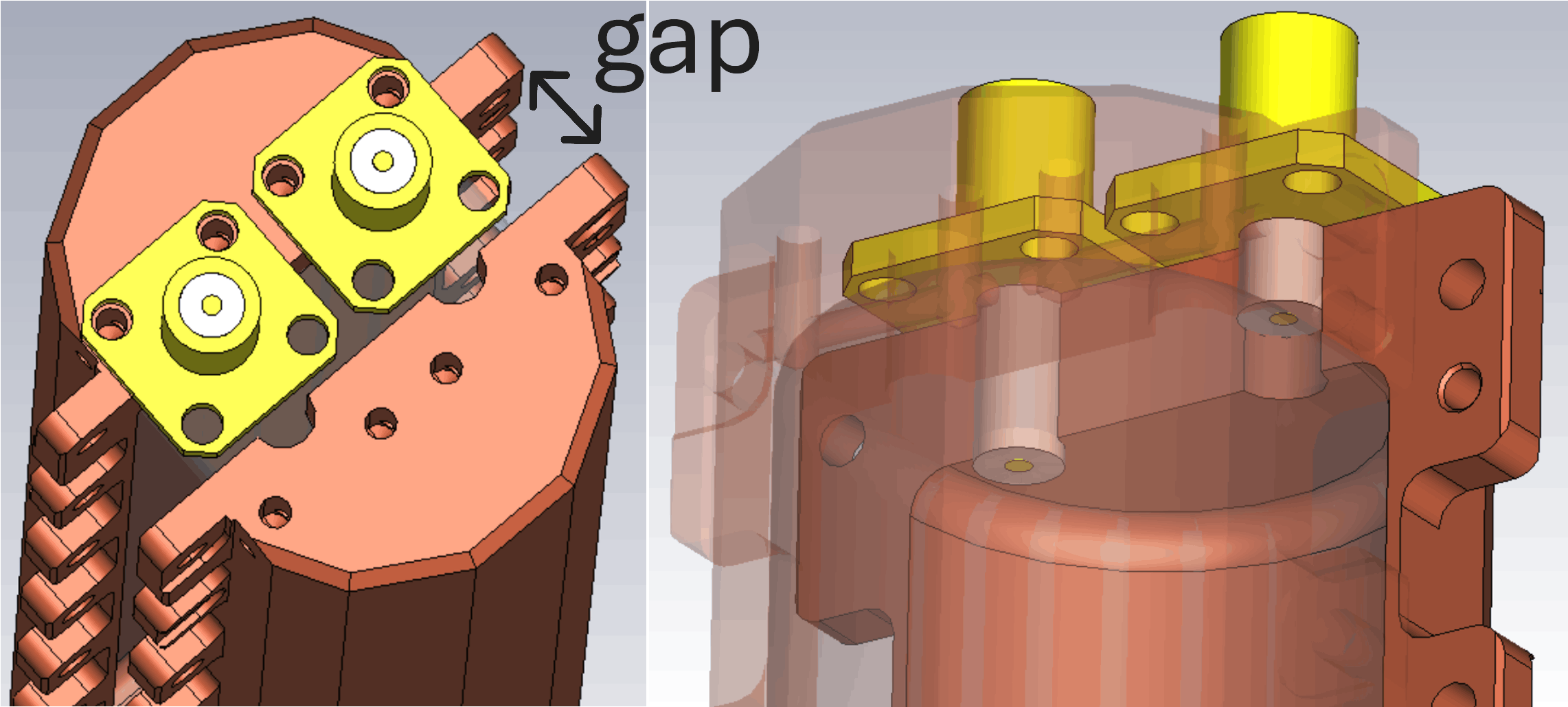}
         \caption{}
         \label{fig:HousingDesign_TopDetails}
\end{subfigure}
\caption{(a) Rods and probe of the FSE system where the cavity can be installed. The inset shows a picture of the rods inserted inside the probe (which limits the experimental space diameter to $53$~mm). (b) 3D model of one-half of the cavity housing design. The screw holes are prepared for installation at the rods. (c) Details at the cavity top showing the two port positions (coaxial SMA panels, in golden colour) and how the tuning gap is implemented.}
\label{fig:Rods_And_CylCavHousingDesign}
\end{figure}

The first step in the system design was to analyse the available space. A housing was designed to fit the cavity in the dilution refrigerator system experimental space, considering the requirements for the vertical cut movement system. In addition to the magnet dimension limitations mentioned in the previous section, a mechanism in this dilution refrigerator called Fast System Exchange (FSE), also known as bottom-loading probe, has been employed where the experimental space is even more limited in the radial axis ($r_\mathrm{FSE}=26.5$~mm, instead of the $38$~mm of the magnet) but allows the insertion/extraction and re-thermalisation (at mK) of the cavity in a much faster and simpler way compared to a standard dilution refrigerator without it. Therefore, the housing design of both cavity halves has been made according to these requirements. In this system, there are two thermal rod links (ladders), shown in Figure~\ref{fig:ProbeAndRods}, to which the cavity can be attached.

Figure~\ref{fig:CylCavHousingDesign_XThick} shows the design of the cavity structure, where the thickness of the outer cavity wall, named $XThick$, has been optimised to minimise the radiative losses and, at the same time, to enhance the compactness of the experiment. Two different versions of the cavity modifying the $XThick$ parameter have been designed and produced for their comparison in measurements. The first version of the cylindrical cavity manufactured has $XThick = 0.75$~mm, to reduce material and thus the thermalisation time when placed in a cryostat. However, it has been seen in simulation that if that thickness is increased to $3$~mm, the unloaded quality factor for an open cavity can increase significantly, especially for large cavity gaps. For example, if it is wanted to implement a relatively large tuning of $700$~MHz, with an $XThick$ value of $0.75$~mm, a $Q_0$ degradation of $36$~$\%$ is obtained (in simulation), while for $3$~mm it is $11$~$\%$. Therefore, another version has been manufactured with the latter option. Although, as will be seen later, some simulation studies and measurements have been carried out for the version with $XThick = 0.75$~mm, the following sections will focus on a comprehensive simulation study of tuning and misalignment for the case that yields the most optimal result, this is, for $XThick = 3$~mm.

Figure~\ref{fig:HousingDesign_TopDetails} shows the gap between the two halves that is used for changing the cavity frequency. The maximum allowable gap depends on the reduction of quality and form factors, which are studied below. This gap limit could also be restricted for mechanical reasons or due to experimental space within the magnet bore, but it has been found that the aforementioned limit is more restrictive.

\subsubsection{Tuning simulation results}
\label{sss:Tuning simulation results}

The frequency tuning of this cavity is achieved by controlling the opening of the gap between the two halves. This technique impacts the quality factor value in two different ways. One is due to the gap without misalignment, and the other is due to misalignment. For the latter, as described in section~\ref{sss:Misalignment effects}, it is essential to maintain alignment between the two halves, as misalignment results in radiation into the gap, which reduces the unloaded quality ($Q_0$) and form ($C$) factors. For the first one, below is a series of plots that illustrate how cavity frequency and quality factor depend on the gap. For the tuning range parameter in $\%$, it has been computed as follows:
\begin{equation}\label{eq:Tuning}
\text{Tuning}\,[\%] = \frac{f_2 - f_1}{f_2}\times 100,
\end{equation}
where $f_2$ and $f_1$ represent the upper and lower limits of the frequency tuning range, respectively.

\begin{figure}[h]
\centering
\begin{subfigure}[b]{0.4\textwidth}
         \centering
         \includegraphics[width=1\textwidth]{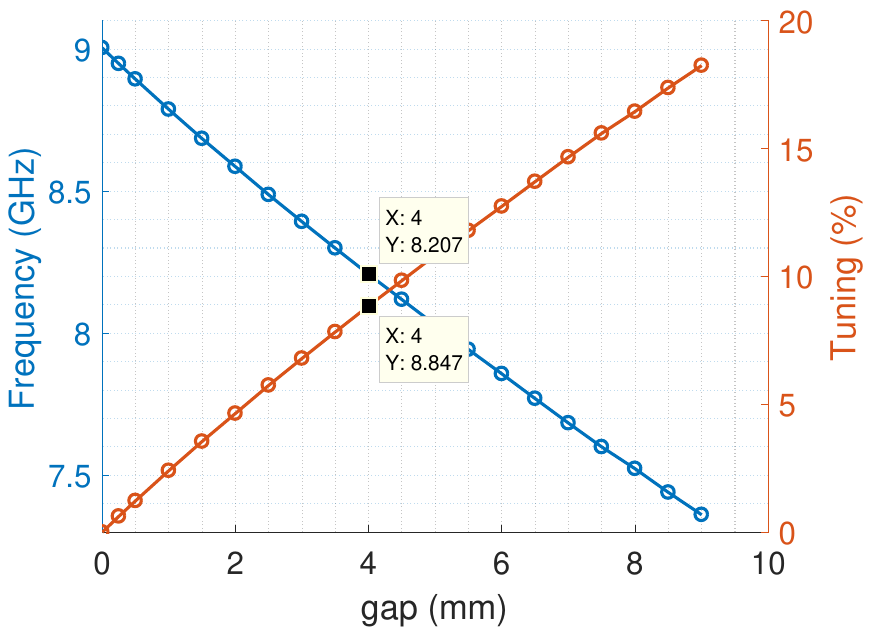}
         \caption{}
         \label{fig:CylCav2_fr_vs_Gap_CST}
\end{subfigure}
\begin{subfigure}[b]{0.4\textwidth}
         \centering
         \includegraphics[width=1\textwidth]{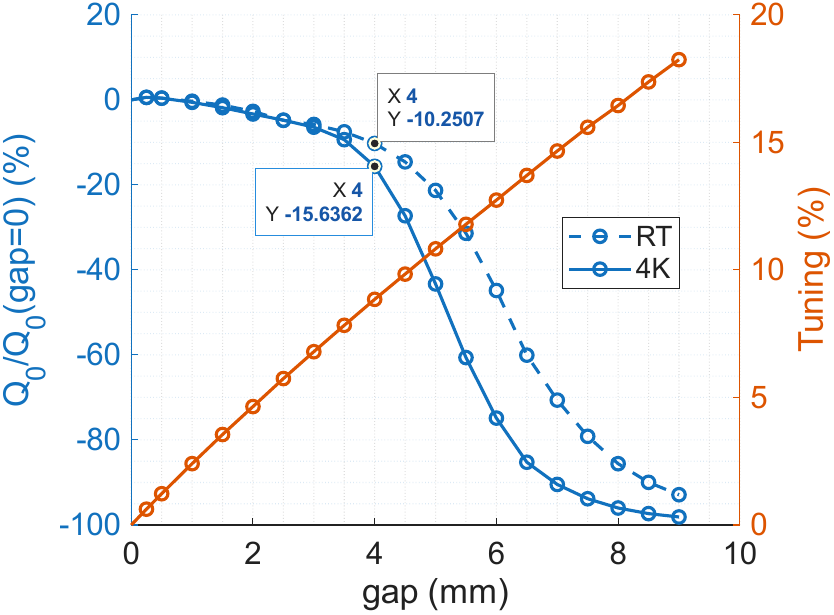}
         \caption{}
         \label{fig:CylCav2_Q0_vs_Gap_CST}
\end{subfigure}
\caption{CST simulation results opening the cylindrical cavity from $gap = 0$ to $9$~mm: (a) resonant frequency shift in GHz (blue line) and (b) unloaded quality factor decrease in $\%$ at room (blue dashed line) and cryo (blue solid line) temperatures for the $TM_{010}$ mode. For reference, the tuning range in $\%$ versus the gap is shown in each plot (red lines).}
\label{fig:CylCav2_fr_and_Q0_vs_Gap_CST}
\end{figure}

Figures~\ref{fig:CylCav2_fr_vs_Gap_CST} and \ref{fig:CylCav2_Q0_vs_Gap_CST} show the simulation results of frequency shift and quality factor reduction, respectively, as a function of gap when opening the cavity depicted in Figures~\ref{fig:CylCavHousingDesign_XThick} and \ref{fig:HousingDesign_TopDetails} (using $XThick = 3$~mm).
A gap opening of $gap=0$ to $4$~mm results in a tuning range of $8.85$~$\%$, $800$~MHz, from $9$ to $8.2$~GHz, with a maximum decrease of $15.6$~$\%$ in quality factor with copper conductivity at cryo temperatures. This range is marked on the plot. For larger gaps, the quality factor drops quickly, thus limiting the tuning range using this technique.

\begin{figure}[h]
\centering
\begin{subfigure}[b]{0.32\textwidth}
         \centering
         \includegraphics[width=1\textwidth]{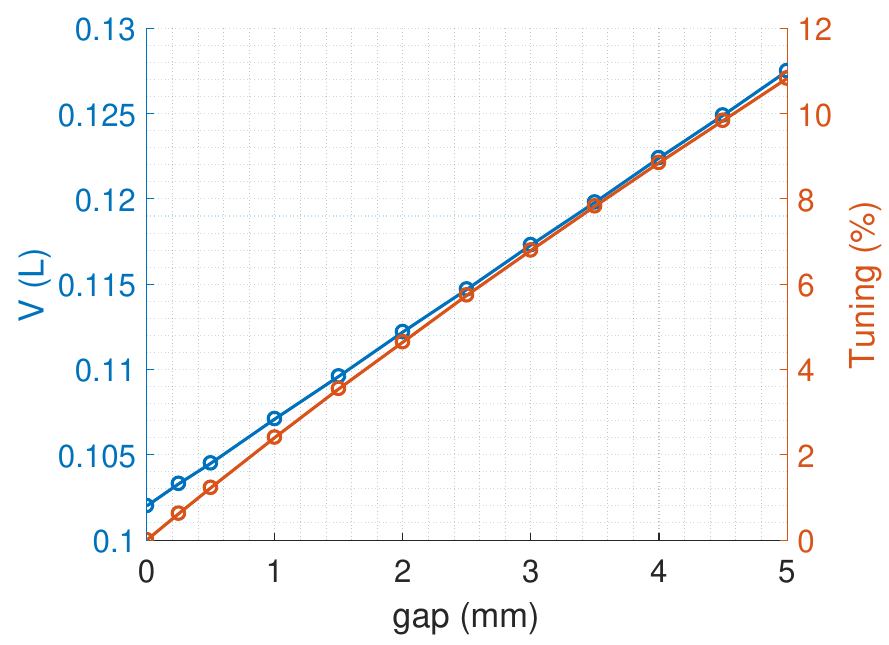}
         \caption{}
         \label{fig:CylCav2_V_vs_Gap_CST}
\end{subfigure}
\hfill
\begin{subfigure}[b]{0.32\textwidth}
         \centering
         \includegraphics[width=1\textwidth]{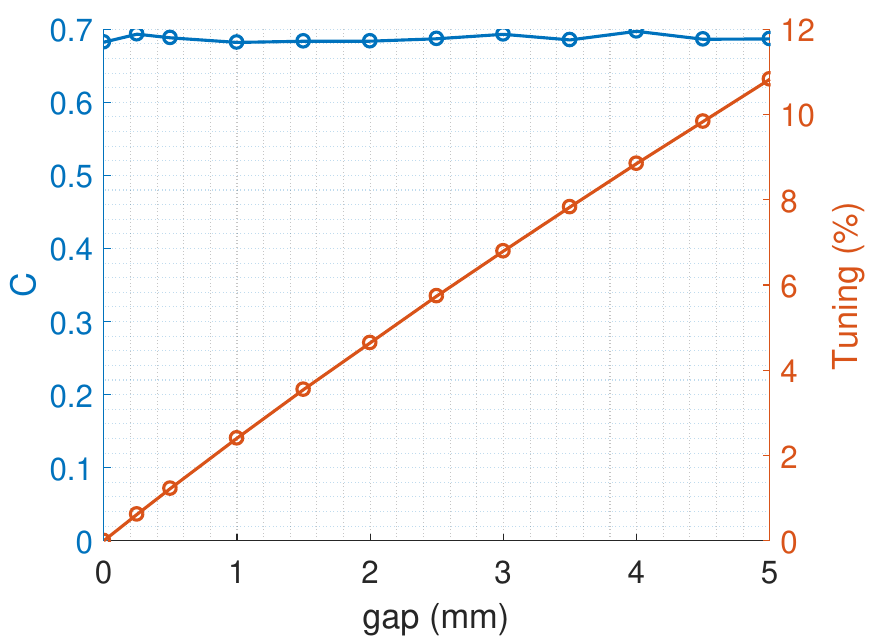}
         \caption{}
         \label{fig:CylCav2_C_vs_Gap_CST}
\end{subfigure}
\hfill
\begin{subfigure}[b]{0.32\textwidth}
         \centering
         \includegraphics[width=1\textwidth]{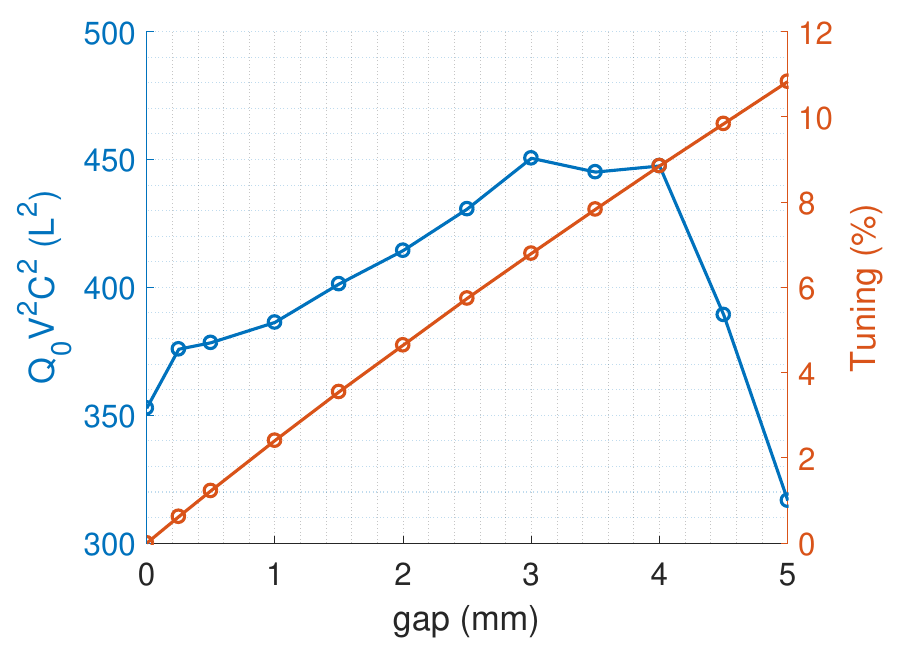}
         \caption{}
         \label{fig:CylCav2_FoM_vs_Gap_CST}
\end{subfigure}
\caption{CST simulation results opening the cylindrical cavity from $gap = 0$ to $5$~mm: (a) volume in L (blue line), (b) form factor (blue line), and (c) figure of merit in L$^2$ (blue line), for the $TM_{010}$ mode. For reference, the tuning range in $\%$ versus the gap is shown in each plot (red lines).}
\label{fig:CylCav2_V_and_C_and_FoM_vs_Gap_CST}
\end{figure}

The volume and form factor parameters have also been analysed, yielding the results shown in Figures~\ref{fig:CylCav2_V_vs_Gap_CST} and \ref{fig:CylCav2_C_vs_Gap_CST}, respectively.
Using the parameters $Q_0$, $V$, and $C$, and Eq.~\ref{eq:FoM}, the figure of merit $FoM$ is calculated as shown in Figure~\ref{fig:CylCav2_FoM_vs_Gap_CST}. By increasing the gap between the two halves of the cavity housing, although the quality factor decreases, the volume increases, so the overall performance remains more or less constant. It can be concluded that a $4$~mm gap provides a $FoM$ parameter that only increases with respect to the fully closed gap case ($gap = 0$~mm), which reflects the potential of this tuning technique in this cavity for that particular resonant mode.

\subsubsection{Misalignment effects}
\label{sss:Misalignment effects}

In a real implementation of vertical cut tuning, there will inevitably be misalignments of the cavity halves relative to each other that will reduce the cavity $Q_0$ and, consequently, reduce the performance of the haloscope.

\begin{figure}[h]
\centering
\begin{subfigure}[b]{0.95\textwidth}
         \centering
         \includegraphics[width=1\textwidth]{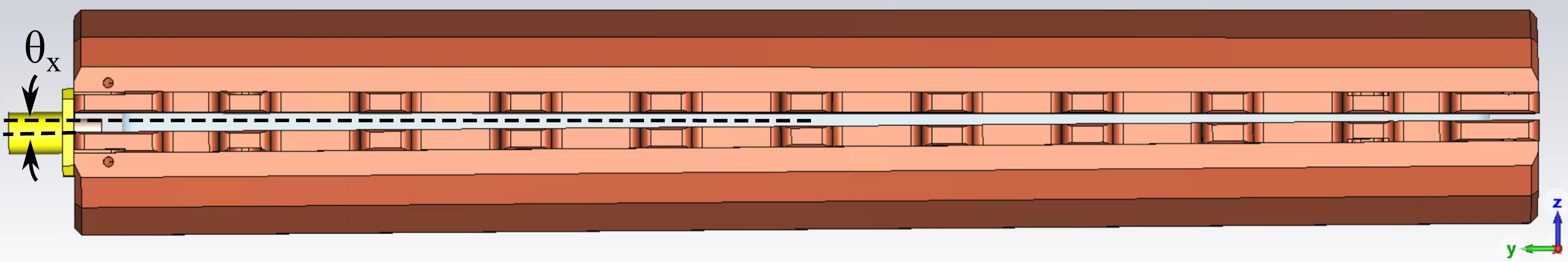}
         \caption{}
         \label{fig:CylCav_ThetaXtilt_3Dmodel}
\end{subfigure}
\hfill
\begin{subfigure}[b]{0.34\textwidth}
         \centering
         \includegraphics[width=1\textwidth]{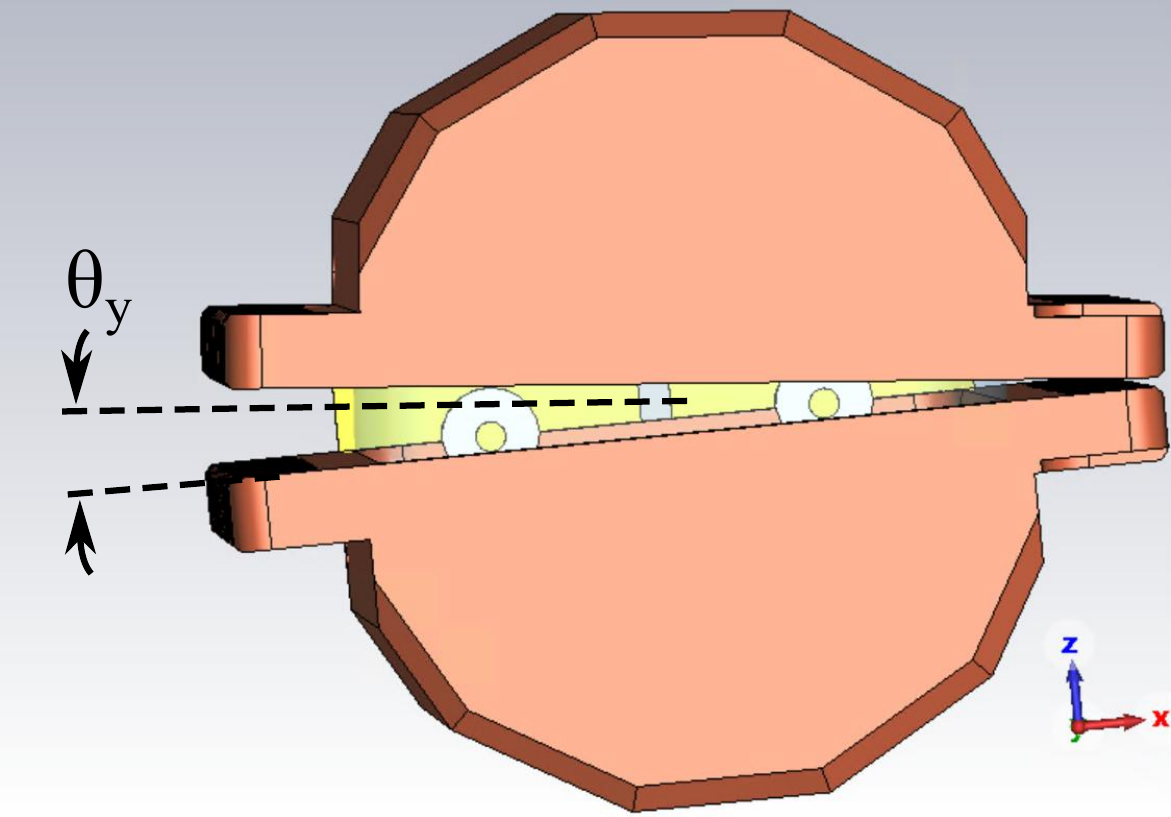}
         \caption{}
         \label{fig:CylCav_ThetaYtilt_3Dmodel}
\end{subfigure}
\begin{subfigure}[b]{0.2\textwidth}
         \centering
         \includegraphics[width=1\textwidth]{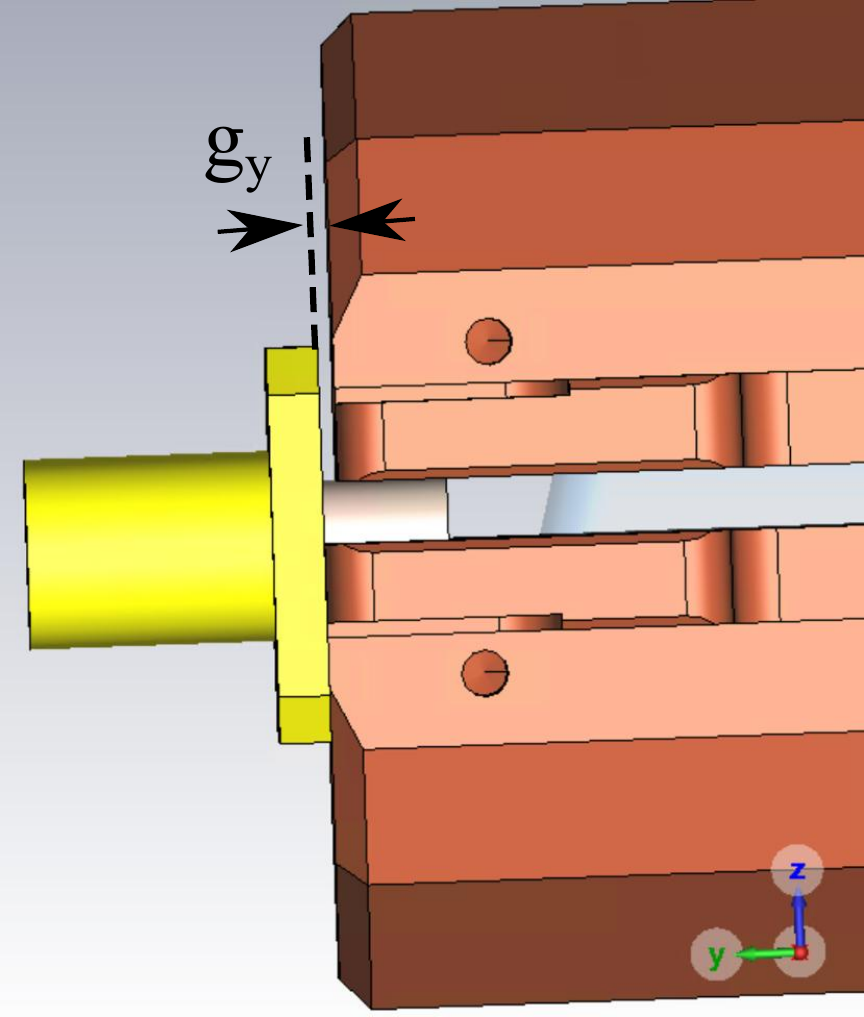}
         \caption{}
         \label{fig:CylCav_gYtilt_3Dmodel}
\end{subfigure}
\caption{3D models of possible misalignments in the analysed cylindrical cavity: (a) angular $x-$axis tilt, (b) angular $y-$axis tilt, and (c) linear $y-$axis displacement.}
\label{fig:CylCav2_Tilt_3Dmodels}
\end{figure}

The following misalignments can occur: angular $x-$axis tilt ($\theta_x$), angular $y-$axis tilt ($\theta_y$), and linear $y-$axis displacement ($g_y$)\footnote{A lateral shift along the $x$-axis ($g_x$) was omitted because it is effectively constrained by the coaxial ports and the tight tolerances of the rail-and-roller guiding system described in the following sections.}. A combination of these three could occur, but it is omitted in this work for the sake of simplicity, since cross-coupling effects between these small perturbations are of higher order and their combined impact can be closely approximated by a linear combination of the independent misalignments. Figure~\ref{fig:CylCav2_Tilt_3Dmodels} shows what each of these misalignments consists of.

To simplify the haloscope design, this tuning method involves keeping the coaxial ports fixed to one half of the cavity. This causes a slight asymmetry between the two halves, which is considered negligible, as numerical simulations confirm a degradation of less than $1$~$\%$ for both quality and form factor over the entire tuning range, similarly to the results obtained in \cite{Golm:2023iwe}.

\begin{figure}[h]
\centering
\begin{subfigure}[b]{0.32\textwidth}
         \centering
         \includegraphics[width=1\textwidth]{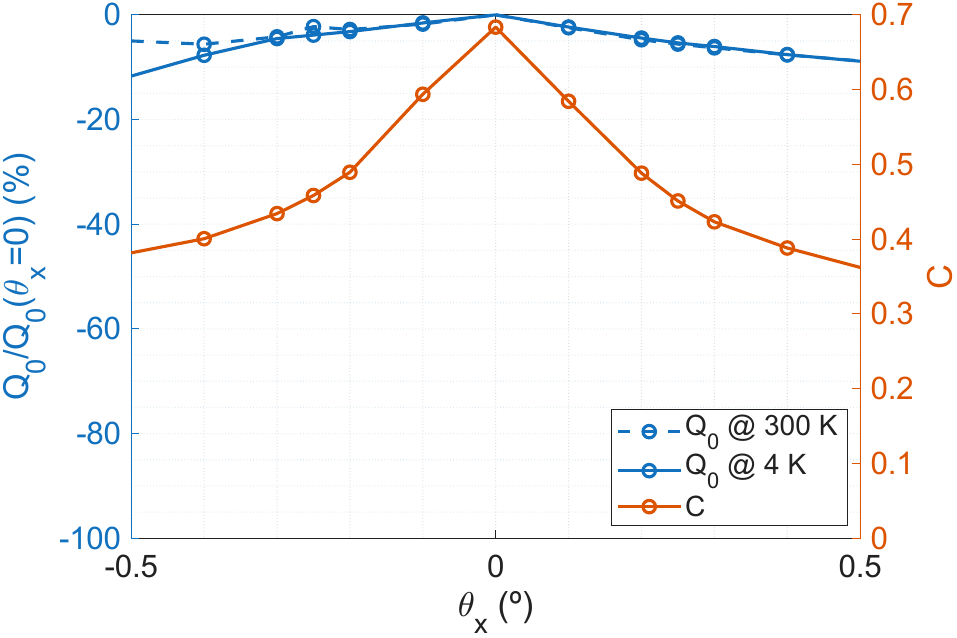}
         \caption{}
         \label{fig:CylCav2_Q0_and_C_vs_Thetax_CST}
\end{subfigure}
\hfill
\begin{subfigure}[b]{0.32\textwidth}
         \centering
         \includegraphics[width=1\textwidth]{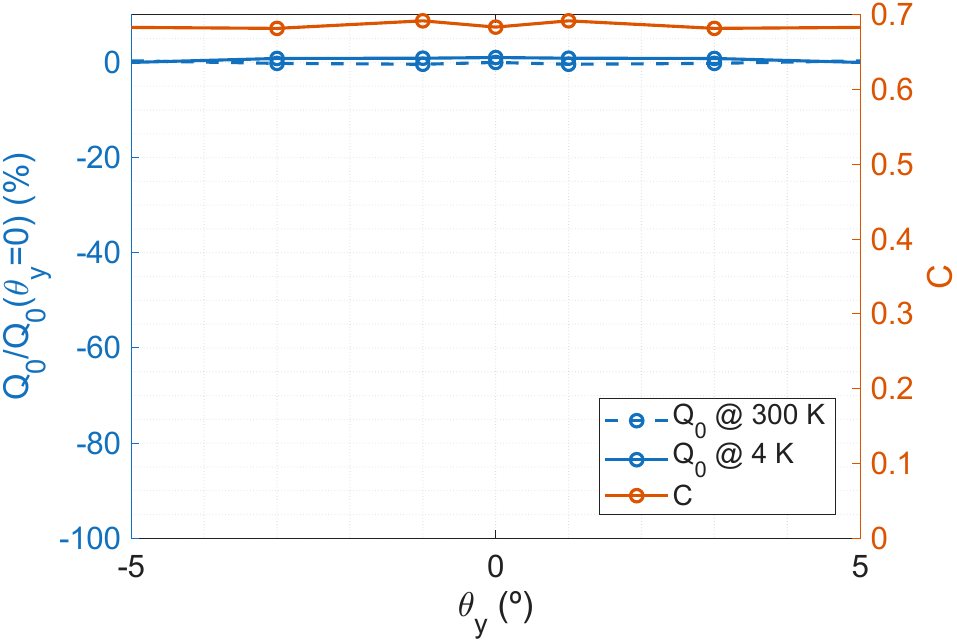}
         \caption{}
         \label{fig:CylCav2_Q0_and_C_vs_Thetay_CST}
\end{subfigure}
\hfill
\begin{subfigure}[b]{0.32\textwidth}
         \centering
         \includegraphics[width=1\textwidth]{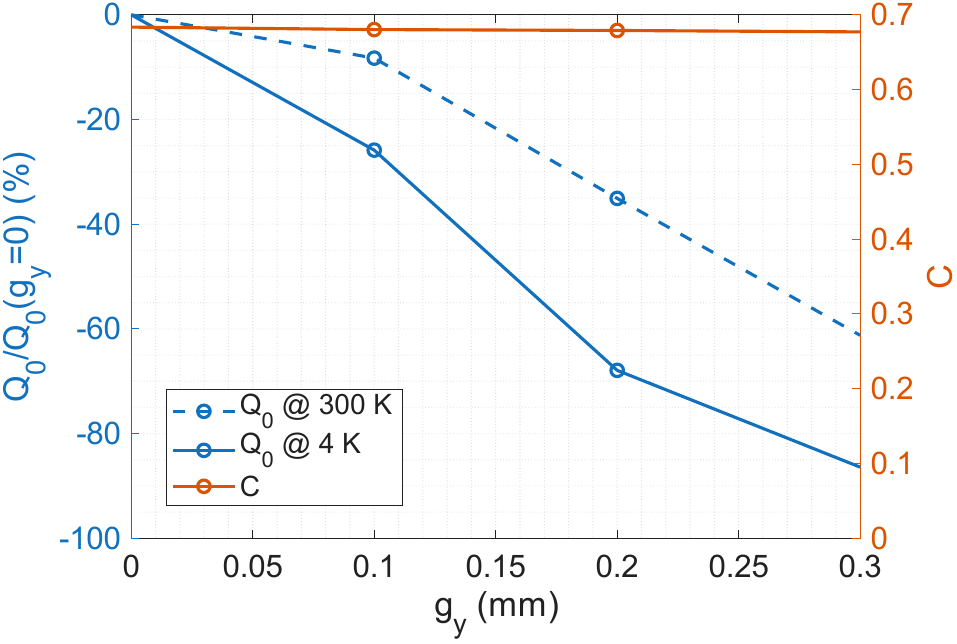}
         \caption{}
         \label{fig:CylCav2_Q0_and_C_vs_gy_CST}
\end{subfigure}
\hfill\begin{subfigure}[b]{0.32\textwidth}
         \centering
         \includegraphics[width=1\textwidth]{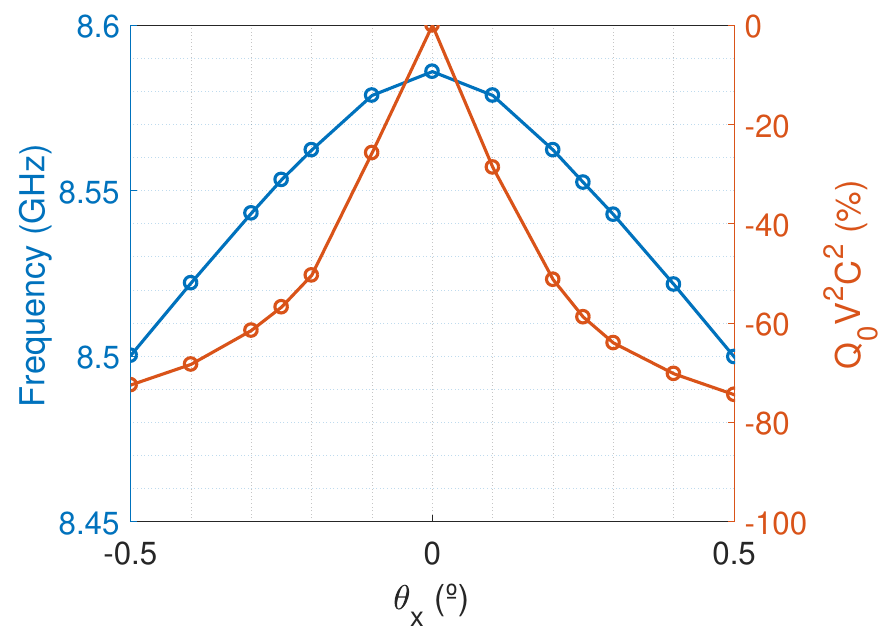}
         \caption{}
         \label{fig:CylCav2_fr_vs_Thetax_CST}
\end{subfigure}
\hfill
\begin{subfigure}[b]{0.32\textwidth}
         \centering
         \includegraphics[width=1\textwidth]{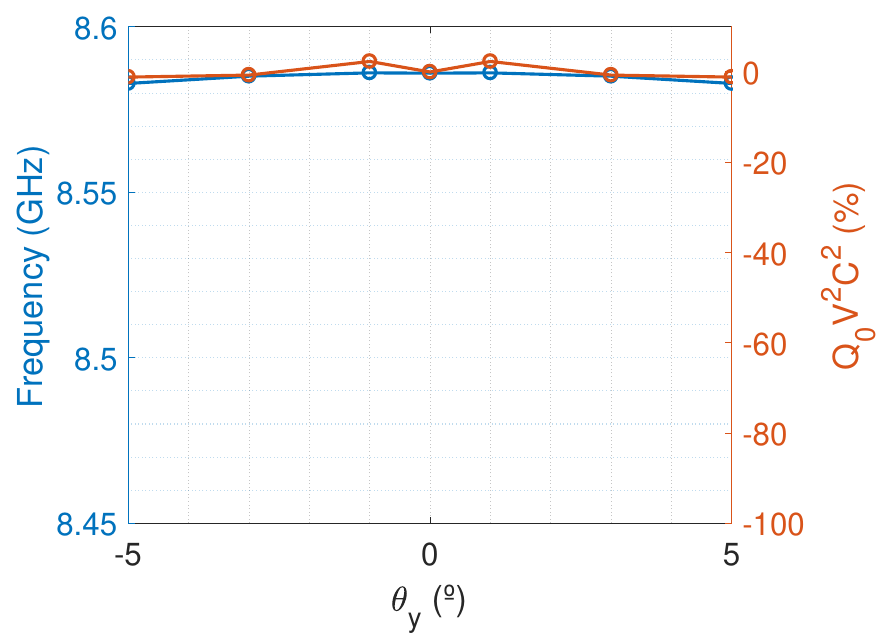}
         \caption{}
         \label{fig:CylCav2_fr_vs_Thetay_CST}
\end{subfigure}
\hfill
\begin{subfigure}[b]{0.32\textwidth}
         \centering
         \includegraphics[width=1\textwidth]{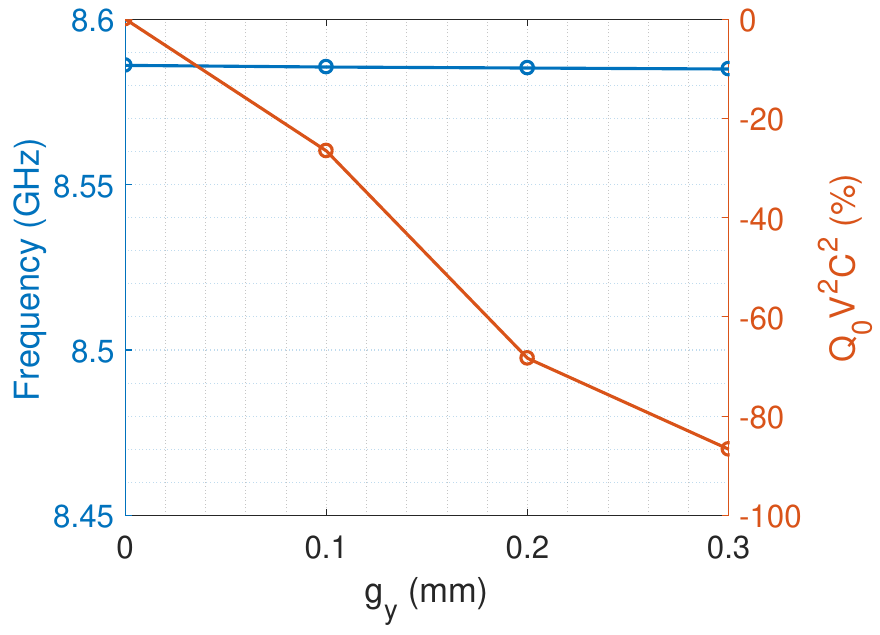}
         \caption{}
         \label{fig:CylCav2_fr_vs_gy_CST}
\end{subfigure}
\caption{CST simulation results for the $TM_{010}$ mode misaligning the cylindrical cavity with $gap = 2$~mm. Top plots: unloaded quality factor decrease in $\%$ at room (blue dashed line) and cryo (blue solid line) temperatures and form factor (red line) versus (a) $\theta_x$ (in $^o$), (b) $\theta_y$ (in $^o$), and (c) $g_y$ (in mm). Bottom plots: frequency shift in GHz (blue line) and figure of merit in $\%$ (red line) versus (d) $\theta_x$ (in $^o$), (e) $\theta_y$ (in $^o$), and (f) $g_y$ (in mm).}
\label{fig:CylCav2_fr_Q0_C_and_FoM_vs_Tilts_CST}
\end{figure}

A simulation study was conducted to investigate the effect of these three misalignments for $gap = 2$~mm, an intermediate point within the tuning range where the cavity maintains good performance for axion data taking, as seen in the previous section. This intermediate point serves as a representative benchmark for the tuning range, as the sensitivity to mechanical misalignments scales monotonically with the gap size, meaning that the tolerances evaluated here provide a scalable estimate that does not restrict the operational tuning range. Figure~\ref{fig:CylCav2_fr_Q0_C_and_FoM_vs_Tilts_CST} shows the results of this study, analysing the frequency shift and the effect on the quality and form factors.
It is important to highlight that while the absolute frequency shift can be mechanically compensated by the tuning mechanism, its evaluation is essential to ensure that the tuning step resolution and tracking stability remain unperturbed by geometric errors. It is remarkable to note that the volume remains unchanged for any misalignment, which, in the case of $gap = 2$~mm, gives $V = 112$~mL. For clarity, the degradation levels are quantified as low ($<0.1$~$\%$), medium ($0.1$~$\%$--$1$~$\%$), high ($1$~$\%$--$10$~$\%$), and very high ($>10$~$\%$). Table~\ref{tab:MisalignmentEffect} provides a summary of the effect produced on each parameter for each misalignment.
\begin{table}[h]
\small
\begin{tabular}{|c|c|c|c|}
\hline
Misalignment & $Q_0$ & C & Freq. \\ \hline\hline
$\theta_x$ & High & Very high & Medium \\ \hline
$\theta_y$ & Low & Low & Low \\ \hline
$g_y$ & Very high & Low & Low \\ \hline
\end{tabular}
\centering
\caption{\label{tab:MisalignmentEffect} Misalignment impact on each parameter of the cylindrical cavity $TM_{010}$ mode.}
\end{table}

For the first misalignment ($\theta_x$), it can be seen how the main problem lies in the reduction of the form factor (it also undergoes a change in frequency, but this could be corrected by readjusting the cavity aperture). This deterioration of the $C$ parameter occurs due to increased inhomogeneity of the RF electric field in the cavity volume, which decreases its coupling with the external magnetic field of the magnet (see Eq.~\ref{eq:C}). Furthermore, for negative $\theta_x$ values, the port coupling coefficient ($\beta$) begins to decrease considerably, although this could be adjusted by modifying the longitudinal position of the coaxial pin inside the cavity. Observing the results for $Q_0$, the effect of $\theta_x$ angular misalignment is mainly a reduction of the C factor. On the other hand, the effect of ($\theta_y$) angular misalignment on the cavity parameters is negligible. Finally, linear misalignments on the $y-$axis ($g_y$) result in a significant reduction in $Q_0$, while the impact on the form factor is almost zero.

As will be seen in the following sections, the tuning mechanism has an assembly system that sufficiently corrects the $\theta_x$ misalignment so as not to lead to large degradation of $C$, $f_r$ and $\beta$. For example, for $\theta_x = 0.5$~$^o$ (with $gap = 2$~mm), it is obtained $gap_{t} = 1.14$~mm and $gap_{b} = 2.86$~mm,  where $gap_{t}$ and $gap_{b}$ represent the distance between cavity halves at the top and bottom sections, respectively. At room temperature, before closing the cryostat, this would be detected by eye and measured easily with a vernier calliper. For $y-$axis misalignments, as will be observed in the next section, the frequency tuning assembly system will maintain minimal alignment to reduce its impact on the quality factor.

\subsection{Mechanical design and implementation}
\label{ss:Manufacturing_details}

\subsubsection{Cavity prototypes}
\label{sss:Cavity_prototypes}

The mechanical design and implementation of the haloscope must simultaneously satisfy stringent RF, cryogenic, and manufacturing constraints. To ensure reliable operation, the engineering architecture of the cavity and its sub-systems is systematically divided into four main functional blocks:
\begin{itemize}
    \item \textit{Fabrication and tolerance control:} The geometric definition of the cavity halves and external flat references required for precise CNC milling and alignment.
    \item \textit{RF tuning mechanism:} The movable assembly is designed to open the cavity gap symmetrically and provide the necessary frequency scanning range.
    \item \textit{Coupling and $\beta$ adjustment:} The coaxial port integration and physical positioning of the coupling pins to control the $\beta$ parameter.
    \item \textit{Cryogenic thermalisation and support:} The anchoring interfaces to the dilution refrigerator rods that guarantee structural stability and rapid cooling to milli-Kelvin temperatures.
\end{itemize}
In the following subsections, the physical realisation and specific features of these integrated components are detailed.

Regarding fabrication and cryogenic integration, a section of screw holes was configured for anchoring to the rods of the dilution refrigerator system (see Figure~\ref{fig:Rods_And_CylCavHousingDesign}), which serve as thermalisation and support points together with the assembly of the tuning system, which will be detailed later. The cavity is also designed not only to be implemented in the tuning system and open both halves in a controlled manner, but also to close it using screws and alignment pins. These screws and pins are removed upon adding the movement structure to allow the sliding. Other alignment strategies are used with the VORTEX setup, as explained in the following section.

On the other hand, as shown in Figures~\ref{fig:HousingDesign_TopDetails} and \ref{fig:CylCav2_Tilt_3Dmodels}, several flat sections have been implemented on the outside of the cavity housing to support it on the milling machine of the workshop during the manufacturing process. Additionally, small venting holes have been implemented at the inner port screw sections to prevent over-pressure sections in the cavity and to keep these holes open for pumping vacuum when the haloscope is installed completely closed (using screws) inside this dilution refrigerator (or in any vacuum environment).

\begin{figure}[h]
\centering
\begin{subfigure}[b]{0.75\textwidth}
         \centering
         \includegraphics[width=1\textwidth]{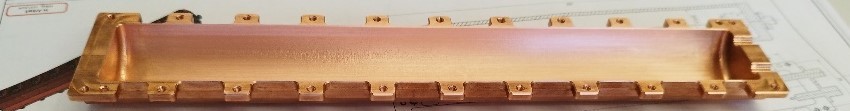}
         \caption{}
         \label{fig:CylCavManufacturedPrototype_v1}
\end{subfigure}
\hfill
\begin{subfigure}[b]{0.75\textwidth}
         \centering
         \includegraphics[width=1\textwidth]{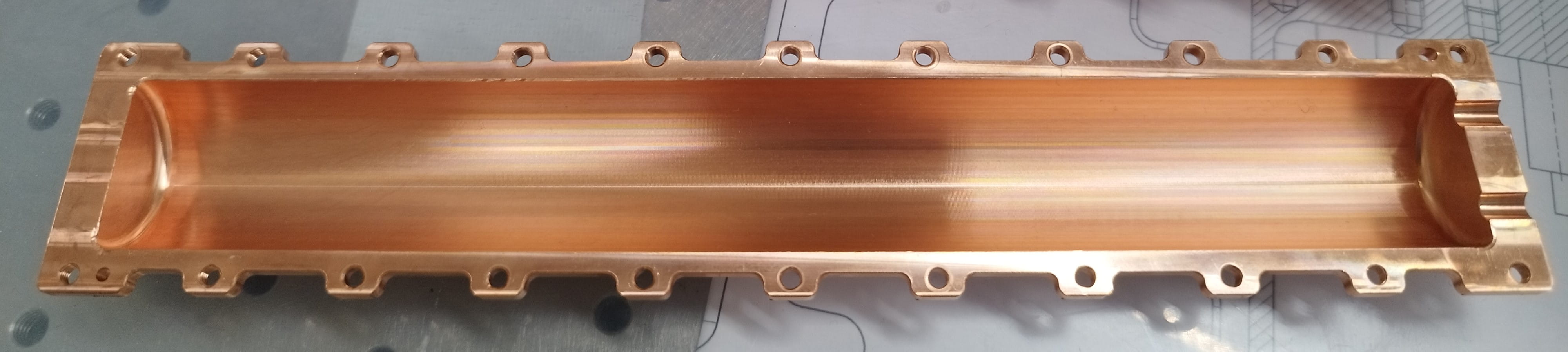}
         \caption{}
         \label{fig:CylCavManufacturedPrototype_v2}
\end{subfigure}
\hfill
\begin{subfigure}[b]{0.75\textwidth}
         \centering
         \includegraphics[width=1\textwidth]{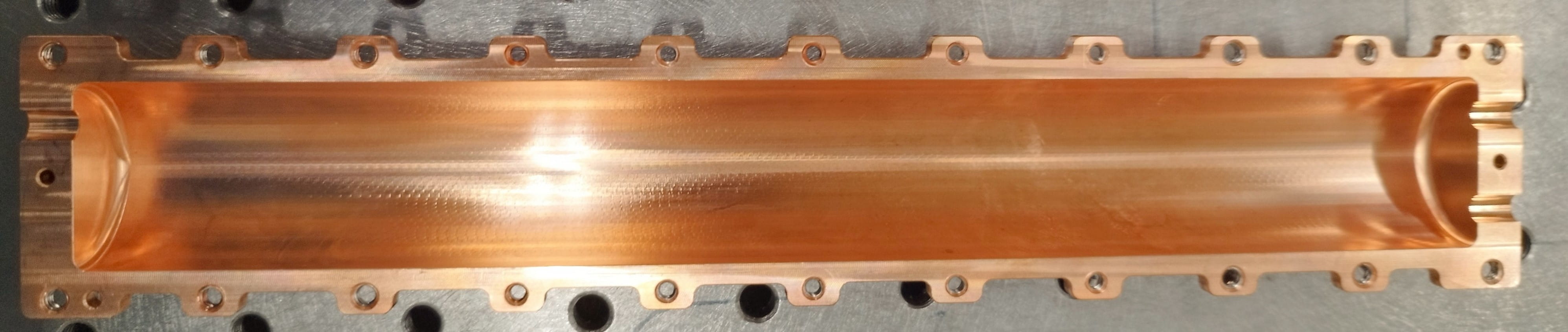}
         \caption{}
         \label{fig:CylCavManufacturedPrototype_v3}
\end{subfigure}
\caption{Versions of the cylindrical cavity: (a) Version $\#1$ (2-port, standard copper), (b) Version $\#2$ (2-port, OFHC copper), and (c) Version $\#3$ (3-port, OFHC copper). For each cavity, only one of the two symmetrical halves is shown.}
\label{fig:CylCavManufacturedPrototypes}
\end{figure}

Three different cavity prototypes, shown in Figure~\ref{fig:CylCavManufacturedPrototypes}, have been designed and manufactured at the mechanical workshop of the \textit{Max Planck Institut für Physik} (MPP) on a 5-axis-milling machine with a ball milling cutter.
The 2-port prototype, manufactured in standard copper, was designed with an $XThick$ dimension of $0.75$~mm. The 2-port cavity, manufactured in Oxygen-Free High Conductivity (OFHC) copper (which helps in the thermalisation process thanks to its higher thermal conductivity in comparison to standard copper \cite{Simon:1992}), has $XThick=3$~mm, improving the quality factor compared to the previous one when a frequency tuning gap is introduced, as explained in previous sections. Although the amount of material in the structure is slightly greater, the difference in thermalisation time is negligible. Finally, a three-port prototype has been manufactured (in OFHC material as well), an evolution of its predecessor, which has certain protrusions on its external walls to properly attach temperature sensors with screws, in addition to incorporating a third port in its lower section for different electromagnetic testing purposes.

\subsubsection{Tuning system assembly}
\label{sss:Tuning_system_assembly}

\begin{figure}[h]
\centering
\begin{subfigure}[b]{1\textwidth}
         \centering
         \includegraphics[width=1\textwidth]{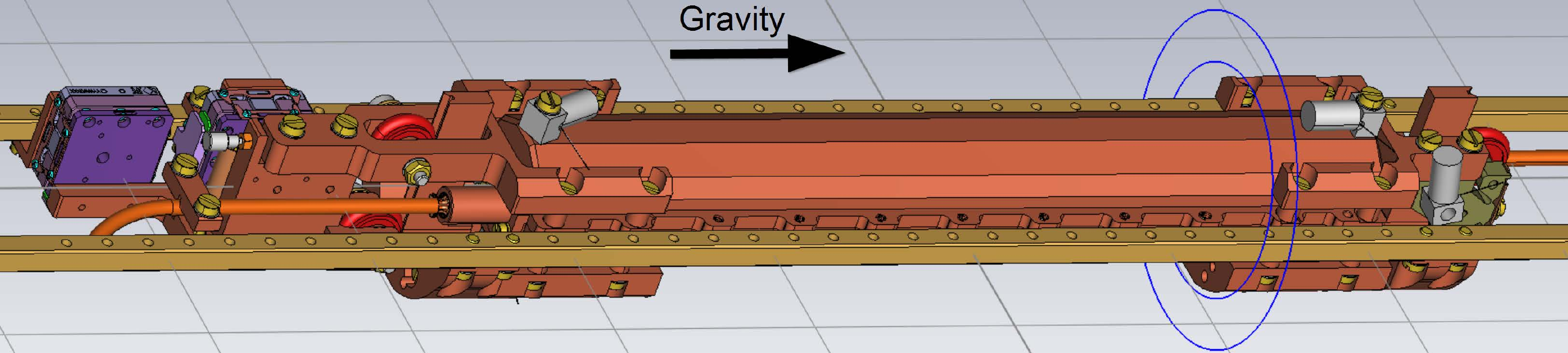}
         \caption{}
         \label{fig:3DmodelTuningAssembly}
\end{subfigure}
\hfill
\begin{subfigure}[b]{1\textwidth}
         \centering
         \includegraphics[width=1\textwidth]{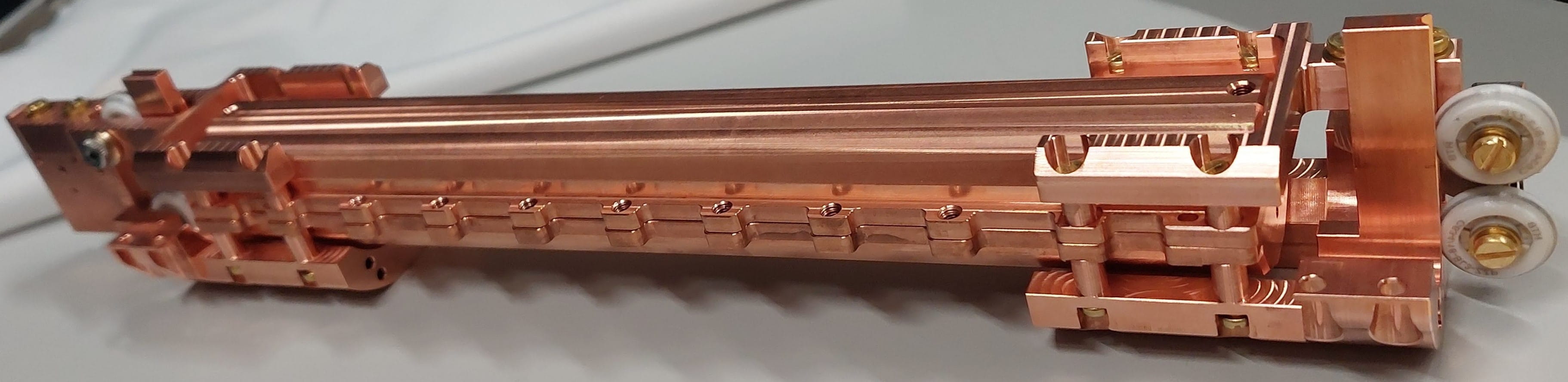}
         \caption{}
         \label{fig:ManufacturedTuningAssembly}
\end{subfigure}
\caption{(a) 3D model of the tuning assembly with two JPE nanopositioners (in purple colour) for the cylindrical cavity. For optimal thermalisation, the main cavity structures (in brown) are manufactured from OFHC copper, while the supporting rods are gold-plated copper. Small non-structural fasteners, such as washers and screws (in yellow), are made of brass. Some parts of the mechanism cannot be seen and are omitted here for simplicity. (b) Manufactured prototype of the tuning assembly.}
\label{fig:3DmodelAndManufacturedTuningAssembly}
\end{figure}

The tuning structure design is based on the use of several mechanisms for the smooth movement of one of the cavity halves, introducing precise control in the gap value in the small experimental space of the FSE dilution refrigerator system and antenna position readjustment for correcting $\beta$ parameter (analysed in the following subsection). In Figure~\ref{fig:3DmodelTuningAssembly}, a 3D model of the tuning assembly with two \textit{CBS7 - CRYO BEARING STAGE 7} nanopositioners from \textit{JPE} \cite{JPE} for the mechanical shift of the cavity and coaxial cable at mK temperatures is shown.
Figure~\ref{fig:ManufacturedTuningAssembly} shows a picture of the manufactured structure. This nanopositioner model has the following parameters: travel range of $7$~mm, step size of $50$~nm, driving force of $1$~N, and load capacity of $2$~Kg and $100$~g at horizontal and vertical motion, respectively. Also, the main body is made of titanium, which allows its operation at high magnetic field levels, along with its compatibility in a mK and vacuum environment. These devices integrate a scanner functionality
for the position feedback (Resistive Linear Sensor or RLS) with a resolution of $100$~nm.

The cavity movement system consists of fixing one half of the haloscope housing to the dilution refrigerator golden rods, while the other half slides on structures attached to rails, thanks to the use of four ceramic rolls (two above and two below, in red in the 3D model), ensuring proper alignment at the same time. The movable structure is pushed by one of the two JPE nanopositioners, which is capable of generating sufficient force to move these structures horizontally (see the direction of gravity in the 3D model). As shown in Figure~\ref{fig:3DmodelTuningAssembly}, temperature sensors are used in both the upper and lower sections to verify an efficient thermalisation gradient between the two areas. This is implemented for both halves of the cavity housing, resulting in a total of four temperature sensors used on the haloscope.

JPE nanopositioners are capable of providing linear movement with a maximum range of $7$~mm, which is more than sufficient both for opening the cavity (where optimal tuning is achieved for a maximum $gap = 4$~mm, as was shown in previous sections) and for the vertical displacement of the coaxial monopole port, as will be explained in the following section. The blue rings of smaller and larger diameters shown on the right side of the 3D model in Figure~\ref{fig:3DmodelTuningAssembly} mark the limits of the experimental space of the FSE and the magnet (without FSE), respectively. The entire system (including the maximum gap opening for right tuning) has been designed to fit within the most restrictive limit, that of the FSE with a diameter of $53$~mm. However, if more space is needed in the future to include new bulky elements not previously considered, this would be possible as long as the limit imposed by the magnet diameter of $76$~mm is met (that is the case shown in Figure~\ref{fig:CavityDesignAtBF}), assuming that the FSE system will not be used.

\subsection{$\beta$ readjustment system}
\label{ss:Beta_readjustment_system}

One of the parameters to be optimised within the equation that regulates sensitivity when taking axion data (Eq.~\ref{eq:ga}) is $\beta$. When frequency tuning is not used, this value should be close to $\beta = 1$ (known as critically coupled). However, when using a tuning system, maximising the scanning rate (Eq.~\ref{eq:dmadt}), $\beta = 2$ (overcoupled case) is ideal \cite{Kim:2020kfo}. Obtaining these values in an axion data acquisition setup, regardless of whether a frequency tuning system is implemented, can be complex. This is because any adjustment in ambient temperatures is altered due to shrinkage at cryogenic temperatures. In the absence of tuning, the thermal contraction of the materials in the experiment causes the $\beta$ value to change from the time the port is installed (properly configured at room temperature) until the cryostat temperature is reduced. Furthermore, if this frequency tuning system is included, adjusting $\beta$ to optimal values becomes even more difficult, since the change in the resonant frequency of the microwave cavity in turn causes a change in this parameter. For this reason, it becomes necessary to use an antenna position adjustment system to ensure optimal $\beta$, especially with the vertical cut frequency tuning system proposed in this work.

$\beta = 2$ is obtained by optimising the coupling-dependent factor in the axion scanning rate (see Eq.~\ref{eq:dmadt}). However, other values could be assumed, not necessarily insufficient for data-taking, whose impact has been studied in simulations. Assuming a frequency tuning range between $8.2$ and $8.9$~GHz, a variation of $\beta$ between $1.5$ and $2.45$ has been found, which corresponds to a loss factor up to $3$~$\%$ in Eq.~\ref{eq:dmadt}. This may seem an acceptable value; however, the use of a port recoupling system becomes very useful for the following reason: above, the assumption has been made that the variation in $\beta$ could be perfectly controlled when shifting from room to cryogenic temperatures, which is not always true in a real experiment, as there are many details of thermalisation and thermal contraction that make control difficult when moving from one temperature to another, thus making the use of a coaxial port positioning system very interesting once again.

In this work, a system has been developed for adjusting the position of the monopole antenna gripping a semi-rigid coaxial cable (in orange colour, at the left side of the 3D model in Figure~\ref{fig:3DmodelTuningAssembly}) connected to the RF cabling of the dilution refrigerator system (to extract the signal) with sufficient play (cable bends) to allow movement, with its termination \textit{stripped} and lying inside the cavity. This grip is achieved by means of an extra structure attached to the second JPE nanopositioner, which is capable of moving the coaxial cable vertically due to its low mass. Additionally, to prevent vibrations and ensure good electrical contact between the external conductor of the coaxial cable and the cavity housing, a system of finger gaskets with a circular shape has been implemented at the movable port section of the cavity by the use of a holder through which the semi-rigid cable passes.

\begin{figure} [htb]
    \centering
    \includegraphics[width=0.43\textwidth]{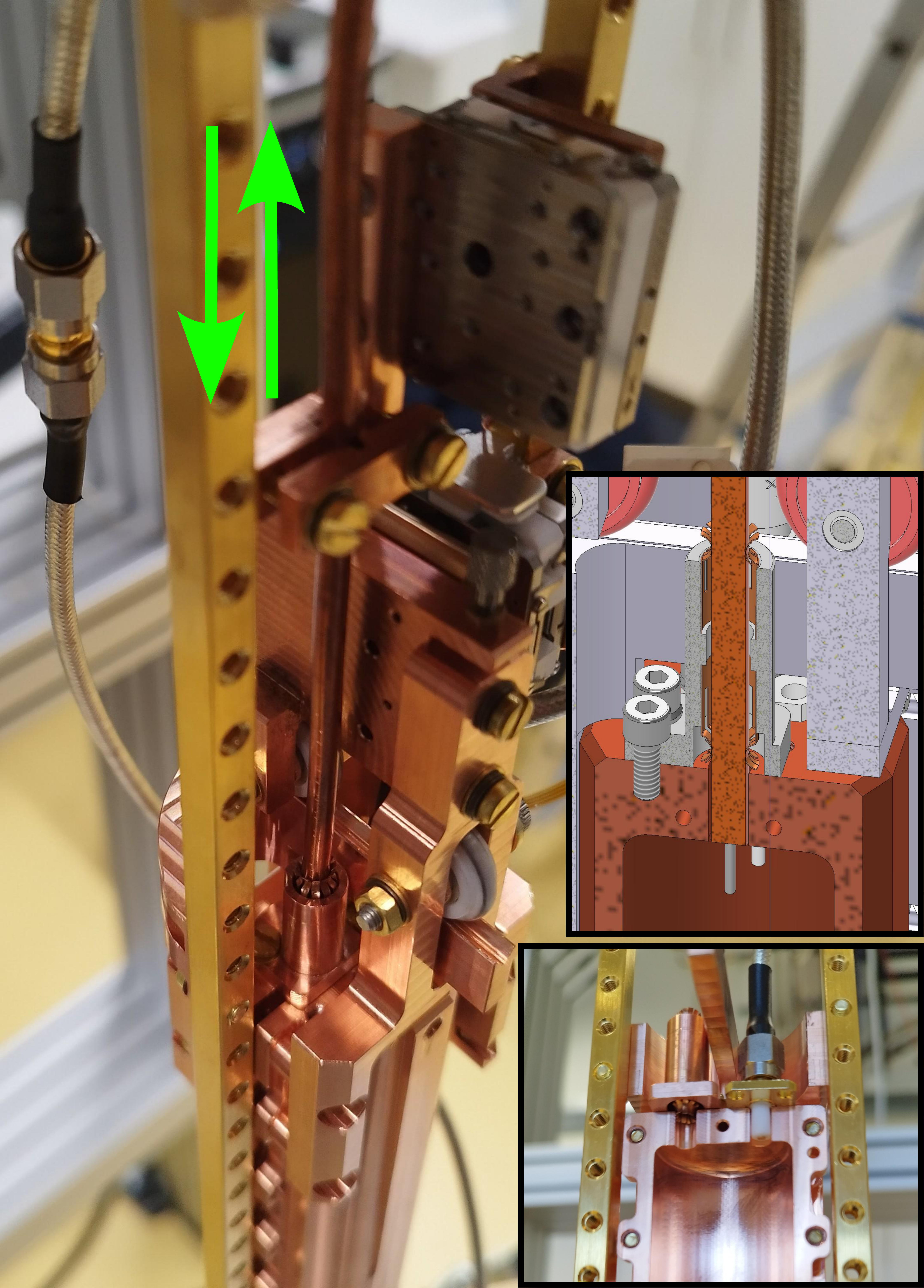}
    \caption{Image of the port input coupling readjustment system with movable coaxial cable. The green arrows depict the direction of movement (provided by the JPE nanopositioner installed vertically). The upper inset shows a 3D model with a longitudinal cut focused on the port section to show internal details, such as the use of finger gaskets. The lower inset shows an image of one cavity half with the coaxial cable holder, next to the secondary port configured with a fixed SMA adapter.}
    \label{fig:BetaReadjustmentSystem}
\end{figure}

Figure~\ref{fig:BetaReadjustmentSystem} shows an image of this system implemented in the cylindrical cavity at the setup of the dilution refrigerator (screwed to the gold rods), in combination with the frequency tuning system, providing additional details of the internal structure and components used for the assembly and movement of the coaxial cable that regulates the $\beta$ parameter.

\section{Cavity measurements}
\label{s:Cavity_measurements}

The cavities were measured to determine the unloaded quality factor value and the change in frequency of the axion resonant mode as a function of the tuning gap at four different temperatures. All of the structures were first tested at room temperature and then at $\sim 77$~K (liquid nitrogen temperature). The second cavity was also installed and characterised at $4$~K. The second and third cavities were finally installed and characterised in the dilution refrigerator at temperatures of a few mK. For each of these measurements, the structures and tuning mechanisms (always based on the gap concept) were implemented in a significantly different manner, as explained below. Consequently, different stepper motors and nanopositioners were strategically selected for each setup to meet the specific thermal, mechanical, and vacuum compatibility requirements of each environment.

\subsection{Room temperature}
\label{ss:Room_temperature}

\subsubsection{First version (2-port, standard copper)}
\label{sss:CylCavv1_RT}

\begin{figure}[h]
\centering
\begin{subfigure}[b]{0.49\textwidth}
         \centering
         \includegraphics[width=1\textwidth]{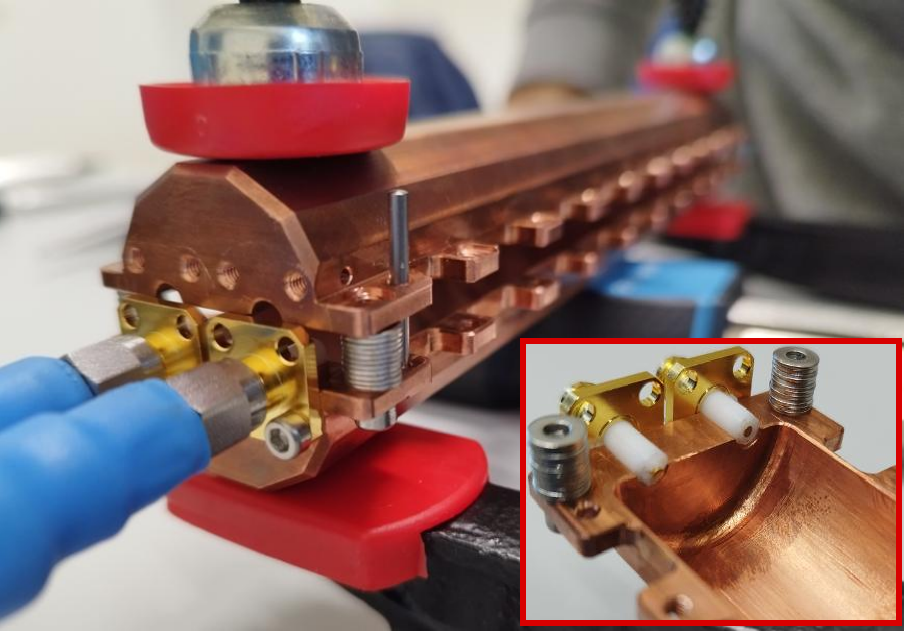}
         \caption{}
         \label{fig:CylCav_TuningAtRTwithWashers_Picture}
\end{subfigure}
\hfill
\begin{subfigure}[b]{0.45\textwidth}
         \centering
         \includegraphics[width=1\textwidth]{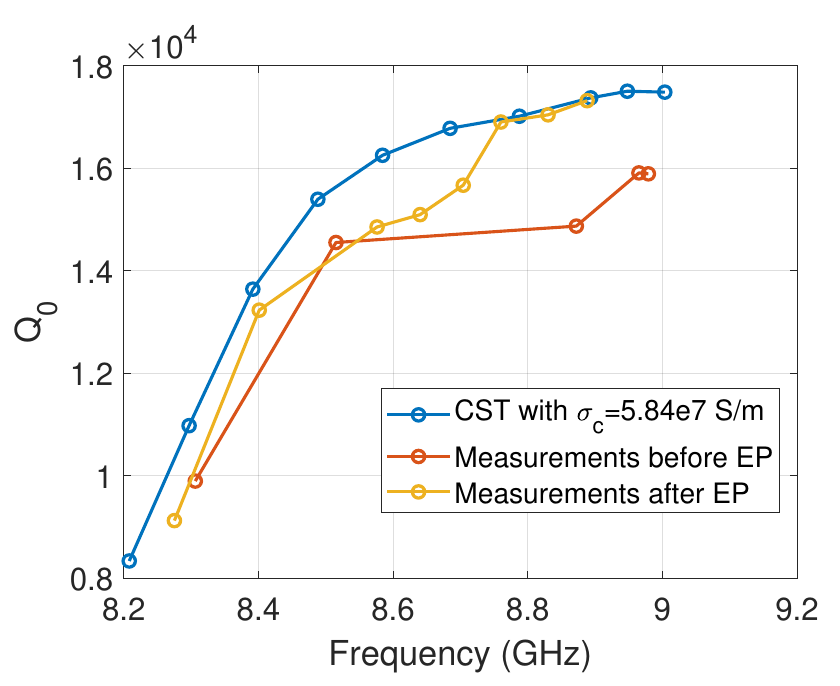}
         \caption{}
         \label{fig:CylCav_TuningAtRTwithWashers_Q0vsFreq}
\end{subfigure}
\caption{(a) Room-temperature electromagnetic characterisation process of the first version of the cylindrical cavity using washers in the corners to ensure a stable gap. The inset shows the section of the cavity, open, with the two coaxial ports and washers installed. (b) Variation of the unloaded quality factor versus the resonant frequency of the mode of interest for several cases: CST simulations (blue line) and measurements before (red line) and after (yellow line) electropolishing.}
\label{fig:CylCav_TuningAtRTwithWashers}
\end{figure}

The first manufactured cavity utilised a standard copper, 2-port configuration. Its room-temperature characterisation served as the baseline procedure, which was later adapted with minimal modifications to test the remaining two cavity versions. These measurements were made at a series of gap thicknesses, obtained by placing a number of washers at the four corners. The washer spacing technique maintained constant and good alignment during measurement. For each frequency tuning gap, the number of washers at the four corners was changed, and the resulting height was measured using an electronic calliper. Figure~\ref{fig:CylCav_TuningAtRTwithWashers_Picture} shows an example of this setup.
Tuning measurements were taken by modifying the gap before and after applying an electropolishing (EP) process to improve the cavity surfaces and, therefore, the quality factor. Figure~\ref{fig:CylCav_TuningAtRTwithWashers_Q0vsFreq} depicts a plot of the obtained $Q_0$ versus frequency. $Q_l$ and $\beta$ values typically have a $5$~$\%$ of systematic uncertainty due to residual mismatch in the cables connecting the Vector Network Analyser (VNA) to the cavity. These measurements are compared in the graph with the simulated values, resulting in a difference around $15$~$\%$ at the highest frequency. A quality factor of $Q_0=17200$ for $gap\simeq0$ has been obtained after electropolishing, an improvement of $+7.5$~$\%$ with respect to the previous case. The final quality factor is approximately $8$~$\%$ lower than that found in simulation over the tuning range. This validates the fabrication of this first version of the cylindrical cavity and demonstrates the potential of electropolishing.

\subsubsection{Second version (2-port, OFHC)}
\label{sss:CylCavv2_RT}

\begin{figure}[h]
\centering
\begin{subfigure}[b]{0.35\textwidth}
         \centering
         \includegraphics[width=1\textwidth]{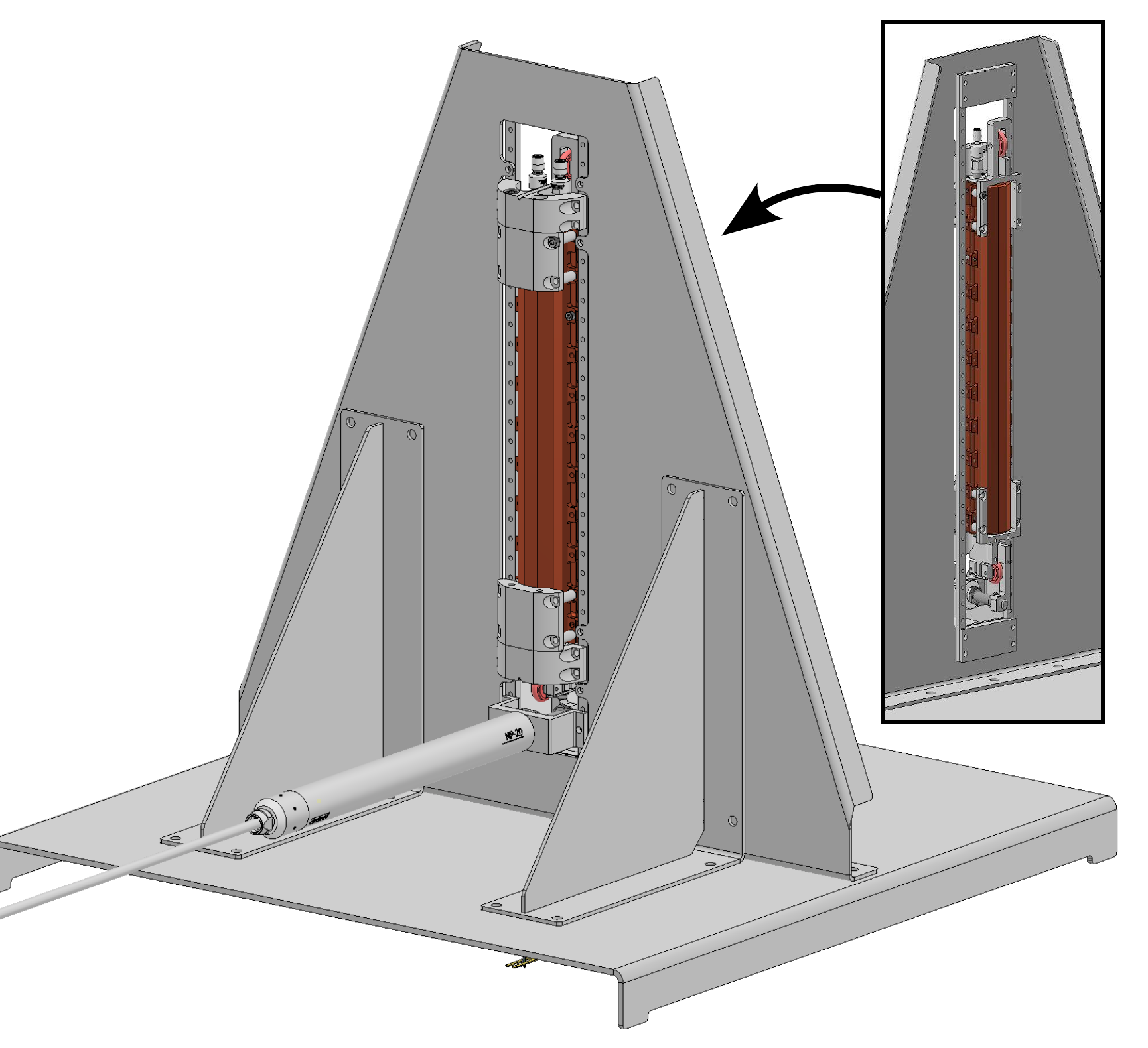}
         \caption{}
         \label{fig:3Dmodel_CylCavSupport}
\end{subfigure}
\hfill
\begin{subfigure}[b]{0.3\textwidth}
         \centering
         \includegraphics[width=1\textwidth]{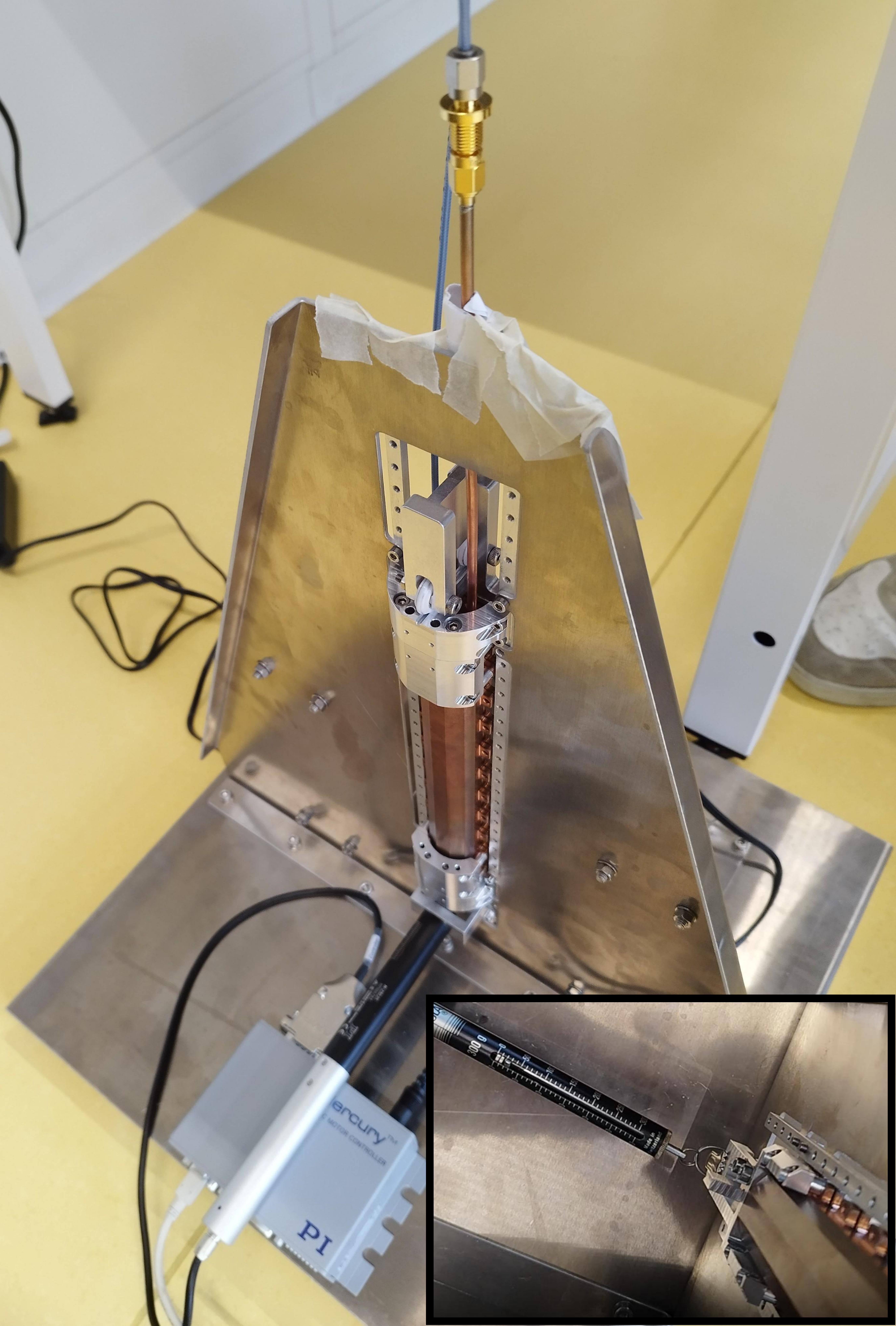}
         \caption{}
         \label{fig:RealCylCavSupport}
\end{subfigure}
\hfill
\begin{subfigure}[b]{0.33\textwidth}
         \centering
         \includegraphics[width=1\textwidth]{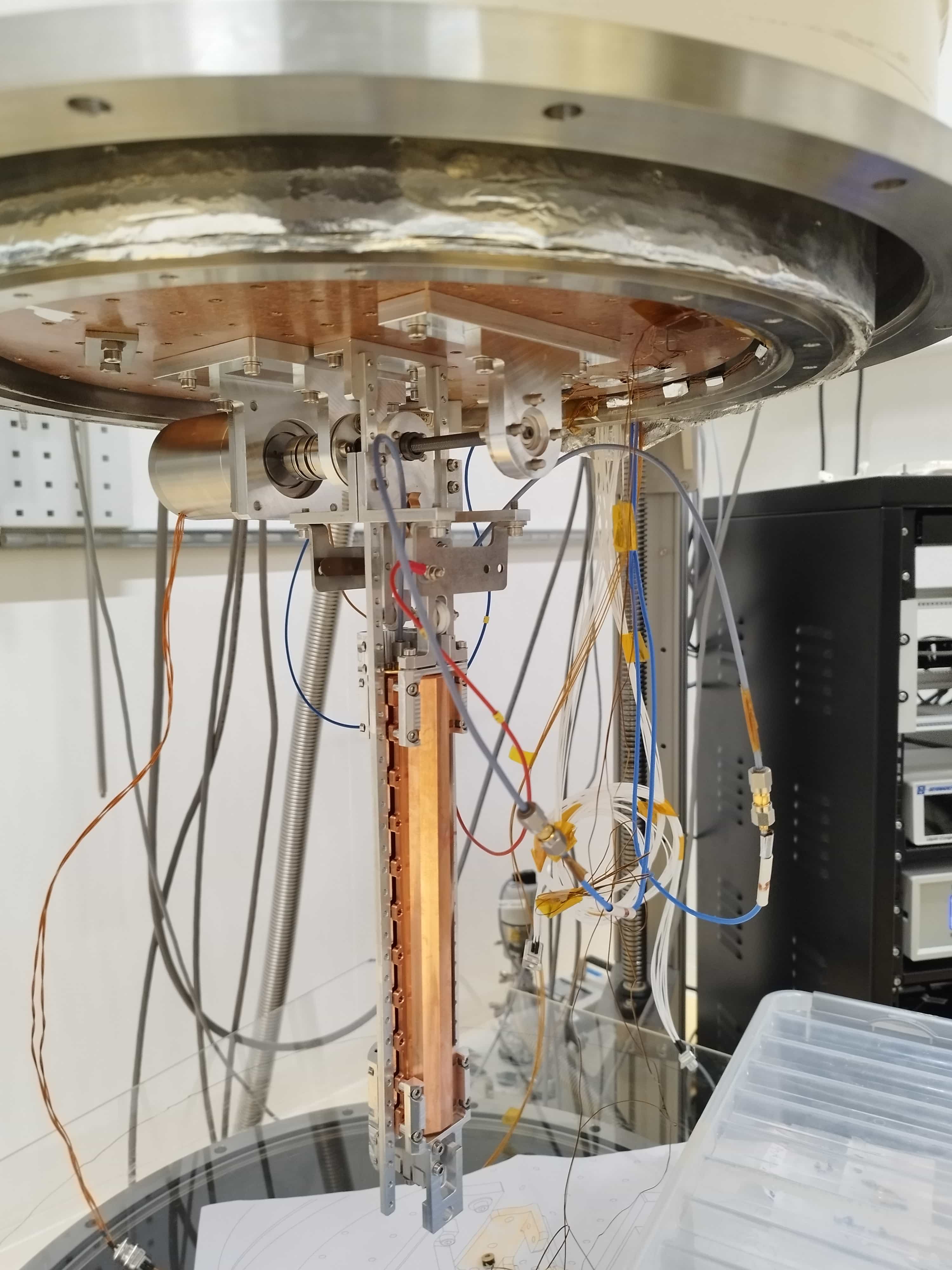}
         \caption{}
         \label{fig:CylCav_v2_atCRYOVAC}
\end{subfigure}
\caption{(a) 3D model of the room temperature support for the haloscope. (b) Picture of this support prototype at the laboratory. (c) Setup of the second version cavity installed at the 4K cryostat.}
\label{fig:CylCav_RT_characterization}
\end{figure}

Next, the electromagnetic characterisation of the second version of the cavity was carried out at room temperature. A study was performed on the electromagnetic behaviour and force required to adjust the gap on a dedicated aluminium support. Such a support is faster, easier and cheaper to produce than one made from copper. The support held the structure vertically on an optical table, simulating the configuration for a cryostat, as shown in the 3D model from Figure~\ref{fig:3Dmodel_CylCavSupport}.
The mounted prototype cylindrical VORTEX cavity is depicted in Figure~\ref{fig:RealCylCavSupport}. The setup also includes a copy of the probe rods from the dilution refrigerator system. For this prototype, a \textit{PI 230.25} stepper motor \cite{PI} was used.  The required specifications are: $25$~mm of travel range, $70$~N of force, step size of $0.05$~$\mu$m, and operating temperature range from $-20$ to $65$~$^o$C. The force to move one of the cavity halves was measured by a spring balance (see inset in Figure~\ref{fig:RealCylCavSupport}). The measured value was $0.1$~Newtons. Consequently, the cryo JPE nanopositioners mentioned in previous sections have sufficient strength for this setup. For a continuous scan of axion parameter space, a small motor step size is required, which depends on the quality factor, such that an overlap of the cavity resonance (e.g., at a $3$~dB fall-off) can be achieved. The resulting needed step size here is 0.5~$\mu$m, which is a value much higher than the accuracy achieved by the JPE nanopositioners. This consideration has been done assuming an increase of the unloaded quality factor from room temperature to cryo of about 2-4 times. A small step size is therefore of vital importance because of the narrowness of the Lorentzian shape.

\begin{figure} [htb]
    \centering
    \includegraphics[width=0.9\textwidth]{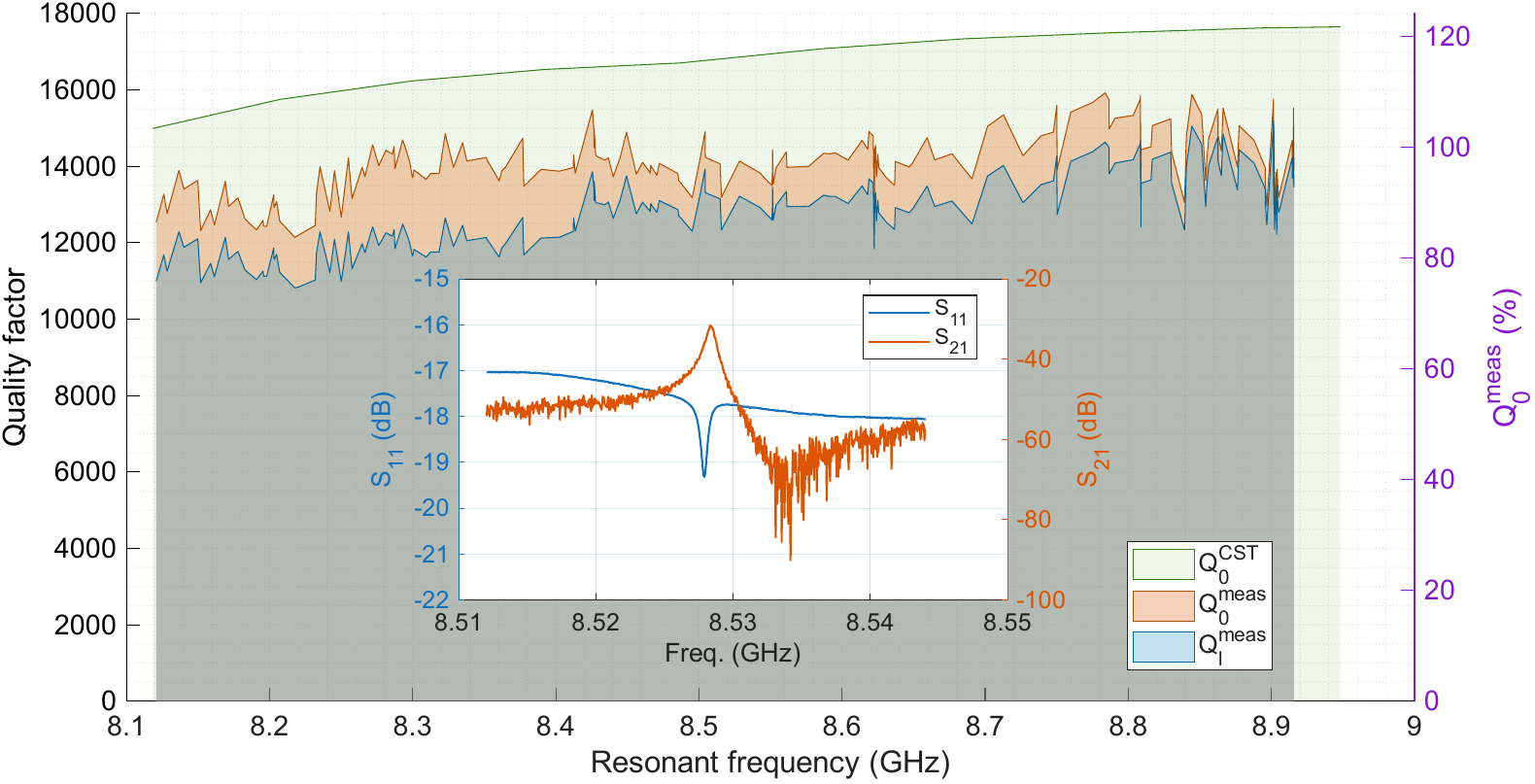}
    \caption{Measurement results at room temperature of the second version of the cavity mounted in the $4$~K cryostat. The plot shows the unloaded ($Q_0$, in red colour) and loaded ($Q_l$, in blue colour) quality factors versus the axion resonant frequency. CST simulation results are also shown, in green colour, for comparison. The right y-axis shows the deviation in $\%$ of $Q_0$ from the maximum frequency value ($gap = 0$~mm case). The inset depicts an example in the middle of the tuning range of the reflection ($S_{11}$) and transmission ($S_{21}$) parameters in $dB$.}
    \label{fig:CylCav_v2_Q0_vs_fr0_RT}
\end{figure}

Once these tests were completed, the cavity was installed in a \textit{Cryovac} cryostat. The same cryostat was later used for haloscope characterisation at $4$~K temperatures. As shown in Figure~\ref{fig:CylCav_v2_atCRYOVAC}, the cavity was installed in the different holding structures of the bottom thermalisation plate (which cools down to $4$~K) attached to a \textit{Phytron VSS-VSH} stepper motor \cite{Phytron}, chosen specifically for its reliable performance and compatibility at $4$~K temperatures. The motor is capable of operating at $4$~K and is adapted to the movement of the cavity housing by means of a small mechanical structure hooked onto the upper section of the assembly. This configuration was chosen because the cavity would be later installed in the dilution refrigerator. During installation various tests were carried out, including a study of the frequency tuning range and quality factor. Port 2 of the cavity was installed with a very weak coupling so that it did not affect the characterisation but, at the same time, allowed the transmission parameter ($S_{21}$) to be extracted. Therefore, only the contribution of the port 1 coupling ($\beta_1$) will be taken into account for the calculation of the unloaded quality factor. The results obtained are shown in Figure~\ref{fig:CylCav_v2_Q0_vs_fr0_RT}.
The value of $Q_0$ has been obtained considering the variation in $\beta_1$ from $0.5$ to $0.15$ along the tuning range, values that are significantly undercoupled at room temperature (hence the reason for obtaining relatively close values between $Q_0$ and $Q_l$). The $Q_0$ deviation is consistent with theoretical predictions at room temperature, showing a maximum reduction of $14$~$\%$ across the frequency range from $8.91$\footnote{Note that frequencies higher than $8.91$~GHz have not been achieved, since forcing $gap=0$~mm (being both halves of the cavity in contact) can cause misalignment due to imbalance of pressure, which would negatively impact the quality and form factors and the resonant frequency value. The implementation of alignment pins, which would ensure proper alignment and allow for a slightly higher frequency range, has been left for future work.} to $8.12$~GHz. The measurement of correct Q-factor and tuning values validates the alignment and range of the tuning mechanism. This conclusion is based on verifying that the $Q_0$ value has not been excessively lowered in certain cavity openings, as was seen to occur in the CST misalignment simulations mentioned in previous sections (see Figure~\ref{fig:CylCav2_fr_Q0_C_and_FoM_vs_Tilts_CST}). Furthermore, comparing simulation results with those of the measurements shows that approximately $85$~$\%$ of the theoretical values have been achieved across the whole frequency tuning range. A further improvement in Q values could probably be achieved by applying an electropolishing treatment, as previously demonstrated for the first version of the haloscope.

An asymmetry nature in the $TM_{010}$ mode resonance was observed, which was produced by the crosstalk between the two coupling ports. Consequently, the \textit{ARPE} calculation tool \cite{Krkotic:2021} was used to calculate the $Q_l$, $\beta$, and $Q_0$ parameters. This tool incorporates a correction technique to account for the crosstalk and thus yields a more accurate calculation than the classic $-3$~dB drop method for calculating $Q_l$. This crosstalk effect occurs due to the proximity of the two coaxial monopole ports, which introduces a significant mutual coupling. This forms a \textit{trisection} topology that introduces a transmission zero near the resonance, forcing the drop of one of the Lorentzian slopes \cite{Aparicio:2024}. For greater accuracy, this tool has been combined with the correction of the coupling losses in postprocessing in order to compensate for the effects of reflections produced by lack of calibration. Cable calibration is not possible at low temperatures inside the cryostat. Also, it accounts for defects in the cables and/or RF connectors which can be identified in the measurement by the separation of the $S_{11}$ baseline (value out of resonance) from the $0$~dB level.

\subsubsection{Third version (3-port, OFHC)}
\label{sss:CylCavv3_RT}

\begin{figure}[h]
\centering
\begin{subfigure}[b]{0.34\textwidth}
         \centering
         \includegraphics[width=1\textwidth]{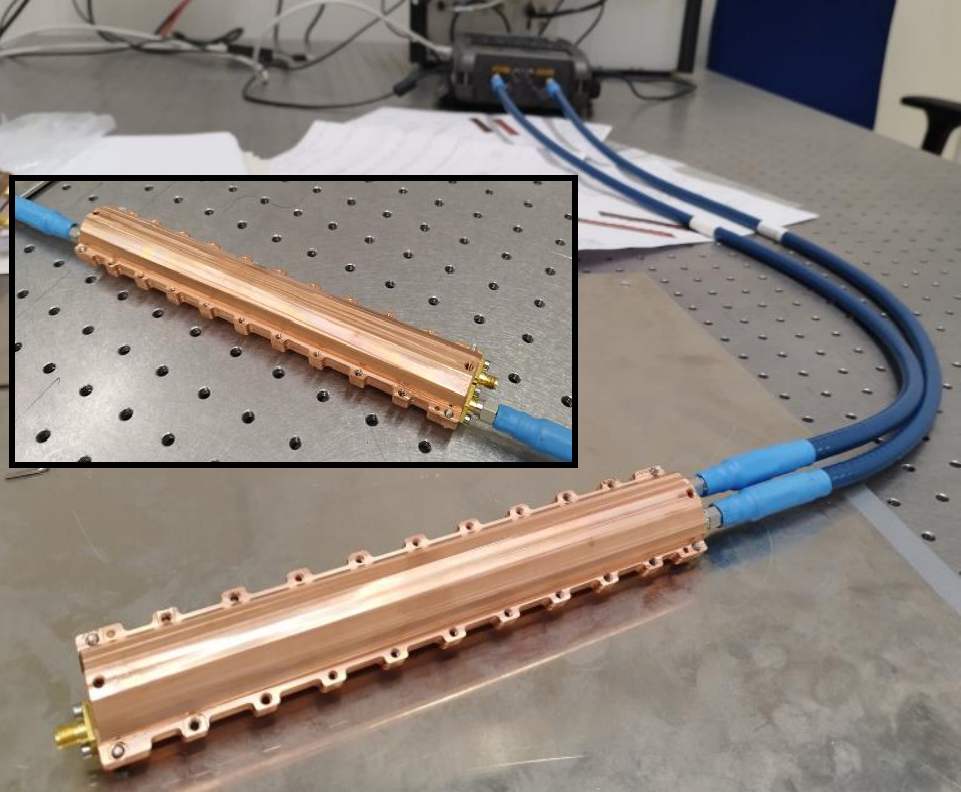}
         \caption{}
         \label{fig:CylCav_v3_RT_measurements_setup}
\end{subfigure}
\hfill
\begin{subfigure}[b]{0.65\textwidth}
         \centering
         \includegraphics[width=1\textwidth]{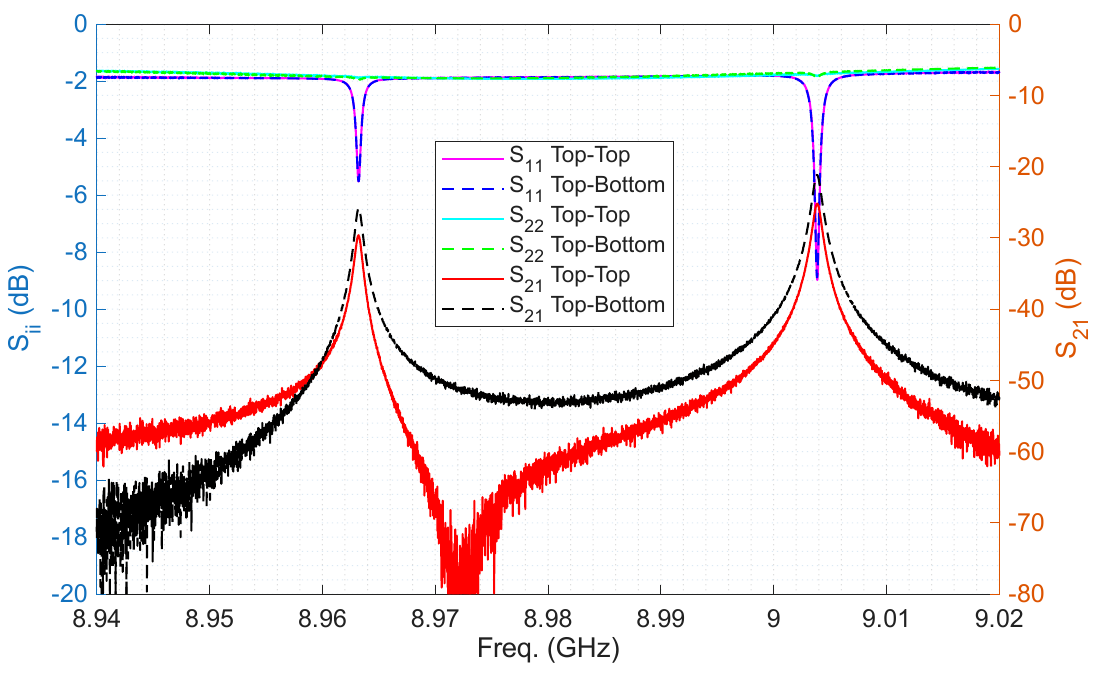}
         \caption{}
         \label{fig:CylCav_v3_RT_measurements_Sp}
\end{subfigure}
\caption{(a) Characterisation setup of the third version of the haloscope at room temperature employing both ports at the top and the ports at the top and bottom (inset). (b) Transmission and reflection parameters (in dB) versus frequency (in GHz) for both cases.}
\label{fig:CylCav_v3_RT_measurements}
\end{figure}

The third version of the haloscope was also measured at room temperature. Figure~\ref{fig:CylCav_v3_RT_measurements_setup} shows a picture of the measurement setup of this cavity, where the behaviour of the transmission parameter has been observed when using different combinations of two of the three cavity ports.
Figure \ref{fig:CylCav_v3_RT_measurements_Sp} details the results obtained when measuring the cavity in two different two-port configurations: one with both coaxial ports at the top, and the other with one port at the top and one at the bottom. For the latter configuration, the top port is located longitudinally opposite to the bottom one.

The results show asymmetric behaviour when the ports in use are located close to each other, as previously observed in cavity 2. This effect has a greater impact on stronger port input/output couplings (i.e., transmissions of higher magnitude).

To calculate the loaded quality factor ($Q_l$), the standard \textit{-3 dB method} method was applied to the measurements featuring a symmetric Lorentzian transmission. Conversely, for the non-symmetric measurements, a more complex process was required utilising the ARPE calculation tool \cite{Krkotic:2021}. Following this procedure, an unloaded quality factor ($Q_0$) of approximately $16800$ was obtained for both measurements at $f_r = 8.963$~GHz. This value represents $95$~$\%$ of the theoretical expectation, matching the performance of the second version of the haloscope when assembled with screws (consistent with their identical internal dimensions).

\subsection{Cryogenic temperatures}
\label{ss:Cryo_temperatures}

Following the measurements at room temperature, various tests at different cryogenic temperatures have been carried out on the three versions of the RADES-VORTEX haloscope.

As a preliminary and quick test at low temperature, the first haloscope prototype, before its electropolishing treatment, was immersed in a liquid nitrogen dewar bath to study the behaviour of the cavity response, primarily the increase in the $Q_0$-factor value. The study is shown in Appendix~\ref{AppA:LN2measurements}.

\subsubsection{Helium cryostat}
\label{sss:He_cryostate}

Tests of the second cavity were then carried out at temperatures of $4$~K using the \textit{Cryovac} cryostat, employing the setup shown in Figure~\ref{fig:CylCav_v2_atCRYOVAC}. Two long RF coaxial cables connected the RF output ports of the cryostat. For this measurement campaign, three calibrated \textit{CU Cernox} temperature sensors \cite{LakeshoreQD} were positioned within the experiment to check correct thermalisation. One of the sensors was placed on the motor ($t_m$), whilst the other two were located in each half of the cavity: one screwed into one of the screw assembly holes at the fixed cavity half ($t_1$) and the other attached to a flat section of the outer surface of the movable cavity half ($t_2$). Also, in this setup, the ports were configured to provide a critical coupling value at Port 1 ($\beta_1 \simeq 1$) and a very weak coupling at Port 2 ($\beta_2 \simeq 0$) at cryogenic temperatures.

\begin{figure} [htb]
    \centering
    \includegraphics[width=0.8\textwidth]{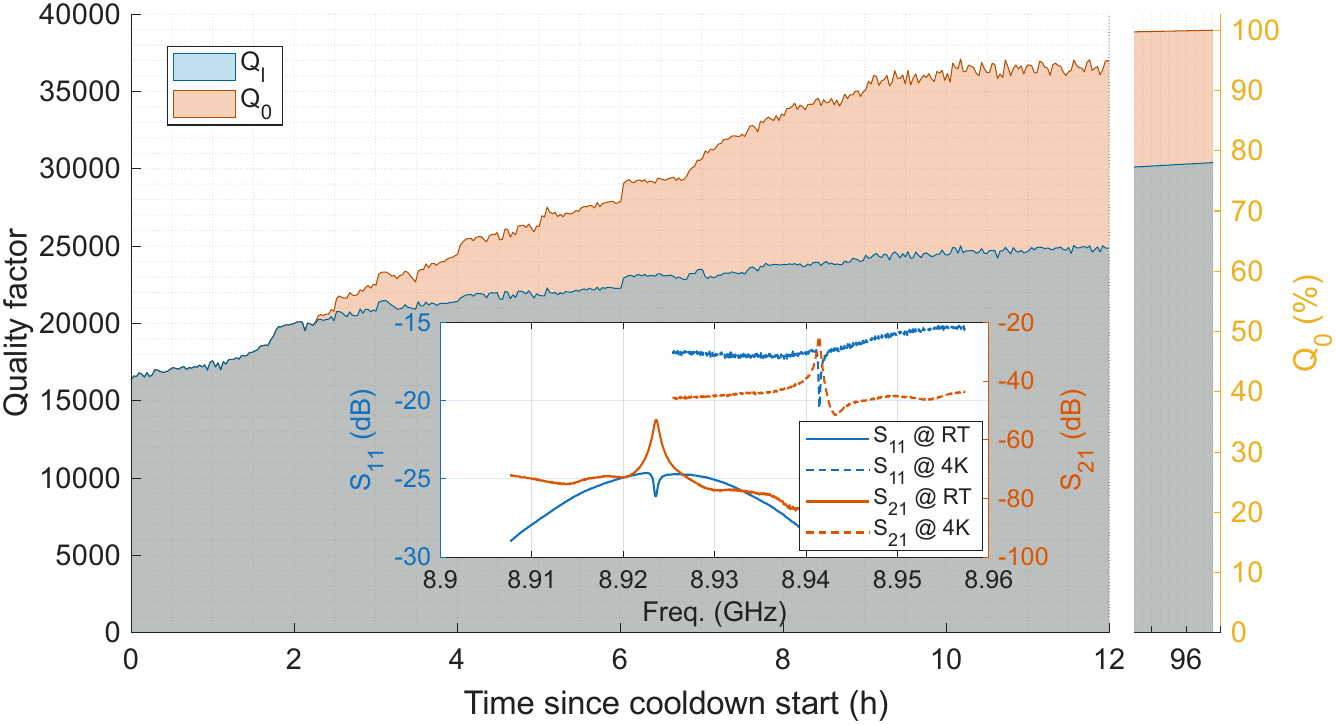}
    \caption{Measurement results during the cooldown of the second version of the cavity mounted in a $4$~K cryostat. The plot shows the unloaded ($Q_0$, in red colour) and loaded ($Q_l$, in blue colour) quality factors versus the number of hours elapsed since the start of the cooldown. The right y-axis shows the deviation in $\%$ of $Q_0$ from the maximum unloaded quality factor value (reached at $4$~K). The inset depicts an example of the Scattering parameters (reflection ($S_{11}$) and transmission ($S_{21}$) parameters in $dB$, in blue and red colours, respectively) before (solid lines) and after (dashed lines) the cooldown.}
    \label{fig:CylCav_v2_Q0_vs_Time_CD}
\end{figure}

Two sets of tests were carried out: during the cooldown (from $300$ to $4$~K), with $gap = 0$~mm, and at the cryogenic base temperature ($4$~K), testing different $gap$ values. For the first one, Figure~\ref{fig:CylCav_v2_Q0_vs_Time_CD} shows the variation of the quality factor ($Q_l$ and $Q_0$) throughout the cooldown.
To illustrate the dynamic behaviour of the cavity during thermal transition, the $Q_0$-factor is plotted against the time elapsed since the start of the cooldown process, where it is assumed a temperature versus time dependence. Furthermore, no data was collected during the cooldown from $77$ to $4$~K (a process different from the cooldown from $300$ to $77$~K), which is why there is a temporal gap between these two temperatures. Several conclusions can be drawn from this cooldown. It can be seen that the quality factor increased from RT to $4$~K, rising from $Q_0=16352$ to $Q_0=38955$ respectively, which represents an improvement of $\times2.38$\footnote{This value is lower than theoretically expected, but a better control of the manufacturing process and electropolishing could improve it.}. It can be seen that, from a certain temperature onwards (two hours after the start of the cooldown), Port 1 clearly becomes overcoupled, causing the loaded quality factor ($Q_l$) to deviate significantly from $Q_0$. This behaviour is driven by the increase in the coupling coefficient ($\beta_1$) as a result of the improved electrical conductivity in the haloscope material and the thermal contraction, where the reduction in the cavity internal dimensions effectively increases the relative penetration depth of the antenna. This occurs because the copper cavity body experiences a greater total volumetric contraction than the rigid stainless-steel support structure of the coaxial line, effectively forcing the antenna pin deeper into the resonant volume at cryogenic temperatures. This increase in the $\beta_1$ value can be confirmed by the Scattering parameters shown in the inset of the plot, where a general increase in the values (in dB) at cryogenic temperatures is also observed due to the reduction in losses in the cryostat internal coaxial cables. These coaxial cable losses decrease at cryogenic temperatures due to the enhanced electrical conductivity of the lines. 

Additionally, analytical calculations considering the thermal contraction of the resonant cavity indicate that the resonant frequency increases by approximately $25$~MHz due to a reduction of the cavity diameter around $75$~$\mu$m, which is close to the result obtained from measurements, where an increase in the resonant frequency of around $18$~MHz is also noted due to thermal contraction of the cavity; see inset from Figure~\ref{fig:CylCav_v2_Q0_vs_Time_CD}.

The lowest temperatures recorded for each temperature sensor were: $t_m = 8.43$~K, $t_1 = 11.1$~K, and $t_2 = 4.52$~K. Although the temperature reached by one half of the cavity is above the expected $4$~K baseline, it can be assumed that the cavity RF properties remain fully representative and stable. This is due to the fact that both the surface resistance (within the anomalous skin effect regime) and the thermal conductivity of OFHC copper asymptotically plateau below approximately $15$~K, meaning that this static thermal gradient does not introduce significant non-uniformities in the cavity behaviour. Furthermore, the quality factor and resonant frequency are experimentally observed to be well-stabilised during the final stages of the cooldown process.

\begin{figure} [htb]
    \centering
    \includegraphics[width=1\textwidth]{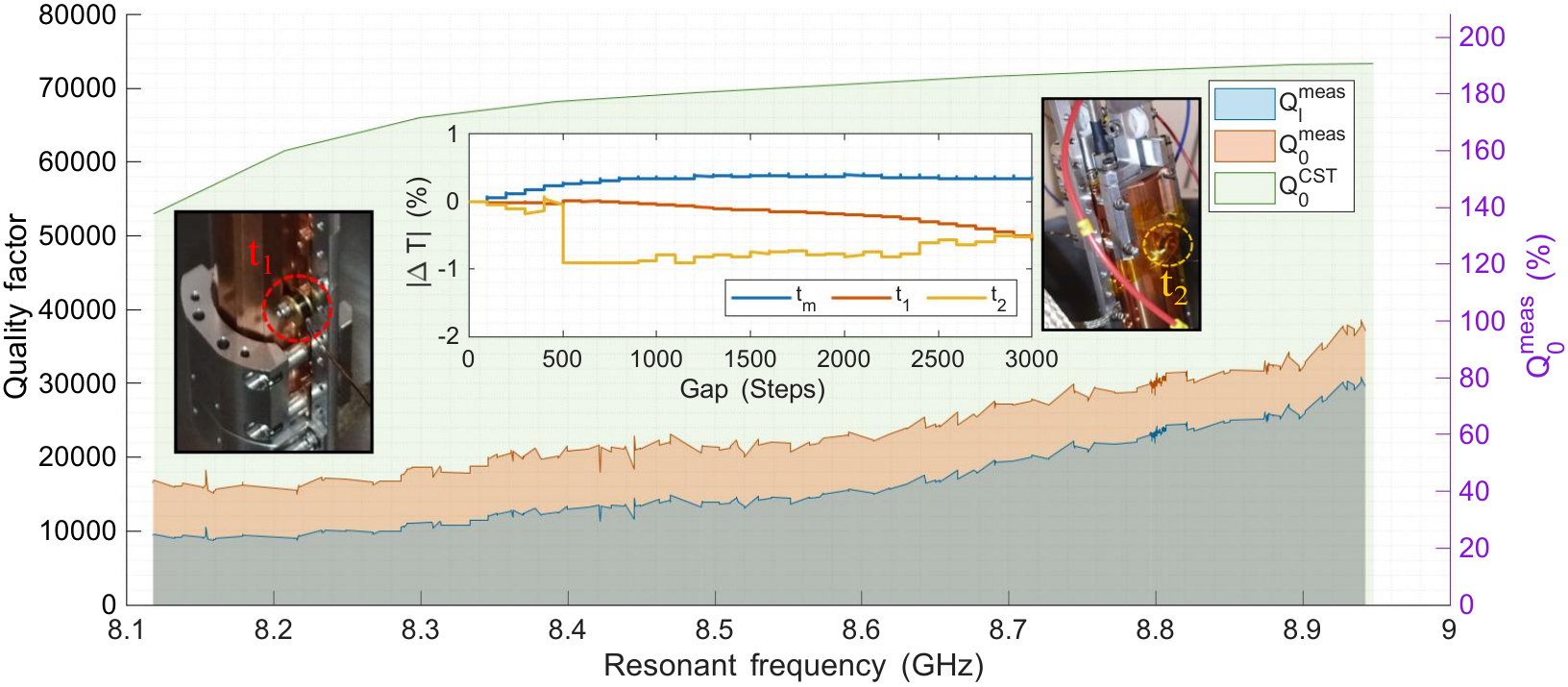}
    \caption{Measurement results at $4$~K of the second version of the cavity mounted at the \textit{Cryovac} cryostat. The plot shows the unloaded ($Q_0$, in red colour) and loaded ($Q_l$, in blue colour) quality factors versus the axion resonant frequency. CST simulation results are shown for comparison as well. The right y-axis shows the deviation in $\%$ of $Q_0$ from the maximum frequency value ($gap = 0$~mm case). The inset depicts the deviation from baseline (in $\%$) on the three temperature sensors to observe the heating when tuning is applied (gap change) at each specific motor step following the mentioned strategy.}
    \label{fig:CylCav_v2_Q0_vs_fr0_4K}
\end{figure}

Once at cryogenic base temperature, various tests were carried out by opening and closing the cavity gap with the motor, recording the values of $f_r$, $\beta$, $Q_l$ and $Q_0$, and continuously monitoring the temperature sensor, as heat sources may arise from the movement generated. The experimental sequence was automated via a Python-based controller to ensure thermal stability throughout the tuning range. This procedure follows an iterative loop: (i) mechanical adjustment of the cavity gap, (ii) a thermalisation standby period lasting until $\Delta T < 5$~$\%$ is restored (being $\Delta T$ the temperature variation of the sensors), and (iii) VNA data acquisition. Figure~\ref{fig:CylCav_v2_Q0_vs_fr0_4K} shows the results of the measurement of the quality factor plotted against the change in resonant frequency of the mode of interest ($TM_{010}$), with simulation results included.
An inset has been added to this plot showing the minimal temperature variation of the sensors attached to the experiment, confirming that this mechanical motion system moves correctly in cryogenic conditions, despite not having used the best materials for thermalisation (aluminium tuning assembly instead of brass, for example). Furthermore, although the results from Figure~\ref{fig:CylCav_v2_Q0_vs_fr0_4K} with the cavity closed differ significantly from those obtained in the simulation, an analysis of these results reveals that the loss of $Q_0$ remains smooth across the tuning frequency range, as is also the case in the simulation, thereby validating this system once again in this regard. It is worth noting that the small dips in the $Q_0$ value in this range are due to experimental noise and the limited frequency resolution of the VNA during data acquisition, not to mode-mixing effects with unwanted modes; the cavity response in the vicinity of the fundamental mode is quite clean throughout the range, since the higher-order modes vary almost in unison with the $gap$ value.

Although the thermalisation time in relation to the heat generated by the motor between one gap and the next has been investigated, a study of the system noise temperature behaviour during active tuning has been left as a task for future work in order to provide a comprehensive analysis of the actual measurement time in future axion searches using the VORTEX cavity.

\subsubsection{Dilution refrigerator system}
\label{sss:BF}

To complete the tests on the VORTEX haloscope at cryogenic temperatures, both version 2 and version 3 were installed in the Bluefors LD250 system, which will be used for data taking. The overall aim was to prepare and test the tuning (both $f_r$ and $\beta$ parameters) system at temperatures in the millikelvin range and at high magnetic fields. The magnet is capable of reaching fields  up to $12$~T.

\begin{figure} [htb]
    \centering
    \includegraphics[width=0.9\textwidth]{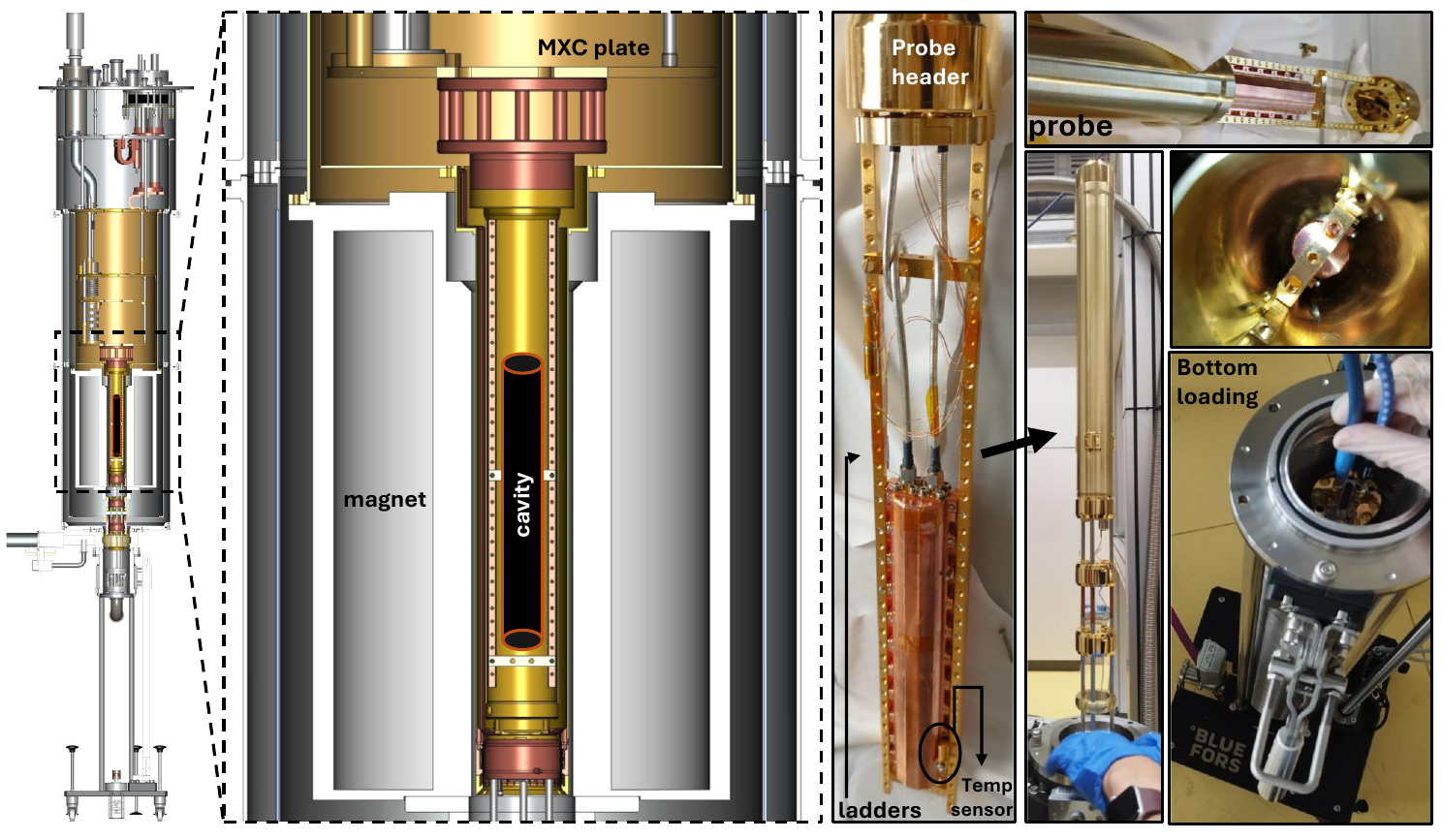}
    \caption{Mechanical integration and experimental assembly of the second cavity prototype in the bottom-loading probe. Left: 3D model cross-section view showing the cavity positioned inside the magnet bore and attached to the mixing chamber (MXC) plate. Right: Photograph sequence of the physical assembly, showing the cavity mounted on the thermalisation rods (ladders) with integrated temperature sensors (left), the fully assembled probe inserted into the cryostat structure (centre), and the bottom-loading insertion mechanism into the BlueFors dilution refrigerator (right).}
    \label{fig:CylCav_v2_AtBF_Bottom-Loading}
\end{figure}

Initially, preliminary tests were carried out on version 2 of the cavity using the FSE system (bottom-loading probe) to characterise its baseline properties. For these measurements, the cavity was mounted on the support/thermalisation rods (or ladders) without tuning, keeping the gap fixed at $0$~mm using screws to close both halves (see Figure~\ref{fig:CylCav_v2_AtBF_Bottom-Loading}).
This preliminary test aimed to verify the mechanical integration and thermal performance of the cavity within the restricted envelope of the bottom-loading probe before implementing the dynamic tuning system. In this case, a simplified RF configuration consisting of two lines was connected to each of the cavity ports with only attenuators in between the thermalisation plates of the dilution refrigerator (i.e., without circulators or amplifiers). The probe header features SMP connectors, so SMP-SMA adapters have been fitted to connect to the coaxial cables. The images show that, once the cavity is installed, there is more space left in the experimental chamber, meaning both the diameter and length could be increased for a future haloscope. Temperatures of hundreds of mK were achieved here, but the characterisation of $Q$ and $\beta$ was left for more advanced setups.

\begin{figure}[h]
\centering
\begin{subfigure}[b]{0.35\textwidth}
         \centering
         \includegraphics[width=1\textwidth]{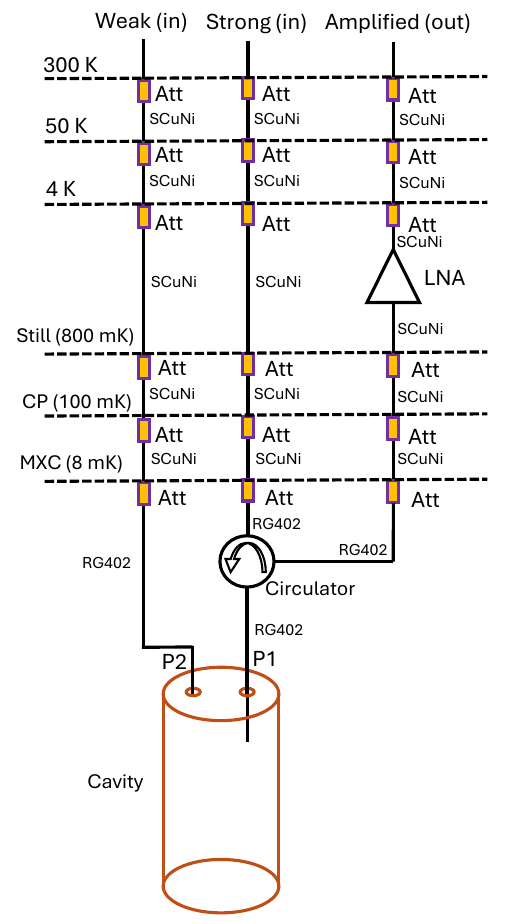}
         \caption{}
         \label{fig:BF_setup_v2}
\end{subfigure}
\begin{subfigure}[b]{0.45\textwidth}
         \centering
         \includegraphics[width=1\textwidth]{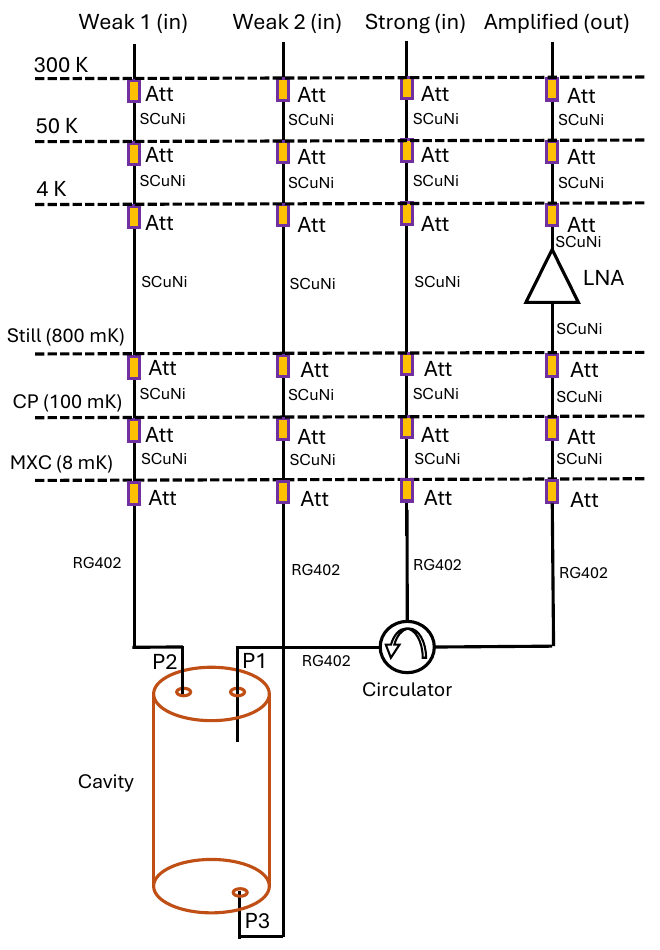}
         \caption{}
         \label{fig:BF_setup_v3}
\end{subfigure}
\caption{Setups in the dilution refrigerator for characterising different versions of the cylindrical cavity: (a) version 2, and (b) version 3. Different attenuators have been installed at the thermalisation plates for all the RF lines to, among other reasons, thermalise the inner conductor of the coaxial cables. The triangle represents a $40$~dB cryo amplifier (Low Noise Amplifier \textit{LNF-LNC4-16B} \cite{LNF}) installed at the $4$~K stage. Also, a circulator \textit{LNF-CIC8-12A} \cite{LNF} has been implemented at the MXC plate. \textit{SCuNi} coaxial cables have been implemented between thermalisation plates. At the experimental space, semi-flexible \textit{RG402} coaxial cables were used.}
\label{fig:BF_setup}
\end{figure}

Following these first verifications, the system was fully instrumented with the dynamic tuning setups described below. Figure~\ref{fig:BF_setup} shows the setups used for these measurements, which are based on the use of three (version 2 of the cavity) and four (version 3) RF lines for transmitting and receiving signals.
In Figure~\ref{fig:BF_setup_v2}, line \textit{weak} is connected to the weakly coupled port \textit{P2} ($\beta\approx0$). This line injects a signal via the VNA for transmission measurements, which are subsequently used for the extraction of the $Q_l$ parameter. On the other hand, line \textit{Amplified} is connected to the critical port \textit{P1}. This port is configured with $\beta=1$ for non-tunable scenarios, or $\beta=2$ to optimise the scanning rate in tunable systems. The signal from this port is extracted using a cryogenic LNA operating at $4$~K. Finally, line \textit{Strong} injects a signal from the VNA into the critical port to extract the configured $\beta$ value again via the amplified output. The use of a circulator allows the signals to be redirected according to the parameter to be measured. On the other hand, in Figure~\ref{fig:BF_setup_v3}, as this cavity introduces a third port far enough apart (\textit{P3}) to prevent asymmetric Lorentzian shapes in the transmission response, a fourth RF line (\textit{Weak 2}) has been implemented so that the $Q_l$ can also be measured in this way via the output line. As both \textit{P2} and \textit{P3} have been configured to provide very weak coupling, these two measured $Q_l$ values will be similar.

\begin{figure}[h]
\centering
\begin{subfigure}[b]{0.38\textwidth}
         \centering
         \includegraphics[width=1\textwidth]{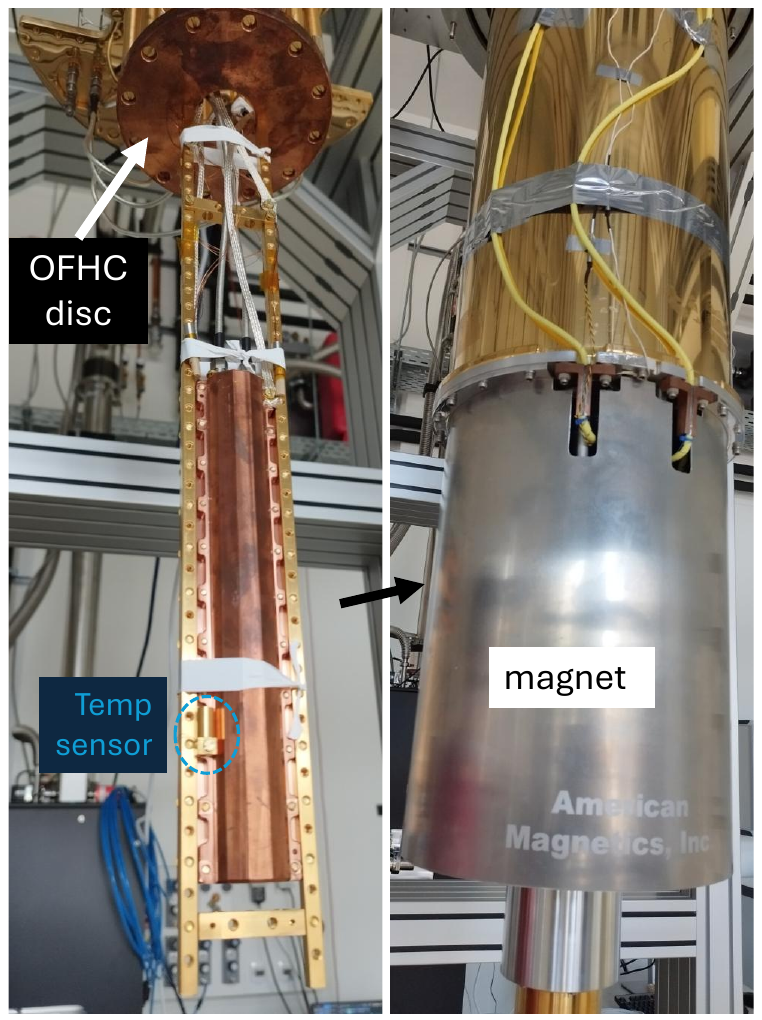}
         \caption{}
         \label{fig:CylCav_v2_AtCuDisc_withMagnet_Picture}
\end{subfigure}
\begin{subfigure}[b]{0.45\textwidth}
         \centering
         \includegraphics[width=1\textwidth]{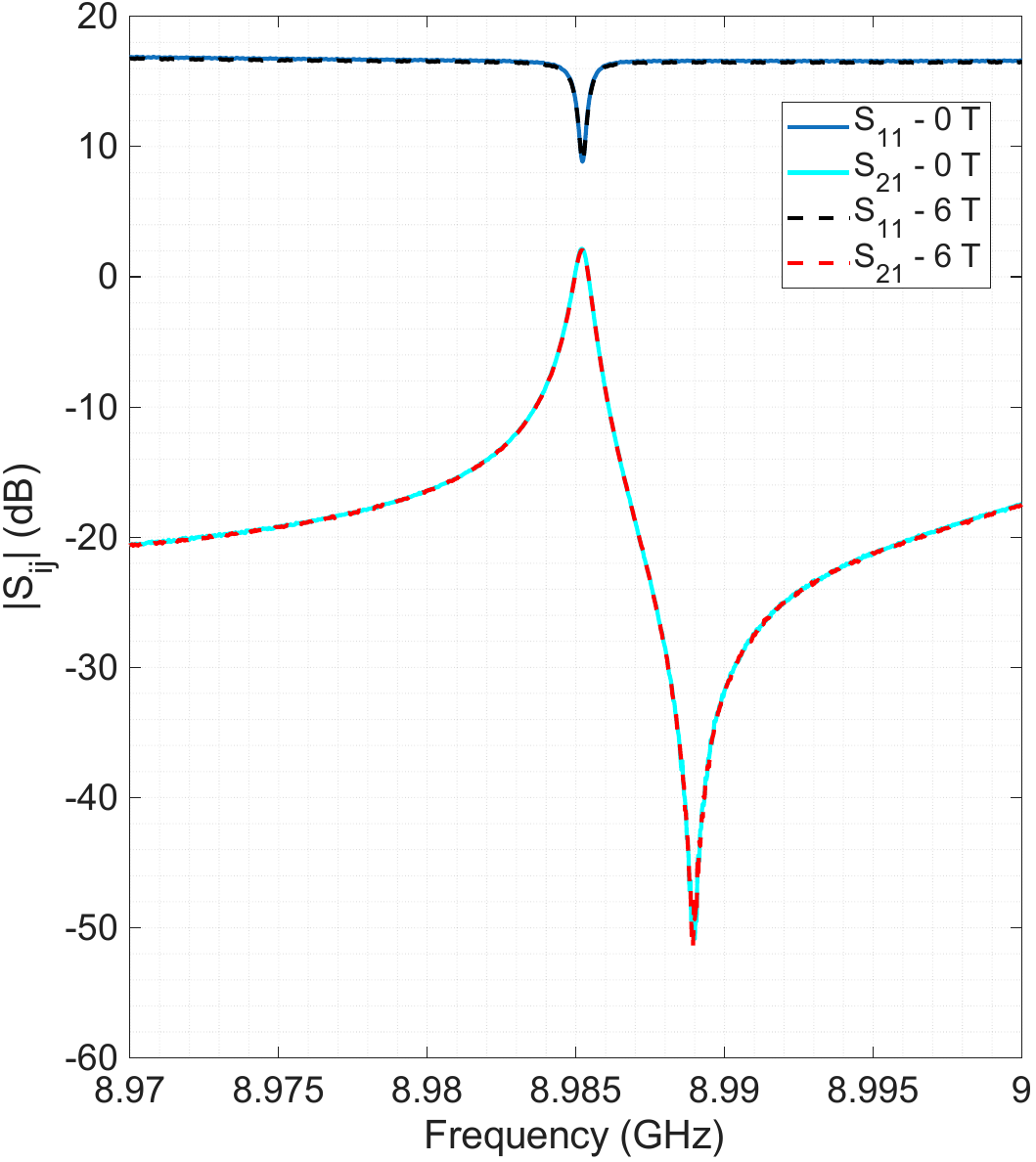}
         \caption{}
         \label{fig:CylCav_v2_AtCuDisc_withMagnet_Sp}
\end{subfigure}
\caption{Measurements with the second version of VORTEX at the OFHC disc with the magnet: (a) picture showing the installation process with the vertical position relative to the MXC plate (left) and the magnet inserted covering the experimental area (right); (b) magnitude of the reflection (blue and black colours) and transmission (cyan and red colours) parameters versus frequency at $0$ (solid lines) and $6$~T (dashed lines) magnetic field values.}
\label{fig:CylCav_v2_AtCuDisc_withMagnet}
\end{figure}

To investigate the influence of the static magnetic field on the cavity response, a dedicated test was carried out using the second version of the cavity and the RF configuration shown in Figure~\ref{fig:BF_setup_v2}. For this experiment, the FSE system was replaced by a circular OFHC copper support anchored directly below the MXC plate, providing a rigid connection for the golden rods (see Figure~\ref{fig:CylCav_v2_AtCuDisc_withMagnet_Picture}).
Once again, in this case, screws have been used to close the cavity completely (without any tuning system, with a gap of $0$~mm). Figure~\ref{fig:CylCav_v2_AtCuDisc_withMagnet_Sp} shows the scattering parameters obtained following a ramp-up from $0$ to $6$~T\footnote{Although this was not carried out due to time constraints, levels of $12$~T could be achieved.}. The cavity response changes only slightly in the presence of the magnetic field, as expected. Although not shown here, measurements were taken during the ramp-up at $0.5$~T intervals, and the behaviour remained the same in all cases. As an example, at $6$~T the quality factor and antenna coupling were computed, obtaining results of $Q_l=15229$, $\beta=2.1691$, and $Q_0=Q_l(1+\beta)=48262$, which corresponds to $66.3$~$\%$ of the value obtained in the simulation (see subsection~\ref{ss:CylindricalCavityDesign}). It has been found that this value could be significantly improved through surface treatments and enhancements in the screw sections to close the cavity. On the other hand, as can be seen, the scattering parameters have very high values, or even exceed $0$~dB. This is due to a lack of calibration of the VNA and to the need to compensate for the amplification provided by the amplifier against the attenuation introduced by the attenuators in the dilution refrigerator RF lines; however, this does not affect the characterisation of the cavity in this particular case.

\begin{figure}[h]
\centering
\begin{subfigure}[b]{0.3\textwidth}
         \centering
         \includegraphics[width=1\textwidth]{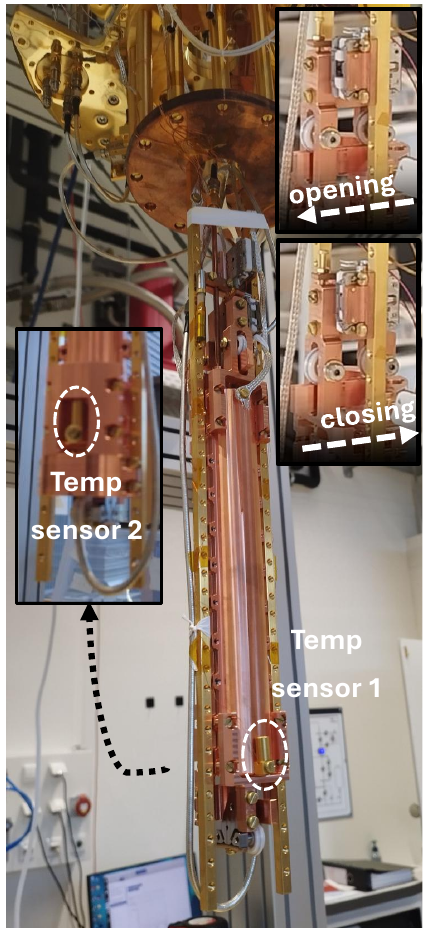}
         \caption{}
         \label{fig:CylCav_v3_AtCuDisc_Picture}
\end{subfigure}
\begin{subfigure}[b]{0.65\textwidth}
         \centering
         \includegraphics[width=1\textwidth]{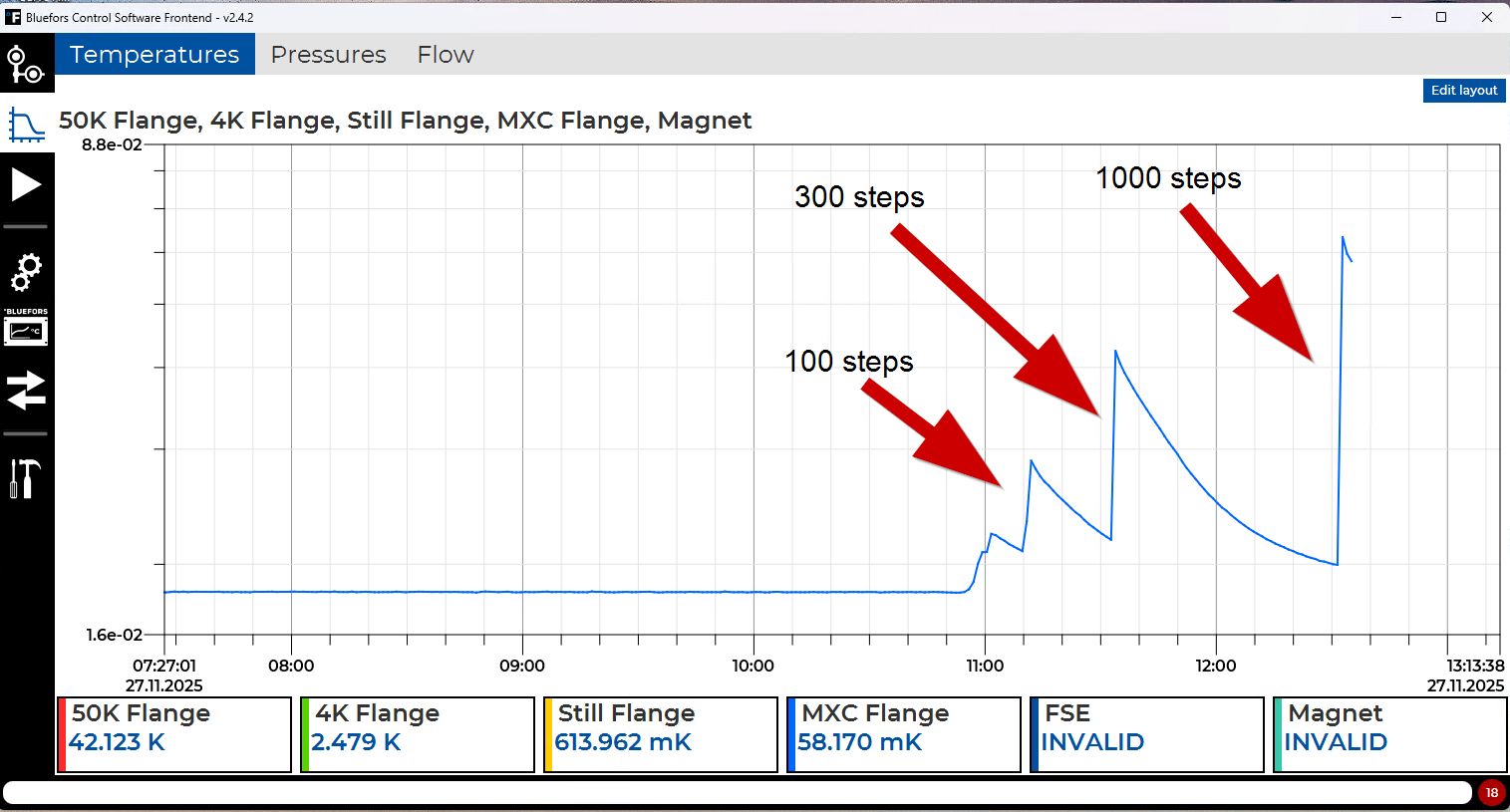}
         \caption{}
         \label{fig:TuningImpactOnMXCTempSensor}
\end{subfigure}
\caption{Experimental evaluation of the third version of the VORTEX haloscope equipped with JPE cryogenic nanopositioners. (a) Photograph of the experimental assembly detailing the placement of temperature sensors on the structure, along with insets illustrating the opening and closing mechanical operation of the tuning system. (b) Real-time monitoring of the MXC plate temperature response during controlled tuning actuation, highlighting the localised thermal dissipation produced by steps of 100, 300, and 1000 movement steps.}
\label{fig:CylCav_v3_AtCuDisc}
\end{figure}

Finally, as the final and most comprehensive test, the latest (3rd) version of the cavity has been installed on this circular copper mount above the gold rods, using the JPE nanopositioners, with the setup shown in Figure~\ref{fig:BF_setup_v3}. For these measurements, the magnet was not implemented for the sake of simplicity. Appropriate DC cabling has been routed from the experimental area to the outside of the dilution refrigerator to bias these positioners and to read the integrated closed-loop system, which provides real-time information on the actual position, serving as a redundancy alongside the resonant frequency and $\beta$ parameter values to determine the current position of both mechanisms. Figure~\ref{fig:CylCav_v3_AtCuDisc_Picture} shows an image of the complete installation.
As can be seen in the image, two calibrated \textit{Rox ruthenium oxide 102B RTD} temperature sensors from \textit{Quantum Design} \cite{LakeshoreQD} have been installed, one for each half of the cavity, securely fastened and thermalised in place using screw holes. On the other hand, although an effort has been made to ensure that this picture shows all the details of the system, readers are encouraged to refer to Figures~\ref{fig:3DmodelTuningAssembly} and \ref{fig:BetaReadjustmentSystem} as well to gain a better understanding of the movable systems ($f_r$ and $\beta$ tunings, respectively).

The temperatures achieved were $44$~mK for the movable half of the cavity (sensor 1) and $134$~mK for the fixed half-cavity (sensor 2). The temperature of sensor 1 is higher because this is where the RF cables are physically connected; these cables can introduce a heating source when injecting signals\footnote{A procedure following the prescription of \cite{Zhu:2022cuw} has also been implemented in this setup.} from the VNA for the characterisation of the EM response. Similarly, it has been observed that there is still room for improvement in the thermalisation points to achieve better performance. For example, for the implementation of a Travelling Wave Parametric Amplifier (TWPA), it has been observed that the temperature would need to be reduced to below $100$~mK, which will be addressed in future work on this haloscope.

Heating tests were carried out by moving both JPE nanopositioners that control the position of the critical antenna (vertical movement) and the cavity gap (horizontal movement). By moving both nanopositioners $100$~steps (far more steps than necessary to correct both the resonant frequency and the $\beta$ parameter), it was found that about $2$~minutes were needed to return to the base temperature. Figure~\ref{fig:TuningImpactOnMXCTempSensor} shows the temperature variation at the sensor attached to the MXC plate of the dilution refrigerator for different steps of the cavity frequency tuning: $100$ ($\Delta f = 4$~MHz, where $\Delta f$ is the frequency shift), $300$ ($\Delta f = 17$~MHz) and $1000$ ($\Delta f = 63$~MHz). Also, lower value of steps were tested: with $50$~steps at the antenna motion (small changes in $\beta$ value), the temperature sensors only suffered a maximum impact of $+10$~mK; while with $10$~steps at the cavity gap ($\Delta f = 250$~KHz), the temperature sensors were raised by around $20$~mK. This result already provides an encouraging perspective for a data-taking campaign with the setup: small thermalisation times are mandatory to achieve an optimal scanning rate and avoid dead time. Given that re-thermalisation can be achieved within a few minutes, we expect the motorised movement of the cavity half to be conceptually employable for the final data-taking setup.

Cavity transmission and reflection responses similar to those shown in Figure~\ref{fig:CylCav_v2_AtCuDisc_withMagnet_Sp} were obtained. By moving the antenna, it was possible to correct the $\beta$ parameter across the entire frequency tuning range.

Table~\ref{tab:CavMeasurementsSummary} provides a summary of all the parameters extracted from the measurements taken in the different versions of the VORTEX cavities.
\begin{table*}[htbp]
\setlength{\tabcolsep}{3.5pt} 
\resizebox{\textwidth}{!}{%
\begin{tabular}{|l|c|c|c|c|c|c|c|}
\hline
\textbf{Meas. results} & 
\textbf{\begin{tabular}[c]{@{}c@{}}Cav. \#1\\ (RT)\end{tabular}} & 
\textbf{\begin{tabular}[c]{@{}c@{}}Cav. \#2\\ (RT)\end{tabular}} & 
\textbf{\begin{tabular}[c]{@{}c@{}}Cav. \#3\\ (RT)\end{tabular}} & 
\textbf{\begin{tabular}[c]{@{}c@{}}Cav. \#1\\ (LN$_2$)\end{tabular}} & 
\textbf{\begin{tabular}[c]{@{}c@{}}Cav. \#2\\ ($4$~K)\end{tabular}} & 
\textbf{\begin{tabular}[c]{@{}c@{}}Cav. \#2\\ (mK)\end{tabular}} & 
\textbf{\begin{tabular}[c]{@{}c@{}}Cav. \#3\\ (mK)\end{tabular}} \\ \hline

$Q_0$-factor & 
$9100$--$17300$ & 
$12250$--$16000$ & 
$\sim16800$ & 
$24200$ & 
$17000$--$39000$ & 
$48300$ & 
- \\ \hline

$f_r$ tuning (GHz) & 
$8.26$--$8.89$ & 
$8.12$--$8.91$ & 
$8.96$ (fixed) & 
$7.32$ (fixed) & 
$8.12$--$8.93$ & 
$8.986$ (fixed) & 
- \\ \hline

Temp. (K) & 
300 & 
300 & 
300 & 
$\sim77$ & 
$4.52$--$11.1$ & 
$\sim0.4$ & 
$0.044$--$0.134$ \\ \hline

B-field (T) & 
- & 
- & 
- & 
- & 
- & 
$0$ to $6$ & 
- \\ \hline

Outcomes & 
\begin{tabular}[c]{@{}c@{}}EP gives $+7.5$~$\%$ in $Q_0$.\\ $90$~$\%$ of $Q_0$ simulated.\end{tabular} & 
\begin{tabular}[c]{@{}c@{}}Reduced $Q_0$\\ degradation.\end{tabular} & 
\begin{tabular}[c]{@{}c@{}}$S_{21}$ symmetry\\ ($Q_0$ more accurate).\end{tabular} & 
\begin{tabular}[c]{@{}c@{}}Good $Q_0$ scaling \\ w/ temperature.\end{tabular} & 
\begin{tabular}[c]{@{}c@{}}Automated \\ tuning control.\end{tabular} & 
\begin{tabular}[c]{@{}c@{}}$Q_0$ stable w/\\ ramping-up/down.\end{tabular} & 
\begin{tabular}[c]{@{}c@{}}Full $f_r$ and $\beta$ control.\\ Good thermalisation.\end{tabular} \\ \hline

Limitations & 
\begin{tabular}[c]{@{}c@{}}Port crosstalk.\\ Thin wall effect.\end{tabular} & 
\begin{tabular}[c]{@{}c@{}}Port crosstalk.\\ No EP.\end{tabular} & 
No EP. & 
\begin{tabular}[c]{@{}c@{}}$LN_2$ boiling\\ ($\varepsilon_r$ fluctuation).\end{tabular} & 
\begin{tabular}[c]{@{}c@{}}Absolute simulation\\ values drift. No EP.\end{tabular} & 
No tuning. No EP. & 
\begin{tabular}[c]{@{}c@{}}Not enough data\\ for plots. No EP.\end{tabular} \\ \hline
\end{tabular}%
}
\centering
\caption{Summary of the cavity performance test results, structured by key measurement parameters across different versions and temperature scenarios. Also indicated are relevant cavity production features such as electropolishing (EP) that impact the performance.}
\label{tab:CavMeasurementsSummary}
\end{table*}

\section{Bead-pull analysis}
\label{s:Bead-pull_analysis}

An accurate determination of the electromagnetic field pattern of the operating resonant mode is essential for the proper characterisation of the axion form factor and, consequently, the sensitivity of the instrument. Small deviations in geometry can result in changes of the field distribution that can directly translate into reduced sensitivity in the form factor $C$, one of the key parameters entering the axion signal power expression.

Although it is possible to obtain field distributions through numerical simulations, these represent idealised geometries and therefore do not fully capture actual imperfections such as mechanical misalignments.

An advantage of the split cavity design is that the resonant modes inside the cavity can be mapped with a bead-pull technique using the space provided by the tuning gap and without introducing dedicated apertures.
The bead-pull method is common and established in accelerator physics applications \cite{SHI201314} and has recently also attracted attention in axion physics. It is especially useful for non-trivial cavity or resonator designs \cite{DiVora:2025,Rapidis:2018BeadPerturbation}. It enables mapping of the spatial distribution of the electric field inside the cavity. The method consists of introducing a small, localised perturbation, typically a dielectric bead, and moving it through the resonant volume. Using Slater’s perturbation theory \cite{1984JAP....55.2648D}, the induced frequency shift, or the change in scattering parameters, can be related to the local electric field amplitude, as follows,
\begin{equation}
 \label{eq:B-P}   
|\mathbf{E_{f_{res}}}(r)| \propto \sqrt{\frac{\Delta f(r)}{f_0}} ~~~,~~~~ |\mathbf{E}(r,f)| \propto \Delta S_{11}(r,f) = |S_{11}^{\text{pert}}(r,f) - S_{11}^{\text{unpert}}(r,f)|,
\end{equation}
where $f_0$ is the unperturbed resonance frequency, $\Delta f$ is the frequency shift caused by the perturbation at position $r$, and $S_{11}^{\text{pert}}$ and $S_{11}^{\text{unpert}}$ denote the scattering parameters obtained with the VNA with and without the perturbation, respectively.

Because of the perturbative character of the approximation in this work, it is important to have an understanding of the influence of the parameters of the bead on this experiment. Perturbation theory \cite{pozar2012microwave} provides a relation between the relative frequency shift for a small dielectric bead and its volume and permittivity. This magnitude scales with bead volume and linearly with its permittivity, as follows:
\begin{equation}
 \dfrac{\Delta f }{f_0} = (\varepsilon_r - 1) r_b^3. 
\label{eq:beadpull_param}
\end{equation}

This technique not only provides a direct validation of the simulated mode field pattern and an experimental validation of the cavity’s electromagnetic performance but also allows identification of mode localisation, or unintended hybridisation, offering valuable insight into how fabrication tolerances or assembly imperfections affect the field distribution. For example, an angular misalignment of the cavity halves $\theta_x$ breaks the symmetry of the $TM_{010}$ field, producing localisation towards the end of the cavity, which can be identified via a bead pull measurement.

\subsection{Description of the concept and setup}
\label{ss:Description_of_the_concept_and_setup}

\begin{figure}[h]
    \centering    \includegraphics[width=0.7\textwidth]{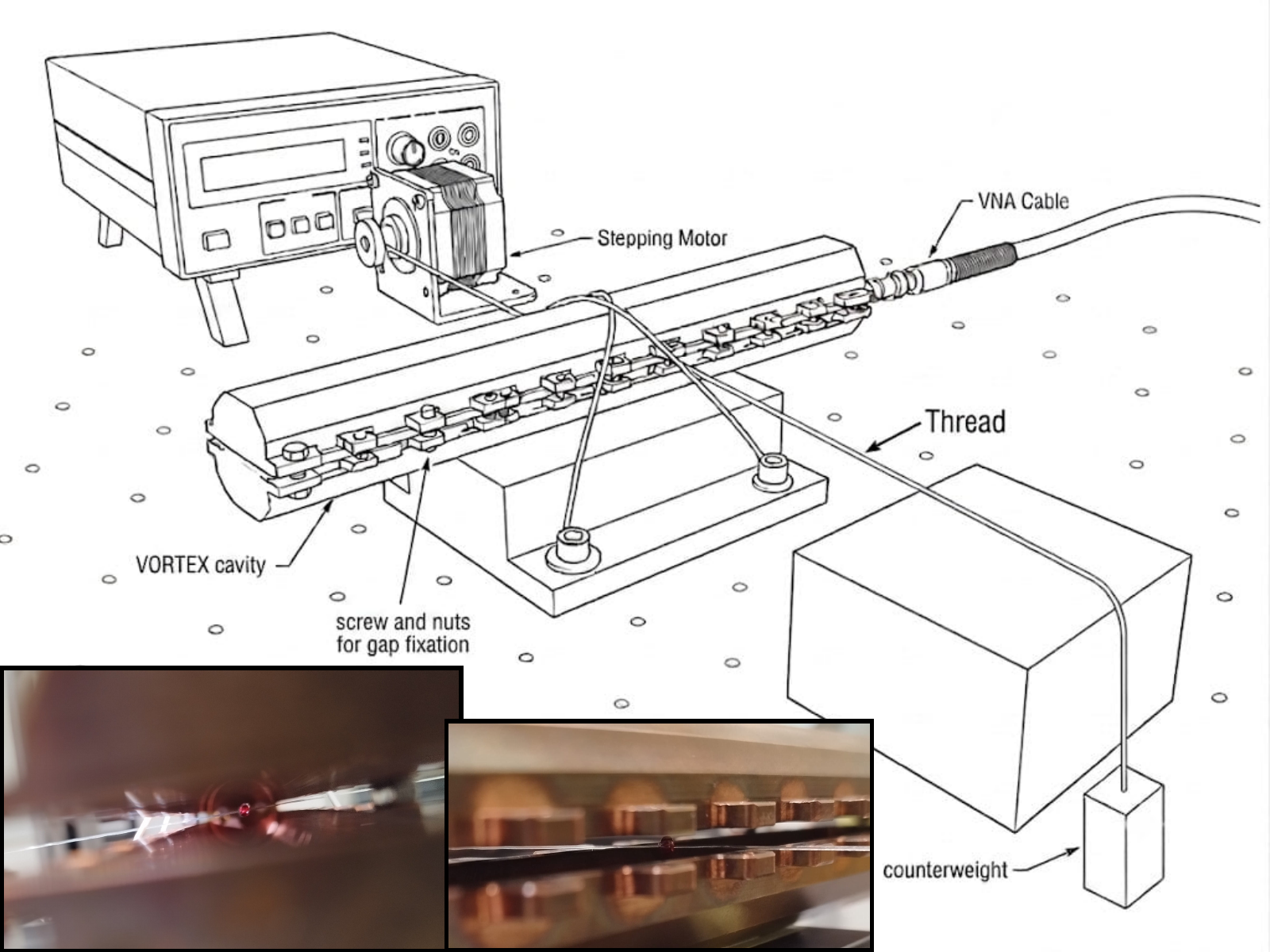}
    \caption{Drawing of the bead pull setup. Two insets show some details from the real implementation focused on the bead position inside the cavity area.}
    \label{fig:Bead_pull_setup}
\end{figure}

A first bead-pull setup was conceived as a proof of principle for field mapping within the first prototype of the cylindrical cavity (2-port, standard copper). The experimental arrangement consisted of a cavity mounted and fixed horizontally on an optical table. A small plastic toroidal bead of approximately $3$~mm in diameter, suspended on a thin nylon monofilament thread, was translated across the cavity aperture using a simple drive composed of a stepper motor and a counterweight system. The position of the bead along each transverse plane was referenced through mechanical screws acting as alignment markers. A scheme of this setup is shown in Figure~\ref{fig:Bead_pull_setup}.

These measurements served primarily to test the feasibility of the technique. They allowed estimating achievable spatial resolution, identifying sources of systematic uncertainty, and determining practical limitations of materials and mechanical stability that could not have been anticipated without an experimental trial. The results of this experimental campaign are discussed in section~\ref{ss:BeadPullMeasurements}, which are subsequently compared with electromagnetic simulations and tuning analyses in sections~\ref{ss:CST_simulation_results} and \ref{ss:CST_simulation_results_misalignment}.

\subsection{Measurements}
\label{ss:BeadPullMeasurements}

\begin{figure}[h]
    \centering
    \includegraphics[width=0.8\textwidth]{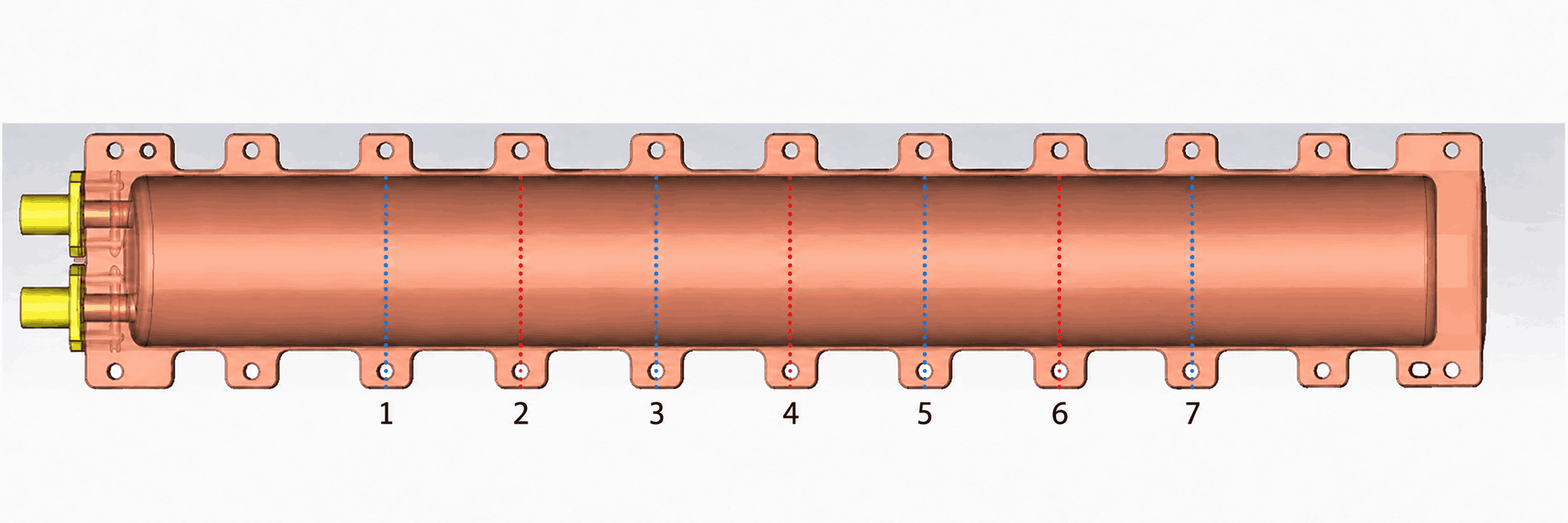}
    \caption{Positions studied along the cylindrical housing for the first version of the VORTEX cavity. The red dotted positions (2, 4 and 6) indicate sections that were analysed but whose results are not presented in this work. The blue dotted positions (1, 3, 5 and 7) indicate sections for which the results are presented below.}
    \label{fig:planes_studied}
\end{figure}

A series of measurements was carried out using the setup described previously for several transverse cavity height positions, corresponding to one-dimensional radial measurements along the centre of the cavity, as described in Figure~\ref{fig:planes_studied}.
A total of seven positions were measured. These were chosen to be equidistant from each other and in the middle section of the cavity. To ensure accurate and equidistant positioning, the measurement planes were aligned with the centres of the mechanical protrusions, utilising the pre-existing screwing holes of the cavity housing as physical references (see Figure~\ref{fig:planes_studied}). This allowed for a characterisation of the field in the central section of the cavity. For simplicity, each of the positions has been numbered; such enumeration will receive the name of the plane index.

\begin{figure}[h]
\centering
\begin{subfigure}[b]{0.35\textwidth}
         \centering
         \includegraphics[width=1\textwidth]{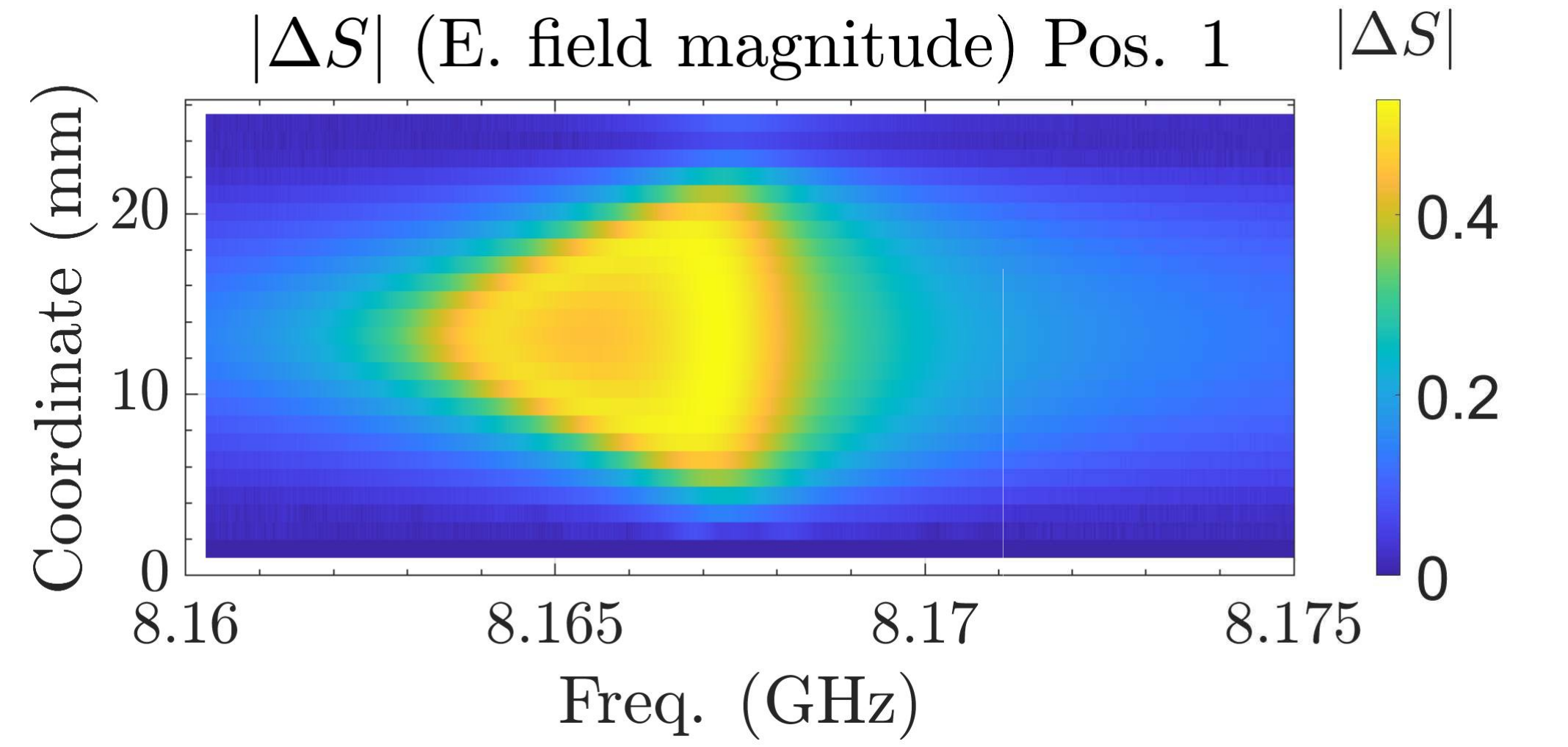}
         \caption{}

\end{subfigure}
\begin{subfigure}[b]{0.35\textwidth}
         \centering
         \includegraphics[width=1\textwidth]{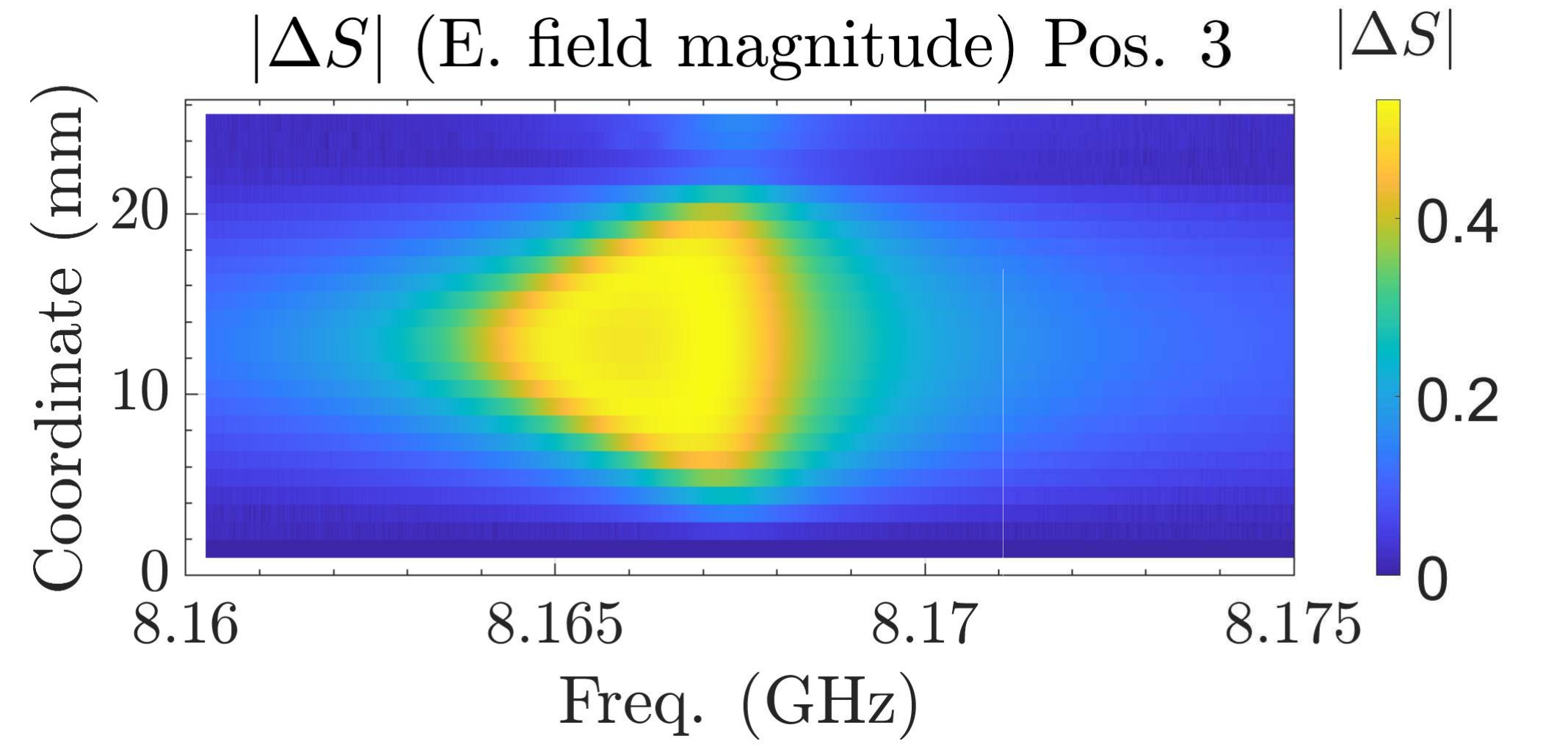}
         \caption{}
         
\end{subfigure} \\
\begin{subfigure}[b]{0.35\textwidth}
         \centering
         \includegraphics[width=1\textwidth]{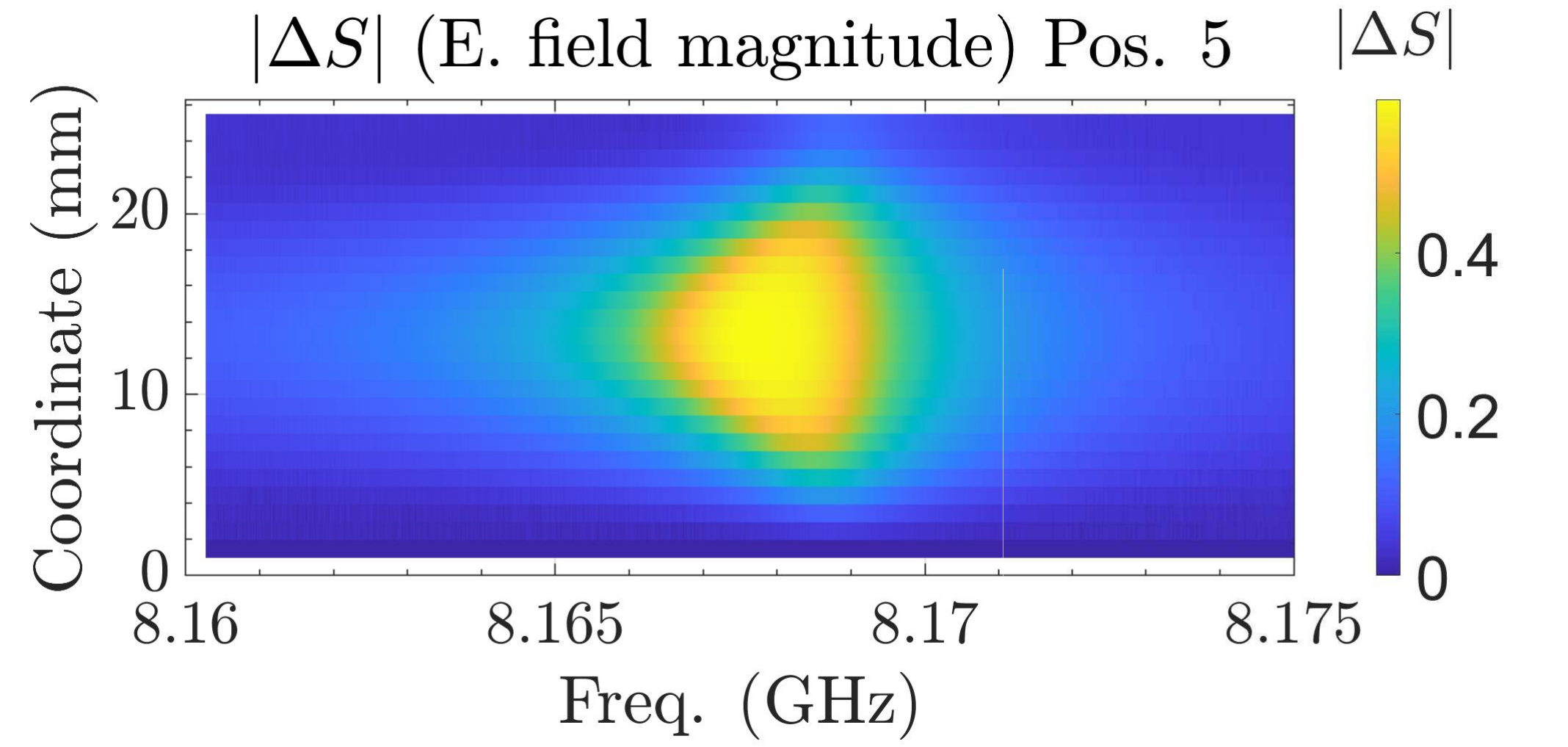}
         \caption{}

\end{subfigure}
\begin{subfigure}[b]{0.35\textwidth}
         \centering
         \includegraphics[width=1\textwidth]{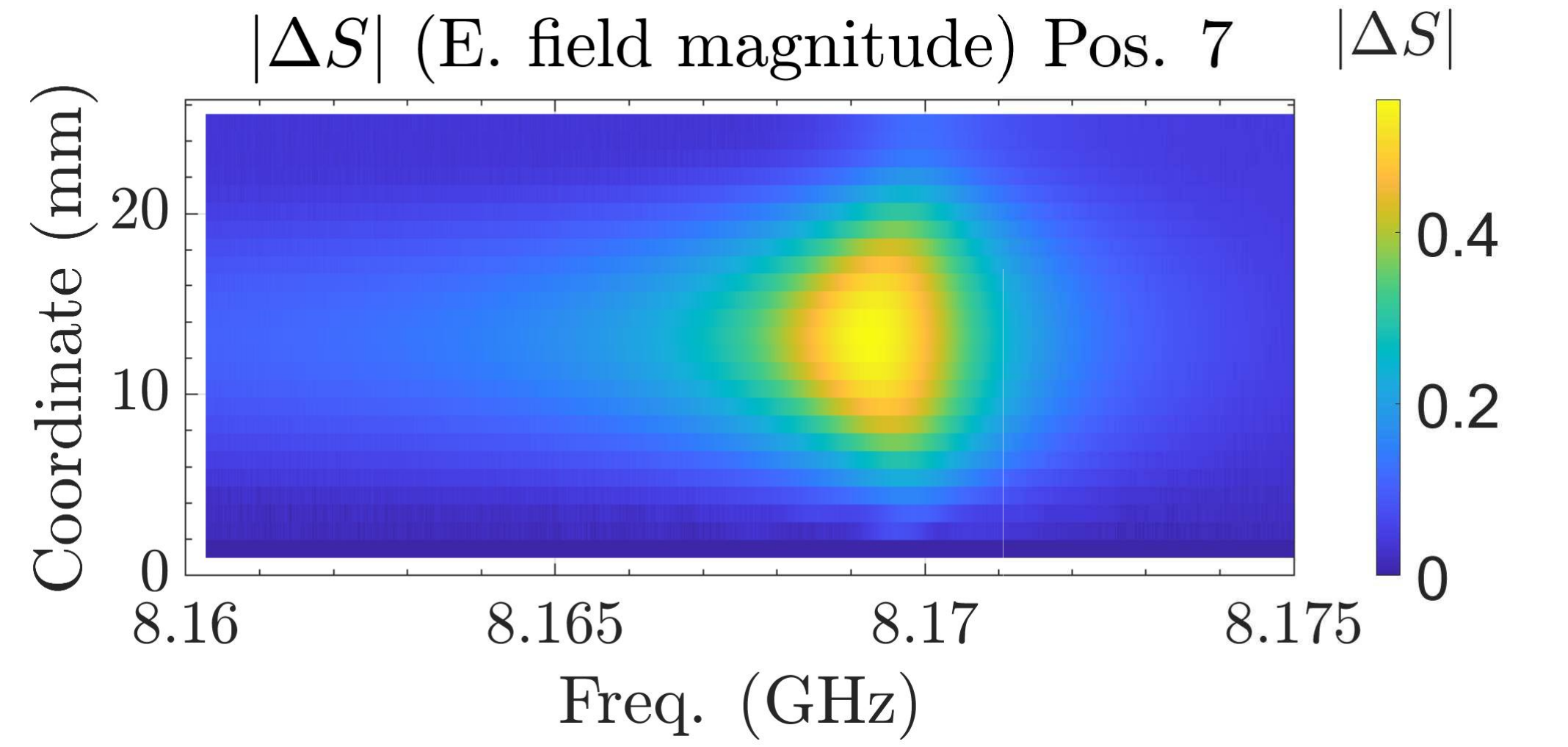}
         \caption{}

\end{subfigure}
\caption{Measured $|\Delta S|$ ($\propto$ E-field) in frequency space for positions (a) 1, (b) 3, (c) 5 and (d) 7 of the cavity depicted in Figure~\ref{fig:planes_studied}.}
\label{fig:BP-Preeliminary1}
\end{figure}

The results of the selected positions are presented in Figure~\ref{fig:BP-Preeliminary1}, which shows the spatial and frequency distribution of the electric field inside the cavity. As shown in these plots, the measured behaviour is far from the simulated distribution of the ideal geometry.
For a perfectly aligned cavity, the electric field amplitude is expected to display a single, well-defined peak at the resonant frequency of the fundamental mode, centred within the cavity. Instead, the first measurement planes show non-negligible field magnitude at frequencies where no field should be present (showing a triangular shape in the plots), and the resonant frequency shifts progressively toward higher values as the plane index increases.

\subsection{CST simulation results}
\label{ss:CST_simulation_results}

\begin{figure}[htb]
    \centering

    \begin{subfigure}{0.32\textwidth} \includegraphics[width=\linewidth]{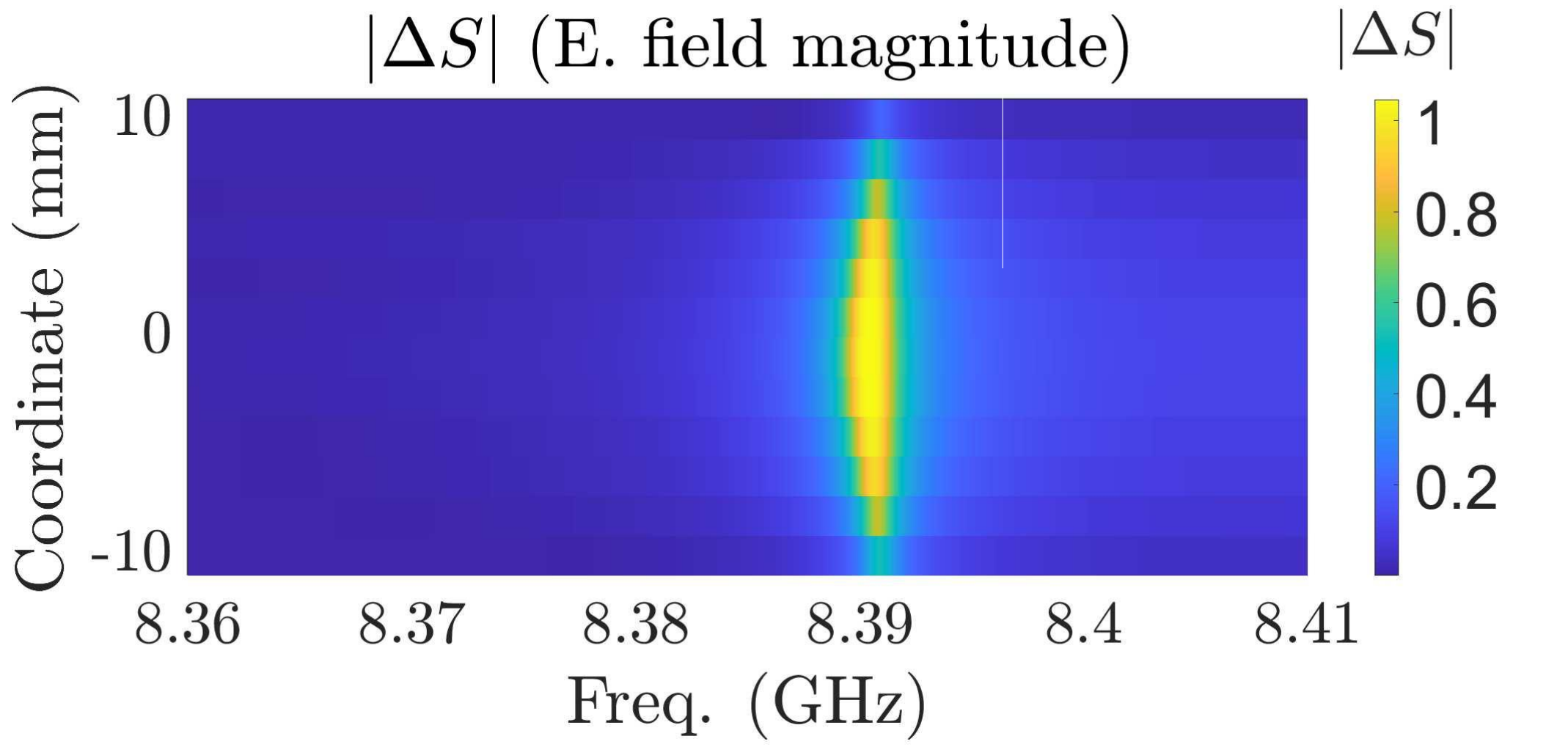}
        \caption{$\varepsilon_r = 9.8$, $r_b = 1.0~\mathrm{mm}$.}
    \end{subfigure}
    \hfill
    \begin{subfigure}{0.32\textwidth}   \includegraphics[width=\linewidth]{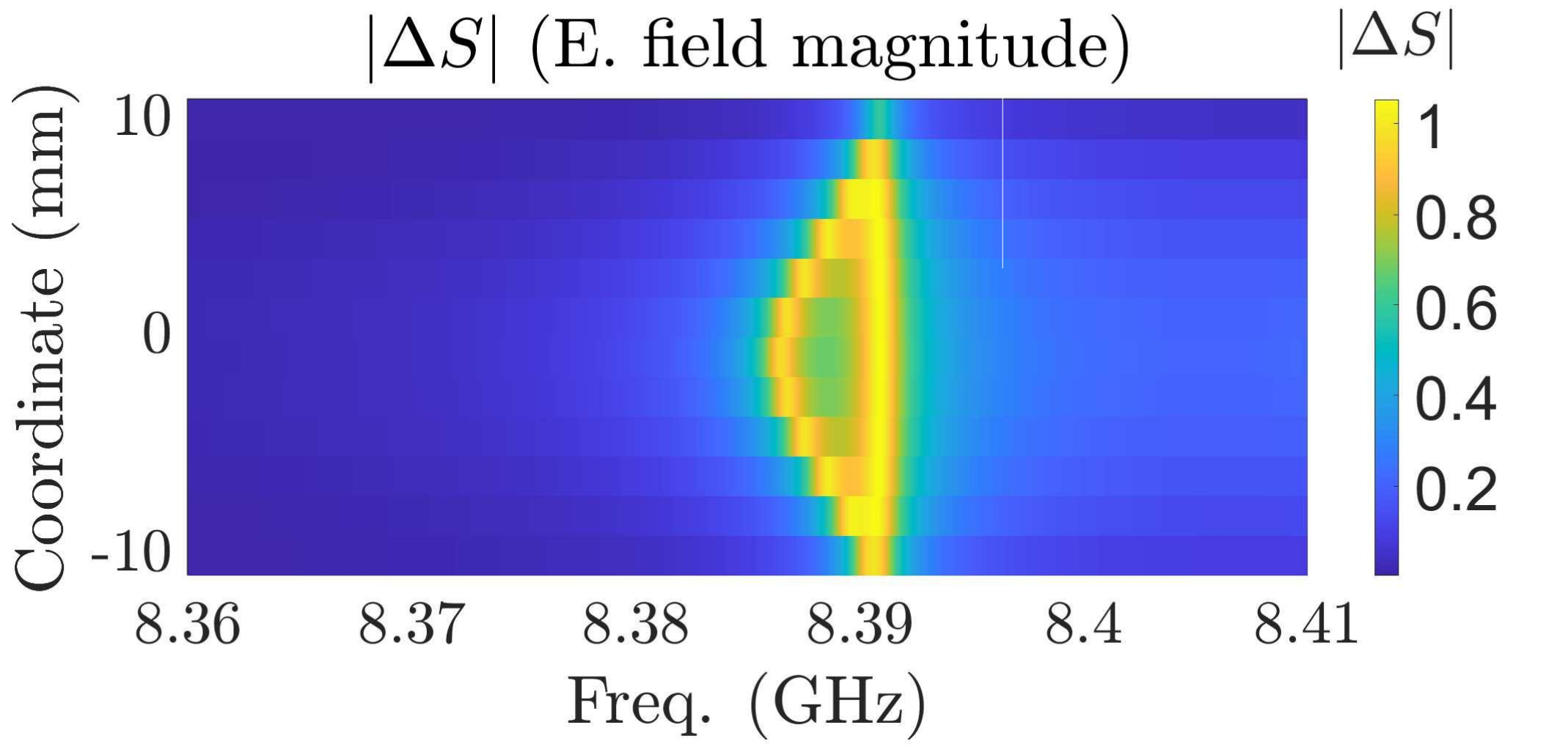}
        \caption{$\varepsilon_r = 9.8$, $r_b = 1.5~\mathrm{mm}$.}
    \end{subfigure}
    \hfill
    \begin{subfigure}{0.32\textwidth} \includegraphics[width=\linewidth]{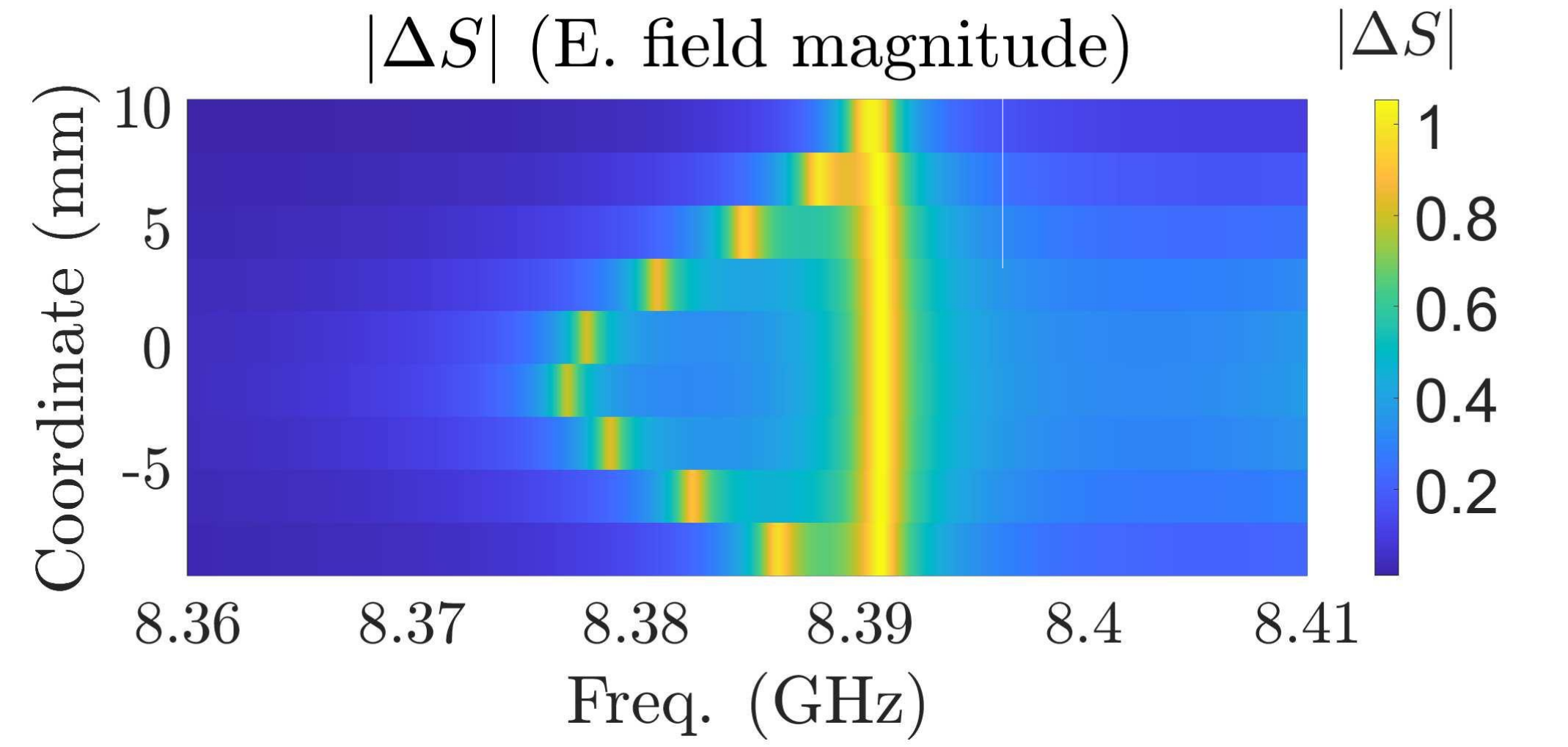}
        \caption{$\varepsilon_r = 9.8$, $r_b = 2.0~\mathrm{mm}$.}
    \end{subfigure}

    \vspace{0.4cm}

    \begin{subfigure}{0.32\textwidth} \includegraphics[width=\linewidth]{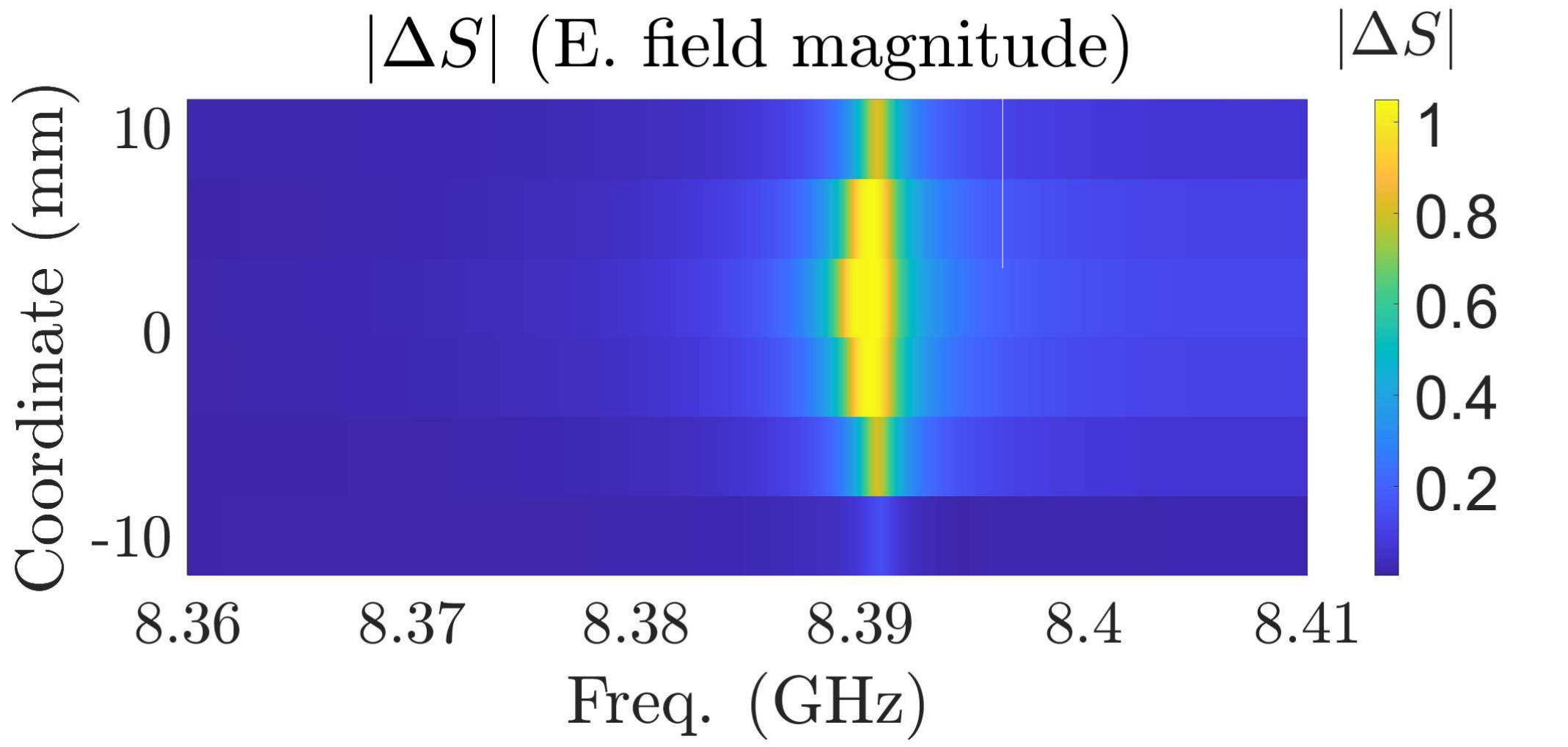}
        \caption{$\varepsilon_r = 25$, $r_b = 1.0~\mathrm{mm}$.}
    \end{subfigure}
    \hfill
    \begin{subfigure}{0.32\textwidth} \includegraphics[width=\linewidth]{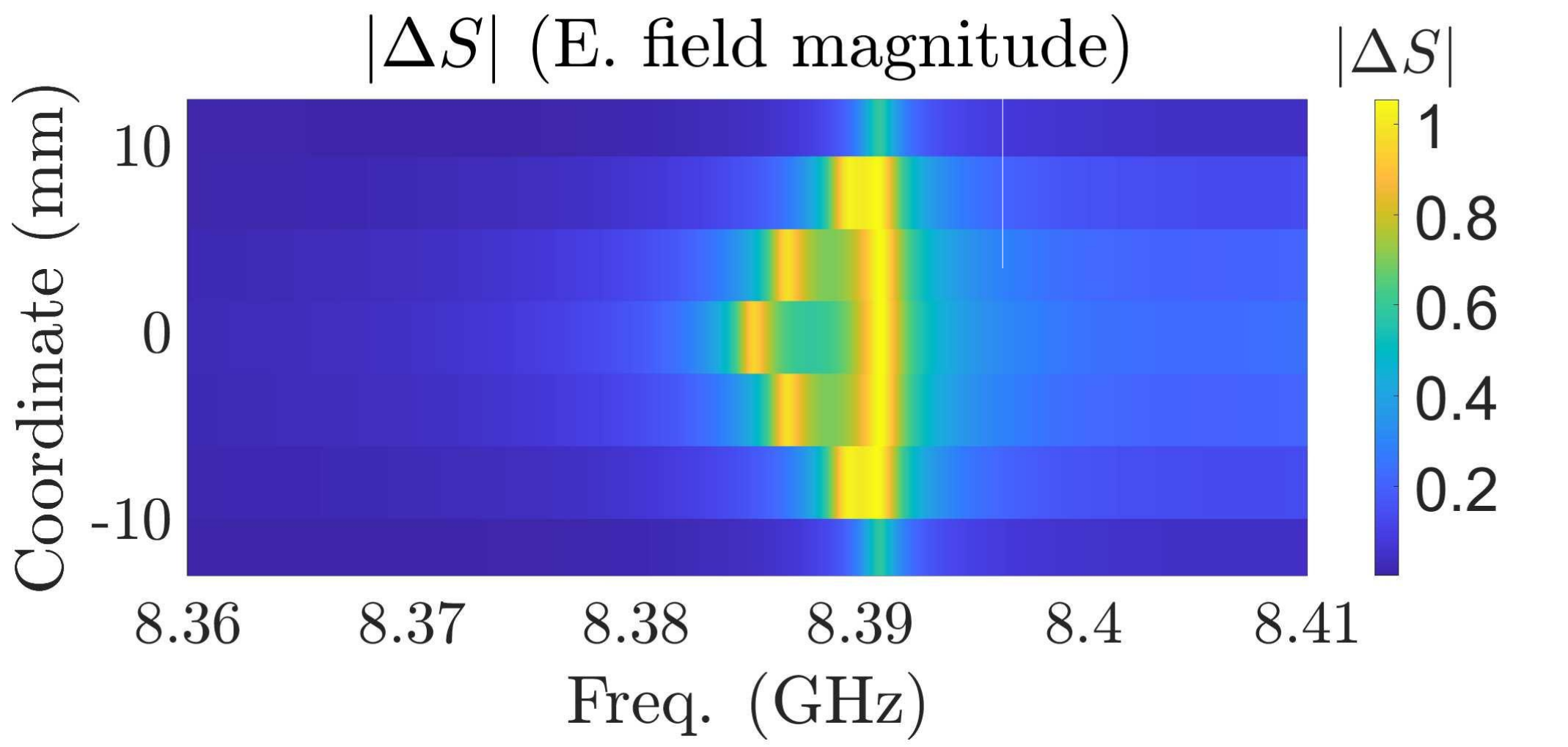}
        \caption{$\varepsilon_r = 25$, $r_b = 1.5~\mathrm{mm}$.}
    \end{subfigure}
    \hfill
    \begin{subfigure}{0.32\textwidth}
    \includegraphics[width=\linewidth]{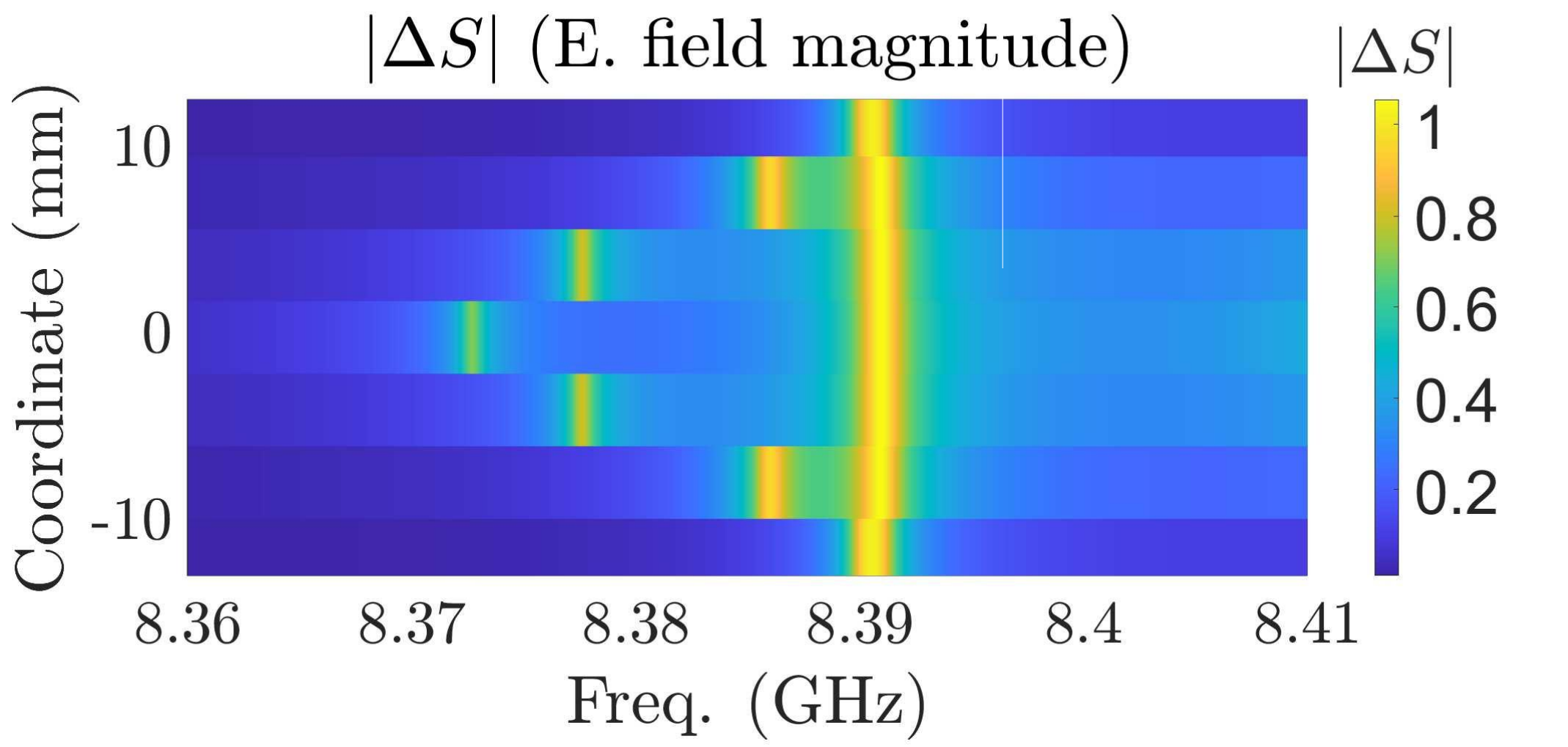}
        \caption{$\varepsilon_r = 25$, $r_b = 2.0~\mathrm{mm}$.}
    \end{subfigure}

    \caption{Simulated $|\Delta S|$ ($\propto$ E-field) in frequency space distribution for different bead radii and relative permittivities.}
    \label{fig:radius_eps_variation}
\end{figure}

To complement the experimental measurements and facilitate the interpretation of the observed field distributions, a comprehensive set of electromagnetic simulations was performed using CST Studio Suite \cite{CST}. Given the complexity of the non-ideal cavity, multiple simulation runs were necessary: first, establishing an ideal baseline; and second, systematically introducing controlled physical perturbations and mechanical misalignments. These simulations serve three purposes: first, to provide a theoretical reference for the expected cavity behaviour in both unperturbed and perturbed configurations and to explain the unexpected performance observed experimentally; second, to assess the sensitivity of the cavity response to key parameters, such as bead size and permittivity, thread dimensions, and possible cavity misalignments, in order to properly prepare the final run; and third, to assess the overall viability of the experiment. Figure~\ref{fig:radius_eps_variation} presents the results of these bead-pull simulations.
From left to right, the bead radius $r_b$ is increased from $1$ to $2$~mm, while from top to bottom, the relative permittivity takes the values $\varepsilon_r = [9,25]$.

The results indicate that the bead radius has a large effect on the field distribution: increasing the radius leads to significantly larger lobes in the field map. While the bead's relative permittivity $\varepsilon_r$ also influences the field, for a fixed radius, higher $\varepsilon_b$ produces slightly larger lobes; its effect is much less pronounced. Consequently, while selecting a bead with low permittivity can be beneficial for being sensitive enough to field changes, the primary factor to control is the bead size.

These observations confirm that the origin of the lobe structure from the measurements (see Figure~\ref{fig:BP-Preeliminary1}) arises from excessive perturbation. This behaviour is fully consistent with theoretical expectations from Eq.~\ref{eq:beadpull_param}.

Examination of the simulated electric field distribution in Figure~\ref{fig:radius_eps_variation}, indicates that the perturbation induced by the simulated bead is smaller than for the experimental results in Figure~\ref{fig:BP-Preeliminary1}. The simulated field patterns are noticeably less distorted than those observed experimentally, supporting the hypothesis that the measurements were affected by an excessively large perturbation, which explains the apparent presence of forbidden modes.

\begin{figure}[htb]
    \centering

    \begin{subfigure}[t]{0.4\textwidth}
        \centering
        \includegraphics[width=\linewidth]{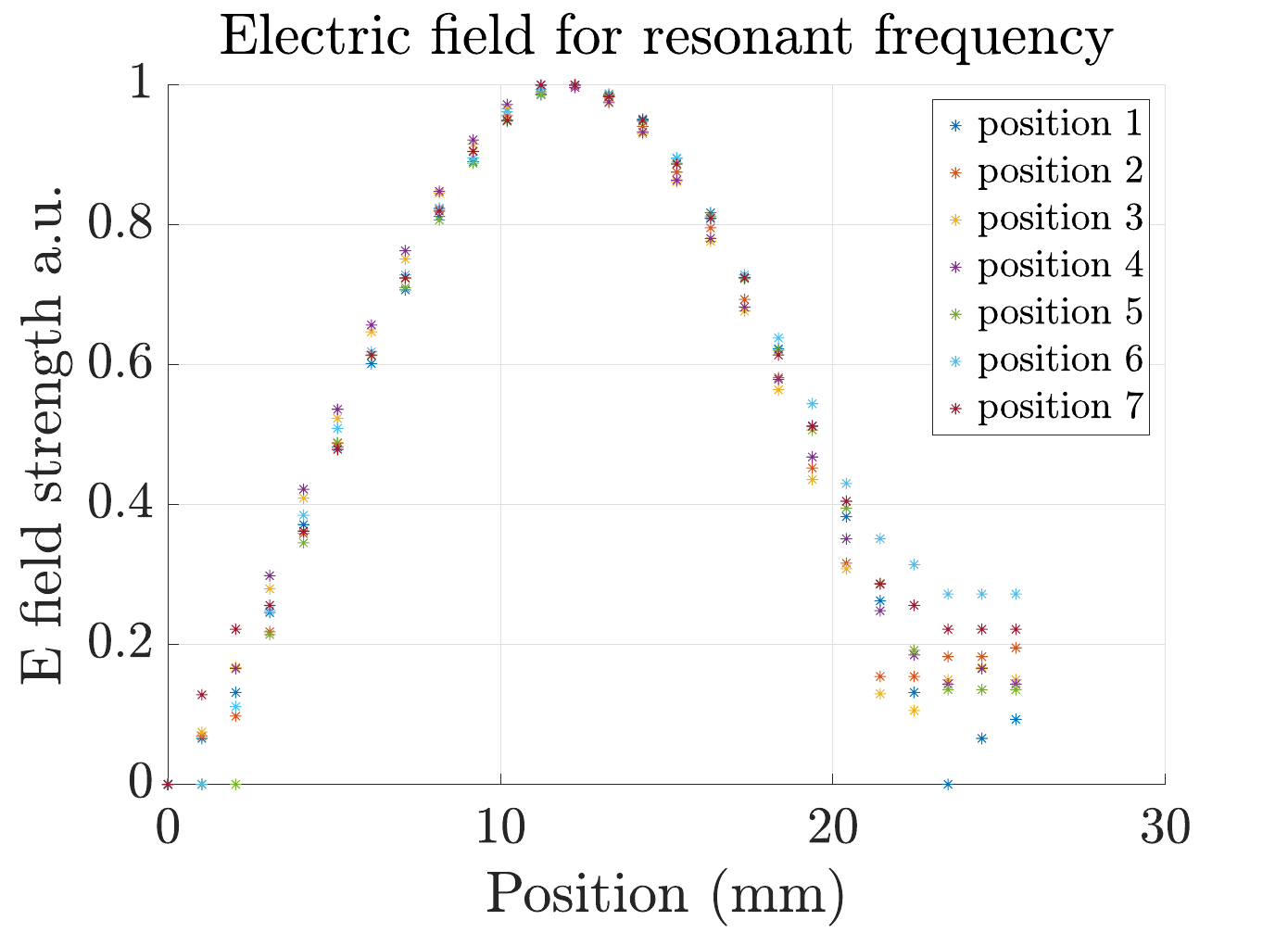}
        \caption{}
        \label{fig:BP-Preeliminary2}
    \end{subfigure}
    \begin{subfigure}[t]{0.4\textwidth}
        \centering
        \includegraphics[width=\linewidth]{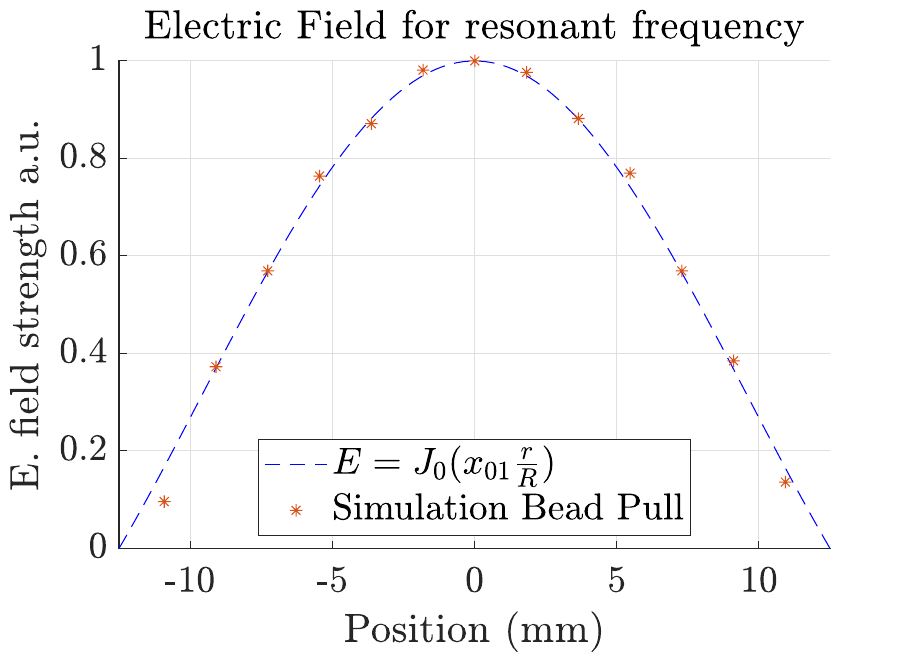}
        \caption{}
        \label{fig:CST-aligned-fshift}
    \end{subfigure}

    \caption{Comparison between the measured and simulated two-dimensional electric-field distributions obtained from the bead-pull technique, both normalised for comparison: (a) Measured two-dimensional E-field distribution in the seven planes shown in Figure~\ref{fig:planes_studied} of bead-pull response. (b) Simulated two-dimensional E-field distribution from bead-pull simulation (red colour points). The analytical solution has been included (dashed blue colour line) to demonstrate agreement with theoretical expectations.}
    \label{fig:comparison-measured-simulated}
\end{figure}

Calculating the electric field via the first equation of \ref{eq:B-P} shows, however, that, for the resonant frequency, a reasonable distribution that seems to be unaffected by the phenomena discussed earlier is still obtained. The results from the bead-pull measurements are shown in Figure~\ref{fig:BP-Preeliminary2} for all the planes at the resonant frequency.

As a matter of fact, the field amplitude at the nominal resonant frequency remains consistent with theoretical expectations, and the profiles across different planes show good overall agreement. Some anomalous and random variations were observed near the cavity edges. Those anomalies can be attributed to the bead coming into contact with the cavity wall at certain positions, producing locally inaccurate measurements.

To properly benchmark the experimentally obtained results, the same bead-pull extraction method was applied to CST simulation data. The results are presented in Figure~\ref{fig:CST-aligned-fshift} including the theoretical expectation of field distribution for a cylindrical cavity. The normalised field profiles observed in Figures~\ref{fig:BP-Preeliminary2} and \ref{fig:CST-aligned-fshift} show good mutual agreement, confirming that the method reliably recovers the expected field distribution.  Comparison of the normalised field values obtained from the resonant frequency shifts with the theoretical predictions for a cylindrical cavity shows overall good agreement. This confirms that the problem associated with the excessive perturbation discussed previously regards primarily the apparent presence of forbidden modes, while the shape of the field at resonant frequency has been recovered and provides an appropriate characterisation. Furthermore, since the analytical solution is derived for an ideal cylindrical cavity, it is important to note that opening the cavity breaks the perfect cylindrical symmetry, which can also contribute to the observed discrepancies between the simulation results and the theoretical expectation in the tails of the distribution in Figure~\ref{fig:CST-aligned-fshift}.

\subsection{Misalignment effects}
\label{ss:CST_simulation_results_misalignment}

After verification of the method on aligned cavity configurations, a series of simulations, including deliberately misaligned configurations, was performed. Such simulations are crucial because small mechanical imperfections or misalignments of the cavity geometry, which are inevitable in practical setups, can significantly influence the field distribution, altering the form and quality factor as discussed in subsection~\ref{sss:Misalignment effects}. By systematically studying unaligned cavities, this study provides valuable insight into mode localisation. In this analysis, two distinct types of misalignment were specifically considered in order to quantify their effects and guide the final experimental setup. In this work, the first and most illustrative scenario considered is the angular misalignment $\theta_x$ (see Figure~\ref{fig:CylCav_ThetaXtilt_3Dmodel}), which has a strong impact on the form factor because of the induction of a mode localisation effect. Rather than being symmetrical with respect to the cavity centre, the electromagnetic mode becomes confined to specific ends of the cavity. Results for the second resonant mode are also present, as these may prove valuable for future experimental investigations.

\begin{figure}[htb]
\centering
\begin{subfigure}{0.3\textwidth}
\centering
\includegraphics[width=\linewidth]{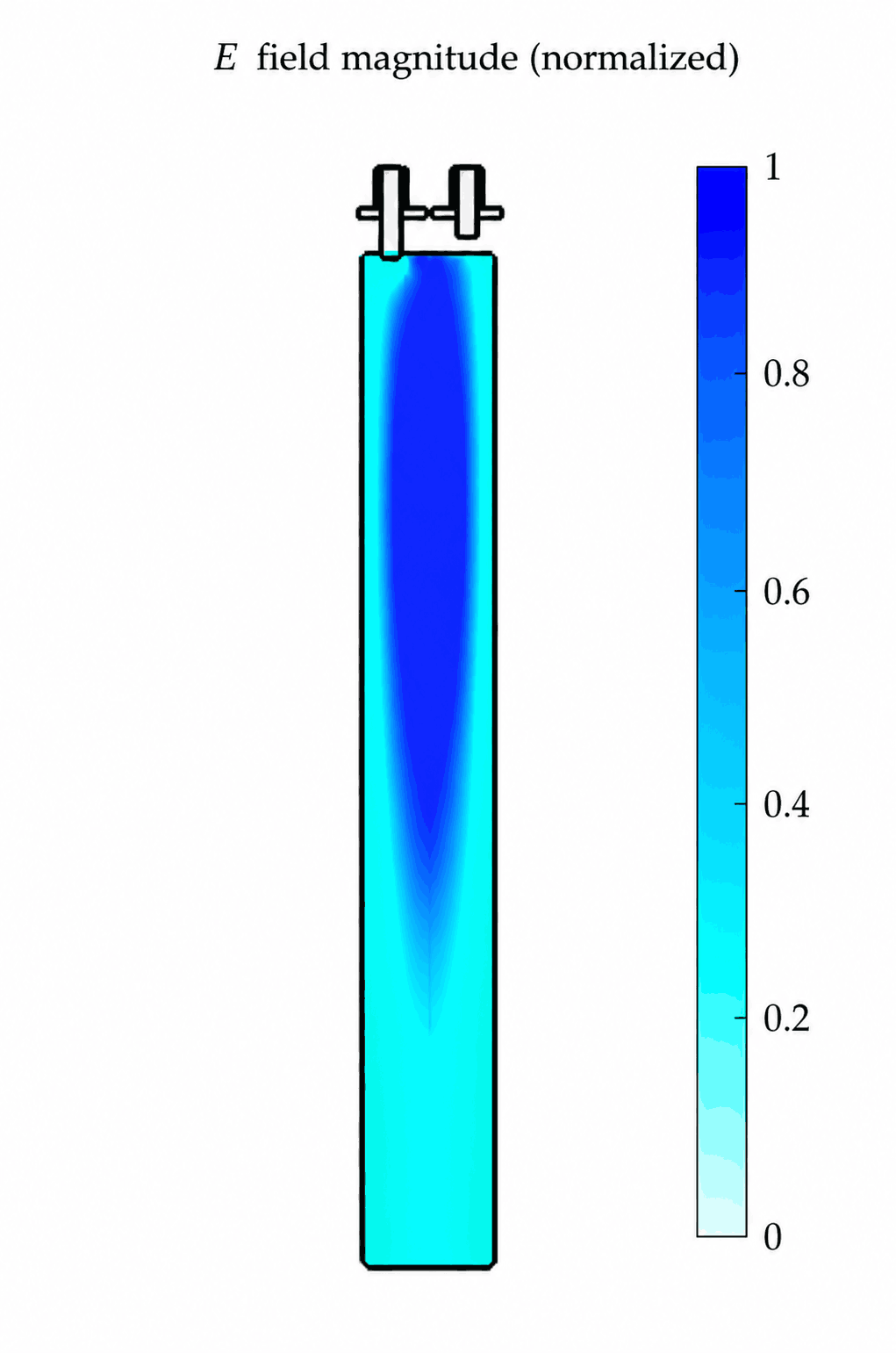}
\caption{}
\label{fig:misalignment_mode0}
\end{subfigure}
\begin{subfigure}{0.3\textwidth}
\centering
\includegraphics[width=\linewidth]{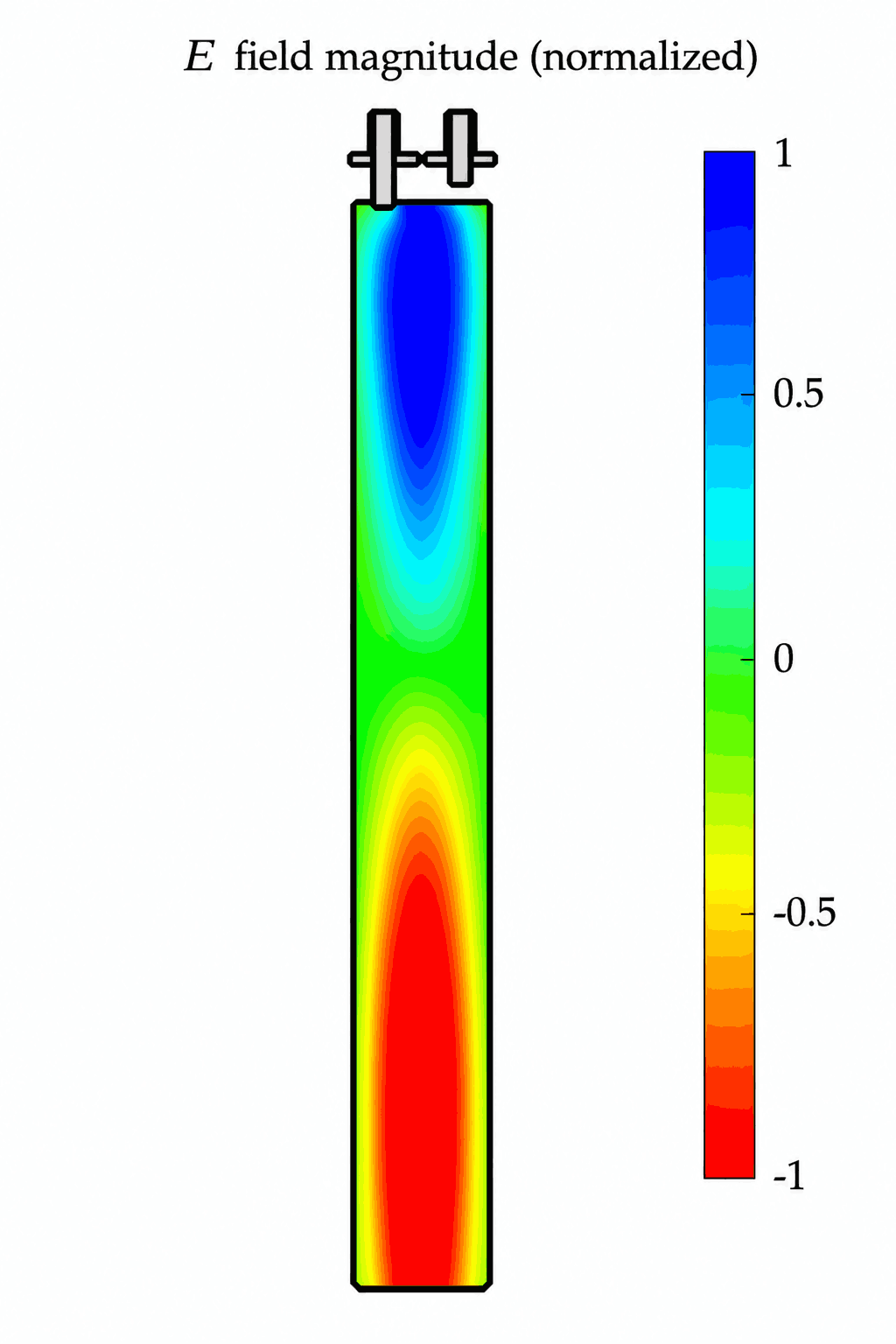}
\caption{}
\label{fig:misalignment_mode1}
\end{subfigure}
\caption{Simulated normalised electric field distributions in the cavity cross-section considering an axial misalignment angle $\theta_x$ with a varying separation gap (refer to Figure~\ref{fig:CylCav_ThetaXtilt_3Dmodel}): (a) Fundamental resonant mode $\text{TM}_{010}$, displaying a unipolar field distribution along the longitudinal axis. (b) Higher-order mode $\text{TM}_{011}$, showing the characteristic field reversal (bipolar profile) with opposing phase regions along the cavity axis.}
\label{fig:misalignment_modes}
\end{figure}

Using the CST software, it is possible to illustrate that for the fundamental mode, a positive tilt, that is where the cavity aperture is narrower near the ports, causes the field to concentrate in that region and diminish almost entirely toward the back of the cavity. This localisation becomes increasingly pronounced as the misalignment grows. In contrast, the second mode exhibits a particularly notable behaviour: the central field node, which remains fixed at the cavity midpoint in an ideal configuration, shifts toward the ends of the cavity as the misalignment increases, as shown in Figure~\ref{fig:misalignment_modes}. 
The field is presented normalised, as the objective is the study of mode localisation, and therefore the main interest is its geometry. 

Exploiting this behaviour, a novel alignment method can be developed. By iteratively adjusting the alignment until the measured field amplitude reaches a minimum for the second mode and simultaneously maximising the field amplitude in the same point for the fundamental mode, the alignment can be determined and corrected in real time without requiring a full bead-pull mapping, significantly reducing measurement time. This represents an original contribution as a potential rapid in situ alignment determination method. This is critical for experimental setups where mechanical adjustments can be validated quickly and reliably.

\begin{figure}[htb]
    \centering
    \begin{subfigure}{0.35\textwidth}
        \includegraphics[width=\linewidth]{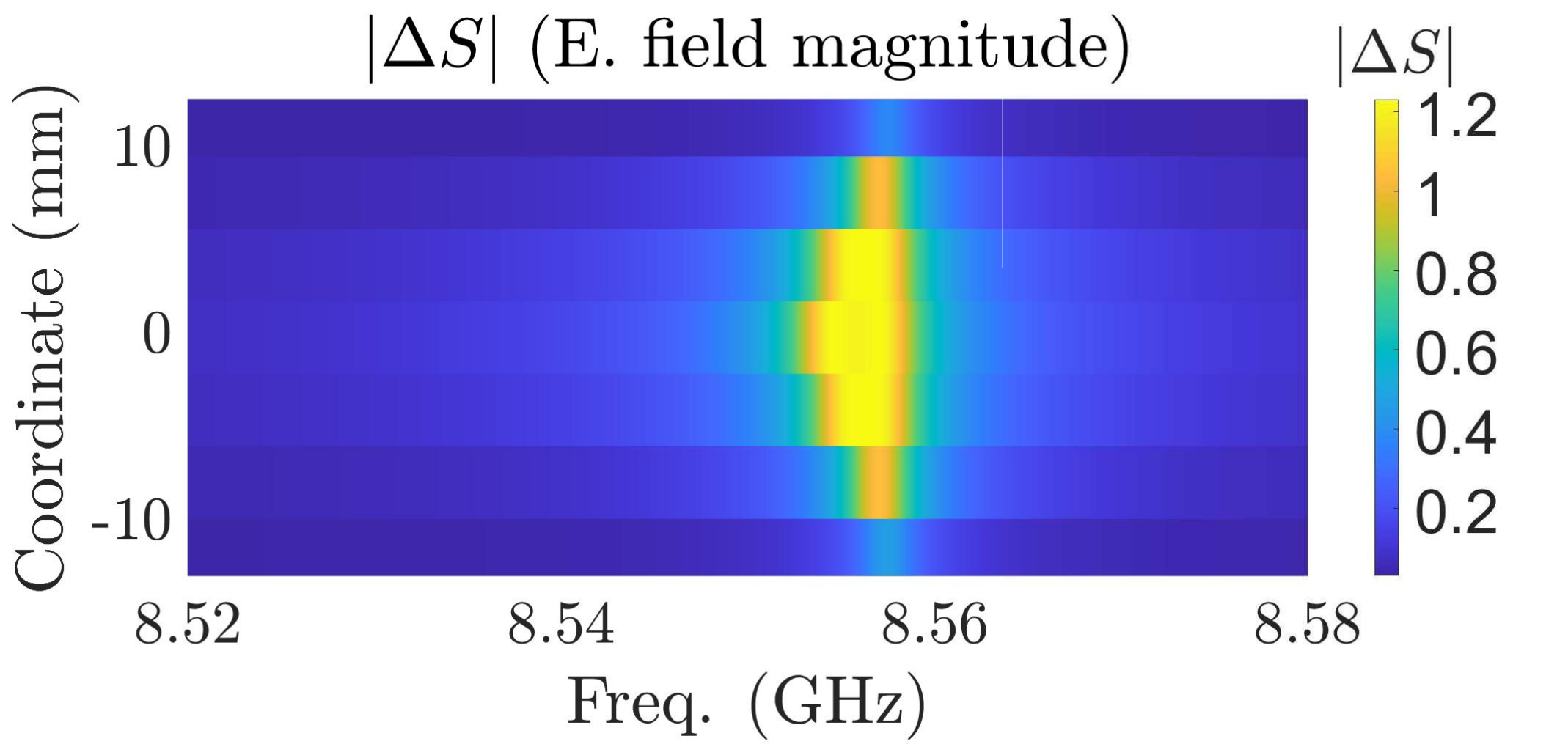}
        \caption{}
    \end{subfigure}
    \begin{subfigure}{0.35\textwidth}
        \includegraphics[width=\linewidth]{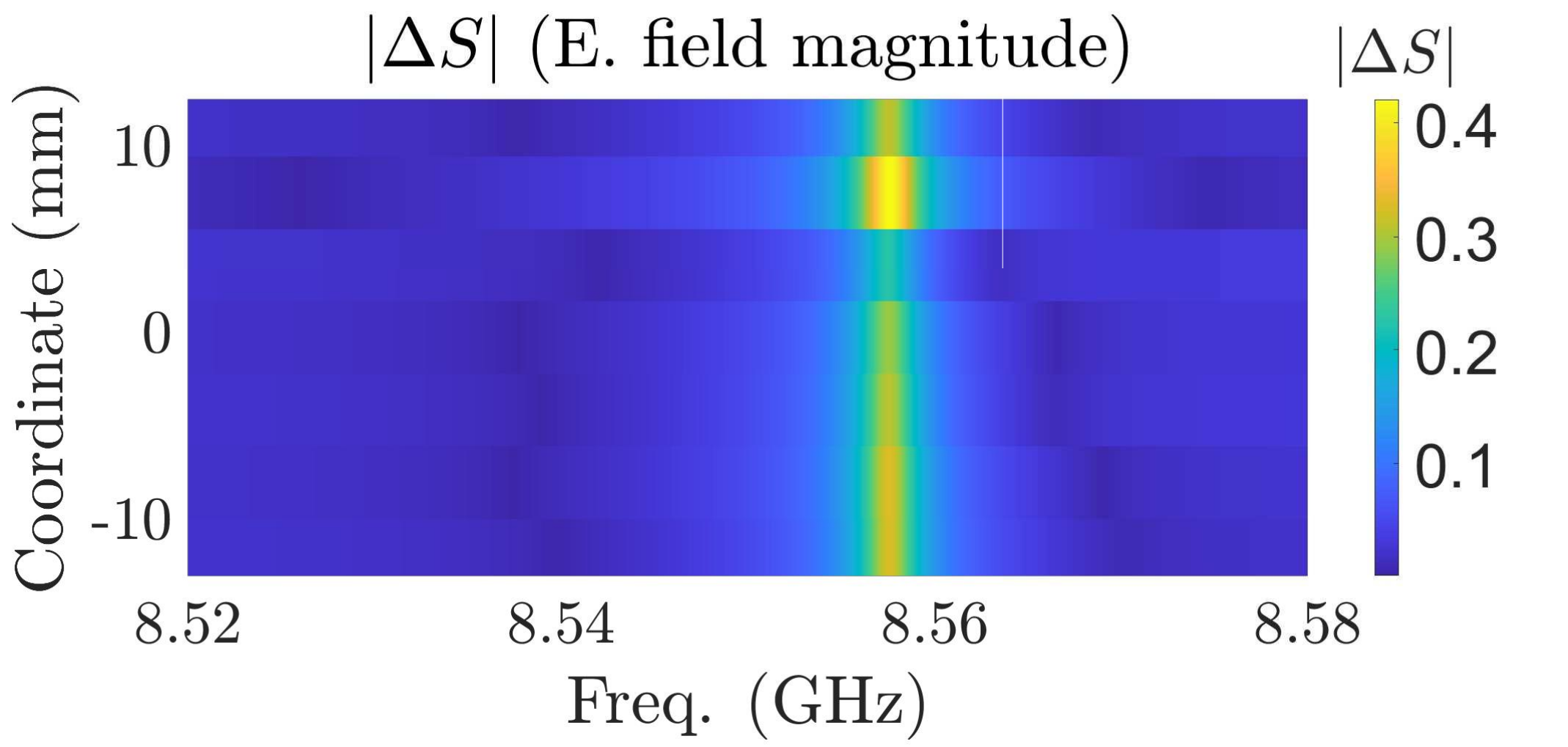}
        \caption{}
    \end{subfigure}
    \caption{Simulated $|\Delta S|$ ($\propto$ E-field) in frequency space for a misaligned cavity with $\theta_x = 2^\circ$ for two different planes described in Figure~\ref{fig:planes_studied}: (a) Position 1 and (b) Position 7.}
    \label{fig:efield_planes}
\end{figure}

To investigate whether a bead-pull measurement could reproduce these effects, two bead-pull simulations were performed for two positions: one close to the ports and one further away, symmetrically distributed around the centre for a tilt of $ \theta_x = 2^\circ$. The results of these studies are shown in Figure~\ref{fig:efield_planes}, where it can be observed that the field with the highest plane index becomes negligible as the tilt increases, while the field near the ports keeps the expected shape but with much higher intensity.
This provides a practical method to determine whether cavity misalignment is present, offering an efficient means to identify misalignment and assess whether mode localisation is occurring within the cavity.

\begin{figure}[htb]
    \centering
    \includegraphics[width=0.6\textwidth]{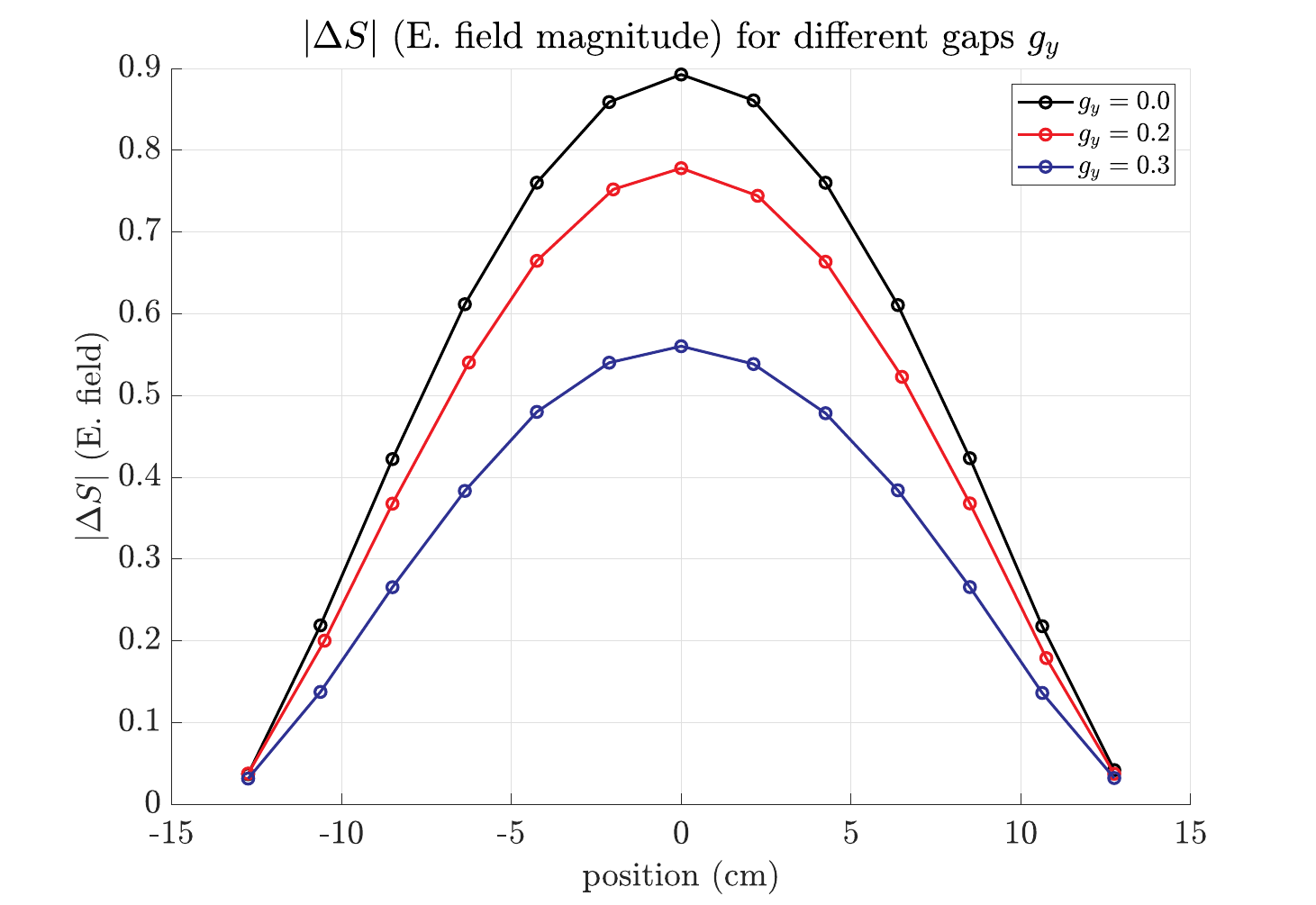}
    \caption{Simulated $|\Delta S|$ ($\propto$ E-Field) as a function of radial position for position 3 in \ref{fig:planes_studied} under different axial gaps $g_y$ (see Figure~\ref{fig:CylCav_gYtilt_3Dmodel}).}
    \label{fig:gy_radial}
\end{figure}

A second misalignment has been studied, corresponding to a relative displacement of one of the cavity halves along the axial direction, denoted by $g_y$ (see Figure~\ref{fig:CylCav_gYtilt_3Dmodel}). The cavity behaviour was analysed for three values of the axial gap, one of which represents the perfectly aligned configuration. This type of misalignment was studied in detail because, as shown in Figure~\ref{fig:CylCav2_fr_Q0_C_and_FoM_vs_Tilts_CST}, even a modest axial offset induces a reduction in the quality factor, making it an important effect to characterise and control. While this degradation can be directly quantified through VNA measurements, bead pull measurements are required to investigate whether the loss of quality factor is associated with modifications of the electromagnetic field distribution. Therefore, the bead pull technique provides insight into the impact of the misalignment on the cavity mode structure and allows the origin of the observed quality factor reduction to be characterised. Simulation results for the electric field in the same plane (Position 3) for several $g_y$ values are presented in Figure~\ref{fig:gy_radial}.

From the simulations, it was observed that the overall field geometry is not strongly affected by small axial displacements. However, for larger values of the gap, a reduction in the relative field magnitude at the cavity centre is observed. This again confirms that this type of misalignment can be determined using bead pull methods by comparing the normalised field distribution against a reference profile obtained from a known, correctly aligned configuration or by monitoring the relative field level at a given position in the centre of the cavity.

\section{Conclusions and prospects}
\label{s:Conclusions}

In this work, the design, mechanical implementation, and comprehensive cryogenic characterisation of the RADES-VORTEX tunable haloscope used in the RADES collaboration have been presented. Built upon the innovative vertical-cut frequency tuning technique, this electromagnetic resonator overcomes one of the most critical bottlenecks in high-frequency axion dark matter searches: achieving a wide, continuous tuning range without degrading the cavity internal quality factor through the insertion of lossy dielectric or metallic components.

The main milestones and results demonstrated in this study are summarised below:
\begin{itemize}
    \item \textit{Loss-less Tuning Mechanism Architecture:} By splitting the elongated cylindrical cavity symmetrically along its longitudinal plane parallel to the $TM_{010}$ mode electric field lines, power radiation into the resulting gap was effectively minimised. This architecture demonstrated an experimental continuous tuning range of approximately $800$~MHz centred around $8.5$~GHz. 
    \item \textit{Cryogenic and Thermalisation Performance:} Structural validation was systematically performed across multiple temperature stages, culminating in millikelvin characterisation inside a dilution refrigerator system. The mechanical sliding mechanism driven by JPE nanopositioners exhibited satisfactory alignment stability. Automated step sequences demonstrated rapid thermal recovery times (approximately $2$~minutes for $100$~steps), proving that the friction and heat dissipation generated by the mechanical movement introduce acceptable dead time for preliminary data-taking campaigns. However, thermalisation must still be improved in a setup in which - instead of a linear amplifier - a quantum-limited amplifier such as a TWPA is employed.
    \item \textit{Dynamic Coupling Recoupling System:} To satisfy the strict sensitivity requirements of the axion scanning rate, a dedicated, automated antenna readjustment system was implemented. Employing a secondary nanopositioner controlling a movable semi-rigid coaxial monopole port, the critical coupling factor was successfully adjusted and maintained near optimal overcoupled conditions ($\beta \simeq 2$) throughout the whole frequency sweep, compensating for any thermal contraction effect.
    \item \textit{Electromagnetic Profile Mapping via Bead-Pull:} Utilising the open nature of the split-housing design, an experimental mapping of the internal electric field distribution using the bead-pull perturbation method was successfully performed. The experimental results show a high degree of agreement with the ideal Bessel distributions modelled via CST Studio Suite simulations. Furthermore, evaluating the frequency-shift estimator against standard scattering parameters allowed us to rule out significant mode-mixing or field localisation anomalies caused by mechanical tolerances or misalignment. Finally, a method based on the electric field study of the second resonant mode has been developed for the identification of misalignments at the cavity housing.
\end{itemize}

The successful validation of the VORTEX haloscope confirms that the continuous vertical-cut mechanism is mechanically robust, thermally compatible with millikelvin environments, and electromagnetically efficient. On the other hand, a few issues have been identified: thermalisation of the cavity needs improvement in order to employ the cavity together with quantum-limited amplifiers, such as TWPAs. Also, superconducting coatings could be added to improve the quality factor of the final VORTEX version.

\begin{figure}[htbp]
\begin{center}
\includegraphics[width=1\textwidth]{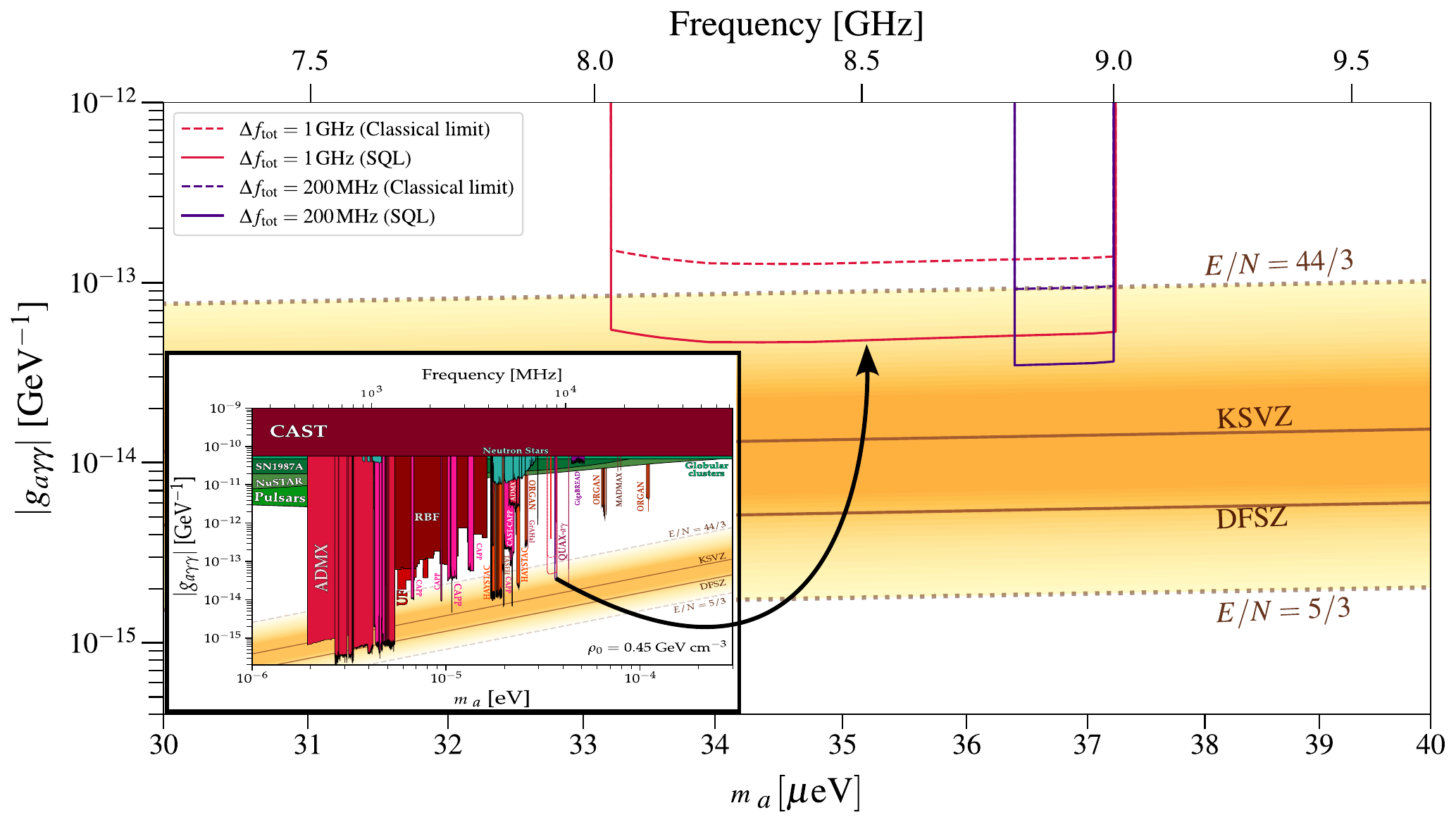}
\caption{\label{fig:sensitivity_projection} Exclusion limit projection and sensitivity reach for the axion-photon coupling constant $g_{a\gamma\gamma}$ as a function of the axion mass $m_a$ derived from the experimental results obtained with the VORTEX tunable haloscope. The projected sensitivity curve is calculated using the measured cryogenic quality factor ($Q_0$), the tuned volume ($V$), and the simulated form factor ($C$) across the $8-9$~GHz operational frequency range. The projection assumes a baseline integration time ($\Delta t$) under a $12$~T external static magnetic field solenoid magnet, compared against standard QCD axion model bands (KSVZ and DFSZ). At the inset, the existing exclusion limits from other haloscope experiments are shown \cite{AxionLimits}.}
\end{center}
\end{figure}

In summary, we have characterised and proved the applicability of this haloscope for a data-taking campaign within the RADES collaboration, utilising a high-field $12$~T solenoid magnet to search for axion dark matter. Figure~\ref{fig:sensitivity_projection} shows a projection of the axion-photon sensitivity considering the following parameters: a confidence level of $90$~$\%$ ($SNR = 1.28$), a total integration time of $\Delta t_\mathrm{tot} = 30$~days, and a coupling factor $\beta = 2$.
For the system temperature, $T_\mathrm{sys} = T_\mathrm{cav}+T_\mathrm{amp}$, it was assumed a cavity temperature of $T_\mathrm{cav} = 20 \, \mathrm{mK}$, consistent with the values achieved for the cavity at base temperature. And two different scenarios for the amplifier noise temperature: a first one implementing a LNA, operating at the classical limit, with $T_\mathrm{amp} = 4.62$~K and another one operating at the standard quantum limit, implementing a TWPA, for which $T_\mathrm{amp} = 452$~mK. For the sake of completeness, a measurement run with the same acquisition time but covering only $200$~MHz has been included (pink lines), showing a clear improvement in sensitivity due to the reduced number of measurement points. Additionally, the sensitivity reach could be increased by the implementation of HTS tapes along the cavity inner walls to enhance $Q$, adopting identical percentage reduction rates in $Q_0$ across the tuning range as experimentally observed in the copper prototype.

\acknowledgments

This work was performed within the RADES collaboration. We thank our colleagues for their support. It is a pleasure to thank the entire workshop team of the Max Planck Institute for Physics for their support in the setup of this experiment. We would especially like to thank D. Kreikemeyer, C. Gooch, A. Ivanov, J. Arcila-Maldonado, G. Obermüller, H. Byun, T. Ortmann, Z. Yang, and E. Gabbrielli for their help during the installation in cryogenic environments.  We also acknowledge a joyful collaboration with the MPP MADMAX team, which gave access to their 4K cryostat infrastructure and stepper motors. The authors acknowledge funding through the European Research Council for grant ERC-2018-StG-802836 (AxScale project), the grant ERC-SYG 101118911  (DarkQuantum), and the Lise Meitner program of the Max Planck Society. This research was also supported by Deutsche Forschungsgemeinschaft (DFG) through Grant No. 532766533 (QUANTERA) and under Germany’s Excellence Strategy – EXC 2094 – 390783311.  
The work also profited from discussions in the framework of the COST Action COSMIC WISPers CA21106, supported by COST (European Cooperation in Science and Technology).

\appendix
\section{$LN_2$ measurements}
\label{AppA:LN2measurements}

\begin{figure}[h]
\centering
\begin{subfigure}[b]{0.2\textwidth}
         \centering
         \includegraphics[width=0.93\textwidth]{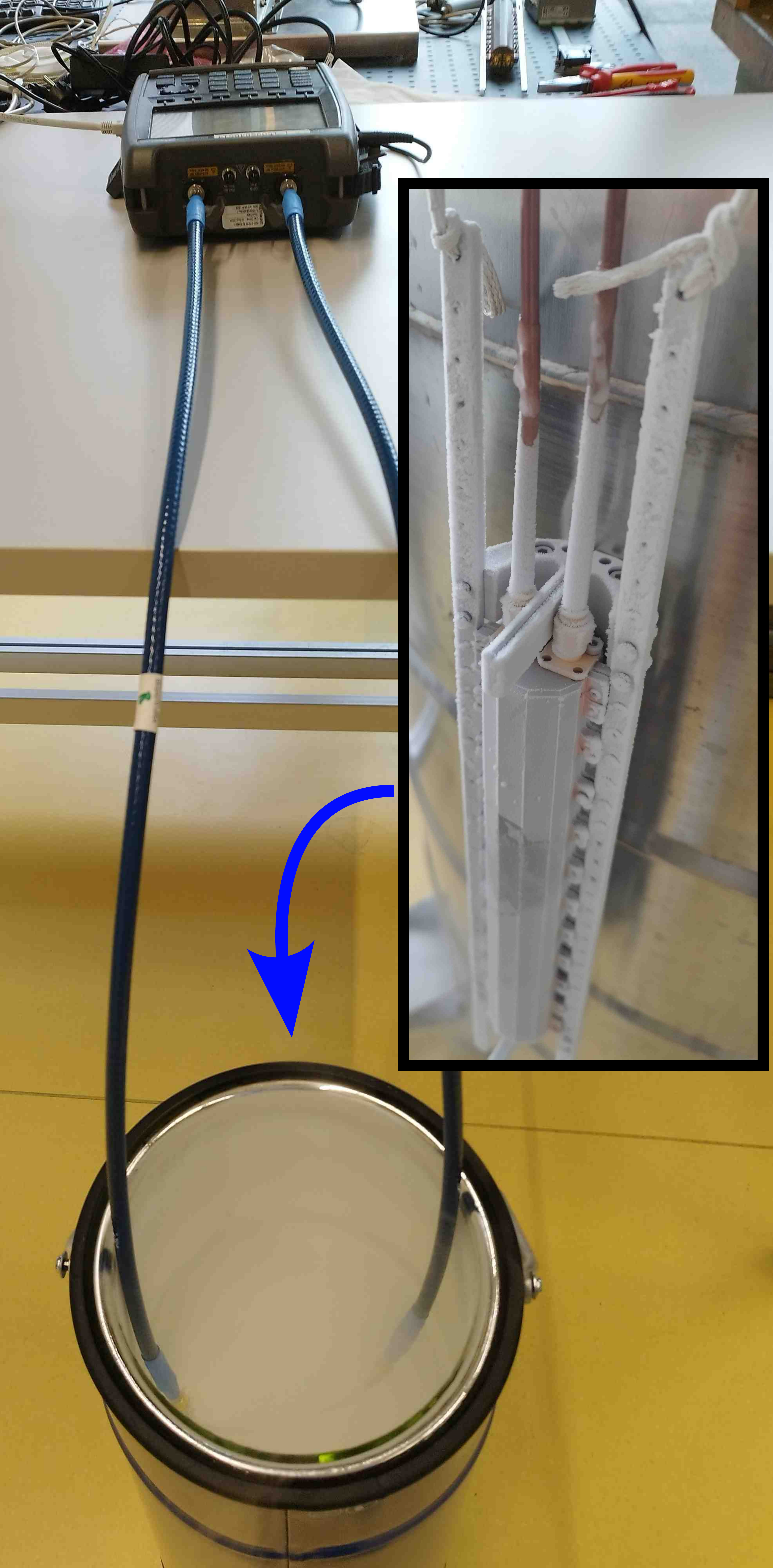}
         \caption{}
         \label{fig:CylCav_v1_LN2_measurements_setup}
\end{subfigure}
\hfill
\begin{subfigure}[b]{0.79\textwidth}
         \centering
         \includegraphics[width=1\textwidth]{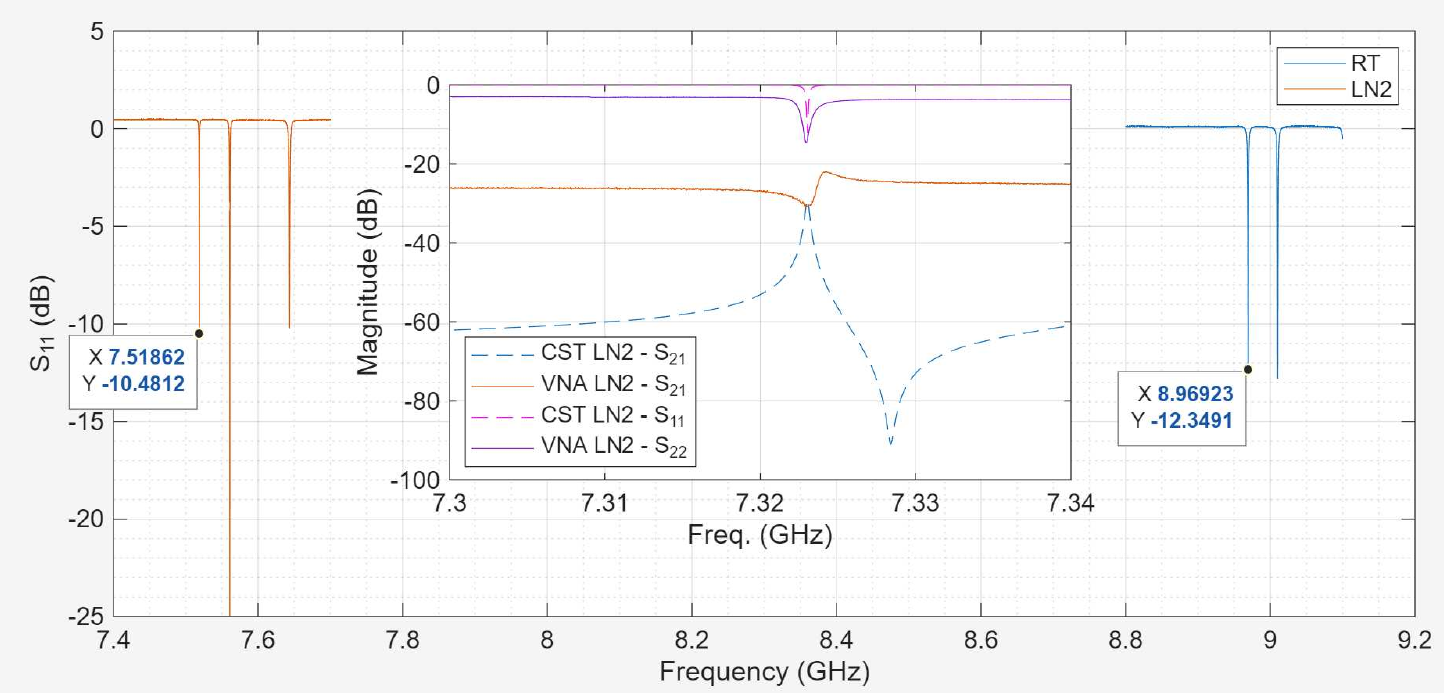}
         \caption{}
         \label{fig:CylCav_v1_LN2_measurements_Sp}
\end{subfigure}
\caption{(a) Characterisation setup of the first version of the haloscope at $\sim77$~K temperatures in a LN$_2$ dewar bath. The inset depicts an image of the cavity when removing it from the liquid nitrogen bath. (b) $S_{11}$ parameter (in dB) versus frequency (in GHz) before (at $300$~K) and after the LN$_2$ immersion (at $77$~K) of the cavity with $gap=0$~mm. The inset shows the transmission and reflection parameters (in dB) versus frequency (in GHz) of the prototype measurement and CST simulation, both at $77$~K conditions with $gap=1.15$~mm.}
\label{fig:CylCav_v1_LN2_measurements}
\end{figure}

Additional measurements were conducted for the first cavity version in a liquid nitrogen dewar bath. Figures~\ref{fig:CylCav_v1_LN2_measurements_setup} and \ref{fig:CylCav_v1_LN2_measurements_Sp} show an image of the measurement process in this LN$_2$ bath and the resulting response, respectively.
The temperature reached in this setting is approximately $77$~K. Liquid nitrogen has a specific relative permittivity of $\varepsilon_r = 1.538 \pm 0.025$ \cite{Hosking:1993}, which has been considered in simulation (involving a reduction in resonant frequency). The electrical conductivity of copper at these temperatures has also been taken into account, which turns out to be $\sigma_\mathrm{Cu}^{77K} = 5\times10^8$~S/m \cite{krainz:1997}. The resonant frequency of the $TM_{010}$ mode in a cylindrical cavity is inversely proportional to the square root of $\varepsilon_r$ ($f_r\propto1/\sqrt{\varepsilon_r}$) \cite{pozar2012microwave}. Therefore, when a resonant cavity with $f_r=9$~GHz is immersed in liquid nitrogen, $f_r$ drops to $7.526$~GHz, which must be the maximum value for the manufactured prototype ($gap=0$~mm) under ideal conditions.

In a first measurement, closing the cavity, $f_r\approx8.969$~GHz was obtained at room temperature. If the above formula is applied to calculate the value at LN$_2$, we find $f_r=8.969/\sqrt{1.538 \pm 0.025} = 7.2321 \pm 0.06$~GHz. Upon immersion in LN$_2$, a value of $f_r=7.519$~GHz was obtained, representing a difference of $+287\pm 60$~MHz. This discrepancy is acceptable and can be attributed to a combination of experimental factors, such as thermal contractions of the structure, minor misalignment during the measurement, and small fluctuations in the effective permittivity ($\varepsilon_r$) induced by the open-air evaporation and bubbling of the nitrogen.

A second measurement was taken using $gap=1.15$~mm, achieved by inserting washers (similarly to the tests depicted in Figure~\ref{fig:CylCav_TuningAtRTwithWashers_Picture}). The result is shown, alongside the simulated case, in the inset of Figure~\ref{fig:CylCav_v1_LN2_measurements_Sp}. The resulting resonant frequency at $77$~K is $f_r=7.323$~GHz, and the unloaded quality factor is $Q_0=24200$, a value $\sim65\%$ higher than that at room temperature for the same gap, although it could be better as observed in the simulation. The value obtained at these temperatures relative to the simulated case is $52\%$, which could be attributed to a lack of thermalisation, minor misalignments, and measurement uncertainty. Furthermore, in this particular case, an antenna with strong coupling has been used, which can increase the measurement uncertainty; this can be easily identified by observing the strong asymmetrical shape of the Lorentzian in the $S_{21}$ measurement parameter (red line in Figure~\ref{fig:CylCav_v1_LN2_measurements_Sp}), which is not the case in the CST simulation curve (blue line), as it has a considerably lower $\beta$ value.

\bibliographystyle{JHEP.bst}
\bibliography{mybibfile.bib}
\end{document}